\documentclass[fleqn,usenatbib]{mnras}

\usepackage{newtxtext,newtxmath}
\usepackage[T1]{fontenc}

\DeclareRobustCommand{\VAN}[3]{#2}
\let\VANthebibliography\thebibliography
\def\thebibliography{\DeclareRobustCommand{\VAN}[3]{##3}\VANthebibliography}

\usepackage{graphicx}	% Including figure files
\usepackage{amsmath}	% Advanced maths commands
\usepackage{color}
\usepackage{tikz}
\usetikzlibrary{positioning}
\definecolor{linkcolor}{rgb}{0,0,0.25}
\hypersetup{
colorlinks=true,        % false: boxed links; true: colored links
linkcolor=blue,    % color of internal links
citecolor=blue,    % color of links to bibliography
filecolor=blue,    % color of file links
urlcolor=blue,      % color of external links
draft=False,
}

\makeatletter
\renewcommand{\@printed}{}
\makeatother

\newcommand{\figurename}{Figure}
\newcommand{\tablename}{Table}
\newcommand{\eqnname}{Equation}
\newcommand{\secname}{Section}
\newcommand{\appenname}{Appendix}

\definecolor{darkgreen}{rgb}{0.0, 0.7, 0.0}

\definecolor{darkblue}{rgb}{0.0, 0., 0.7}

\newcommand{\gaia}{\emph{Gaia}}

\newcommand{\tmass}{\emph{2MASS}}

\renewcommand{\vec}[1]{\ensuremath{\mathbf{#1}}}
\newcommand{\teff}{\ensuremath{T_\mathrm{eff}}}
\newcommand{\logg}{\ensuremath{\log g}}
\newcommand{\xh}[1]{\ensuremath{[\mathrm{#1/H}]}}

\newcommand{\um}{\ensuremath{\mu m}}

\newcommand{\plx}{\ensuremath{\varpi}}
\newcommand{\mas}{\ensuremath{\mathrm{mas}}}
\newcommand{\uas}{\ensuremath{\mu\mathrm{as}}}

\newcommand{\bp}{\ensuremath{G_\mathrm{BP}}}
\newcommand{\rp}{\ensuremath{G_\mathrm{RP}}}
\newcommand{\bprp}{\ensuremath{G_\mathrm{BP}-G_\mathrm{RP}}}
\newcommand{\jh}{\ensuremath{J-H}}
\newcommand{\jk}{\ensuremath{J-K}}

\newcommand{\approptoinn}[2]{\mathrel{\vcenter{
  \offinterlineskip\halign{\hfil$##$\cr
    #1\propto\cr\noalign{\kern2pt}#1\sim\cr\noalign{\kern-2pt}}}}}

\definecolor{spec2paramcolor}{HTML}{007FFF}
\definecolor{param2speccolor}{HTML}{00994D}
\definecolor{spec2speccolor}{HTML}{FF0000}
\definecolor{param2paramcolor}{HTML}{0060BF}
\definecolor{empiricalcolor}{HTML}{6600CC}

\def\figheadfontsize{15}
\def\figboxcorner{0.3cm}
\def\figboxlw{0.5mm}
\def\figboxsep{3pt}
\title[Towards an astronomical foundation model for stars]{Towards an astronomical foundation model for stars with a Transformer-based model}

\author[Leung \& Bovy]{
Henry W. Leung$^{1}$\thanks{E-mail: henrysky.leung@utoronto.ca} \&
Jo Bovy$^{1,2}$
\newauthor
\\
$^{1}$David A. Dunlap Department of Astronomy and Astrophysics, University of Toronto, 50 St. George Street, Toronto, Ontario, M5S 3H4, Canada\\
$^{2}$Dunlap Institute for Astronomy and Astrophysics, University of Toronto, 50 St. George Street, Toronto, Ontario, M5S 3H4, Canada
}
\date{Accepted XXX. Received YYY; in original form ZZZ}

\pubyear{2023}

\begin{document}
\label{firstpage}
\pagerange{\pageref{firstpage}--\pageref{lastpage}}
\maketitle

\begin{abstract}
Rapid strides are currently being made in the field of artificial intelligence using Transformer-based models like Large Language Models (LLMs). The potential of these methods for creating a single, large, versatile model in astronomy has not yet been explored. In this work, we propose a framework for data-driven astronomy that uses the same core techniques and architecture as used by LLMs. Using a variety of observations and labels of stars as an example, we build a Transformer-based model and train it in a self-supervised manner with cross-survey data sets to perform a variety of inference tasks. In particular, we demonstrate that a \textit{single} model can perform both discriminative and generative tasks even if the model was not trained or fine-tuned to do any specific task. For example, on the discriminative task of deriving stellar parameters from \gaia\ XP spectra, we achieve an accuracy of 47 K in \teff, 0.11 dex in \logg, and 0.07 dex in \xh{M}, outperforming an expert \texttt{XGBoost} model in the same setting. But the same model can also generate XP spectra from stellar parameters, inpaint unobserved spectral regions, extract empirical stellar loci, and even determine the interstellar extinction curve. Our framework demonstrates that building and training a \textit{single} foundation model without fine-tuning using data and parameters from multiple surveys to predict unmeasured observations and parameters is well within reach. Such ``Large Astronomy Models'' trained on large quantities of observational data will play a large role in the analysis of current and future large surveys.
\end{abstract}


%%%%%%%%%%%%%%%%%%%%%%%%%%%%%%%%%%%%%%%%%%%%%%%%%%

%%%%%%%%%%%%%%%%% BODY OF PAPER %%%%%%%%%%%%%%%%%%

\section{Introduction}

Astronomy is on a path of ever-expanding data sets collected by large surveys like \gaia\ \citep{2016A&A...595A...1G} and Sloan Digital Sky Survey (SDSS; \citealt{2017AJ....154...28B,2017arXiv171103234K}) now, and Rubin Observatory Legacy Survey of Space and Time (LSST; \citealt{2019ApJ...873..111I}) and Euclid \citep{2011arXiv1110.3193L} in the future, across multiple areas such as spectroscopy, photometry, and time-domain observations. Due to the rapidly expanding size of the data sets, data-driven analysis has become increasingly popular among astronomers. But so far, bespoke data-driven models are created for every separate task and data-driven models that focus on cross-survey and/or cross-domain analyses (like the work of \citealt{2023MNRAS.522.4577L}) are only trained on the intersection of the relevant data instead of their union, due to the lack of flexibility in model inputs and outputs. Data-driven science using deep neural networks in particular requires big data sets for training and it would be ideal if we can train such models on most of the available data to truly create a synergistic understanding of multiple surveys.

A Transformer is a neural network architecture that relies on the attention mechanism \citep{2017arXiv170603762V}. Currently, there is ongoing rapid development of Transformer-based models in the guise of Large Language Models (LLMs). LLMs like OpenAI's GPT series of models \citep{2023arXiv230308774O} and Google's BERT \citep{2018arXiv181004805D} employ instruction fine-tuning and they have shown be able to do some tasks previously thought to only be possible with a general intelligence model \citep{2023arXiv230312712B}. Science communities have been critical of these big natural language models due to the problem of hallucinations (i.e., the model returns false information that sounds convincing) and because they can easily fail at simple math and logic problems. These issues mean that naively applying existing natural-language focused models to science is difficult. Moreover, they focus on natural language applications such as chat-bots, which involve completely different kinds of data from the floating point astronomical data. Nevertheless besides some uses of the attention mechanism (e.g., \citealt{2020PASP..132d4503Z, 2023AJ....165...18P, 2023arXiv230615703R, 2023arXiv230806404M, 2023arXiv230405350D}), LLMs have already been used in astronomy using its natural language focus to build, e.g., assistants for literature review \citep{2023RNAAS...7..193C} and for creating human-readable summaries of data in a data science platform for time-domain astronomy \citep{2023ApJS..267...31C}. The big commercial state-of-the-art models are trained and controlled by big companies, where data quality and access are not guaranteed.

Nevertheless, Transformer-based models have great potential for building a model that learns general knowledge about scientific areas such as observations of stars, the example we focus on in this paper. Such a model would obviously be useful in many scientific applications. We propose here that such models can be built by adopting the same core technologies and ideas of LLMs using Transformers, but applying them to tasks that are not focused on natural language.

Instead of giving a model sentences like in a LLM, we give the model floating-point data directly along with the type of the data and the model then vectorizes this input to floating-point vectors (the model's embedding). In our star-model application, the input can be a list of known parameters of an astronomical object, such as the flux over a certain wavelength range, to build a context of that particular object. Because the model sees a large amount of data during training and learns how different parameters describing stars are related to each other, we can then ask the model about other unmeasured parameters. Hence, in an implicit way, the model acquires embodied knowledge of stars.

There are a few key advantages specifically to adopting and adapting the technology behind Transformer-based models and the ideology behind LLMs and using this to train science-specific models rather than simply fine-tuning existing models:

\textbf{Expert knowledge:} Big, commercially-trained models are powerful but usually focus on natural languages or commercial domain data which are inherently different from astronomical data. Fine-tuning existing state-of-the-art models does not fundamentally solve the issue that existing models contain little astronomy knowledge that cannot be augmented without significant re-training with an astronomy focus.

\textbf{Big Data across surveys:} The flexibility of Transformer-based models like LLMs in the input and output nodes allows us to train on a significant portion of astronomical data, because we can train on all data even when the data set is highly incomplete due to, e.g., different survey footprints or photometric bands. Unlike the usual data-driven methods, we can use the union set of all surveys instead of their intersection. Transformer-based models provide ways to handle variable length input data by masking empty spaces with special masking tokens.
%(to simulate this in the application below, we randomly sample available data from each celestial objects as input and output during training).

\begin{figure*}
\centering
\includegraphics[width=\textwidth]{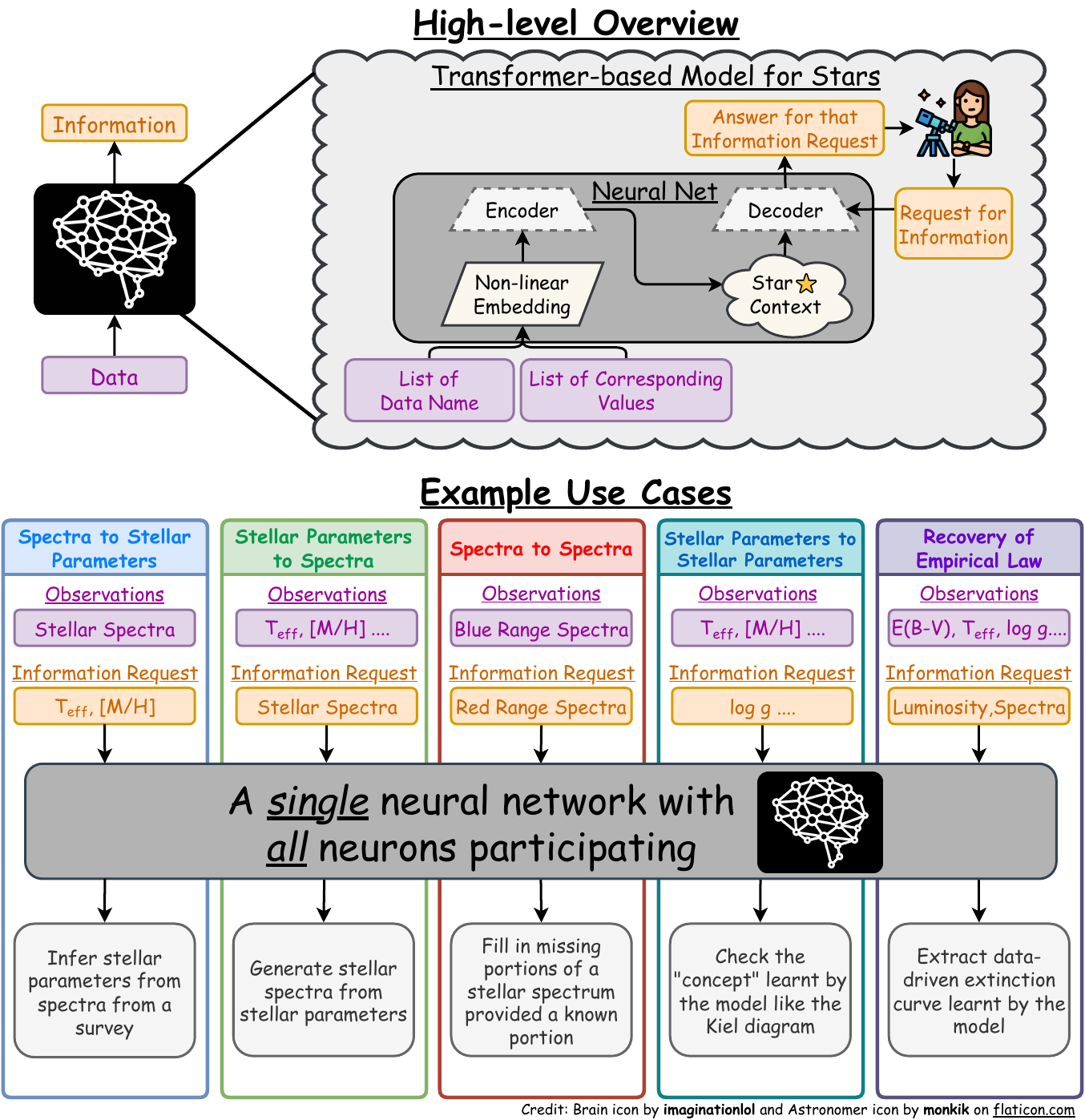}
\caption{High-level overview of our foundation model for stars. The top part of the figure displays the overall goal of training a big single neural network for astronomy that can turn a combination of data into useful information in the left part, while in the right part the high-level architecture of the Transformer-based model for stars in this paper is shown (for architectural details see \figurename\ \ref{fig:model_spec}).
%The network architecture closely resembles that of a typical Transformer-based encoder-decoder model commonly used by natural-language neural network models. It consists of an encoder, which consists of multiple Transformer encoder blocks that encodes information from the inputs to get a "context" of a star, and a decoder, which also consists of multiple Transformer decoder blocks that decodes the "context" of a star based on a request given by a human (or another agent) and outputs the answer to the request. The inputs to the encoder come from a non-linear embedding of the astronomical observations that combines the type of observation and its value (see \eqnname\ \ref{eq:embedding}). 
The bottom part of the figure gives an overview of the versatility of our model on multiple use cases, showing how a \textit{single} neural network without any fine-tuning can perform many useful tasks, such as ``generative'' and ``discriminative'' tasks in the traditional machine learning framework, with all weights in the model participating in those tasks. For each of these example use case, we will discuss the performance later in \secname\ \ref{sec:result}.}
\label{fig:model_overview}
\end{figure*}

\textbf{Interpretability:} Because our Transformer-based model that we have implemented (see below) has generative capabilities (as shown in \figurename\ \ref{fig:model_overview} and will be demonstrated in \figurename\ \ref{fig:spec_recon_teff_mh} and \figurename\ \ref{fig:spec_recon_logg}), the model is as interpretable as traditional generative models used in astronomy (indeed, it is even \emph{more} interpretable, because it can generate both data and parameters). In the future, these kinds of Transformer-based models should carry similar capabilities as LLMs, where you can ask for reasoning even if you have no way to know exactly how a deep neural network comes up with the answer in terms of neural pathways. In the end, an expert (such as an astronomer) who is using the model should be the one who decides whether to trust the result from data-driven models trained on a large amount of data. However, we emphasize that this is no different in current data-driven models.

\textbf{Versatility:} Transformer-based models have demonstrated a versatile range of solutions to perform downstream tasks like few-shot learning, rapid fine-tuning (e.g., \citealt{2021arXiv210609685H}), and agents (e.g., \texttt{LangChain} \citealt{Chase_LangChain_2022}). With a Transformer-based foundation model for astronomy, astronomers can fine-tune it to specialized tasks or new data sets (we give an example of this in \appenname~\ref{sec:foundationdemo}).

\textbf{Portability:} The deep learning community is currently putting a large amount of research effort into Transformer-based models such as LLMs and their role in artificial intelligence. There are new developments every day on LLMs and their applications. If we can rephrase our machine learning problem into a setting where Transformer-based models can excel, current and future developments like successors to Transformers using the attention algorithm can potentially be easily ported to our foundation model for astronomy.

In this work, we present a novel perspective on the use of Transformer-based models in astronomy by constructing a model that utilizes the core ideas and technologies of LLMs without involving natural language. We train a proto-foundation model that is not trained on specific input/output pairs for specific supervised and unsupervised tasks, but rather is trained in a self-supervised manner with a big data set to contain general knowledge of the data set, in our case observations and properties of stars in the Milky Way. From this context of a star, one can later request information from the model about other properties or observations of the same star with predictive uncertainty from the model. As a proof-of-concept of this new research area, we specifically build a Transformer-based model for stars using cross-survey, cross-domain data from APOGEE, \gaia, \tmass\ and dust maps. Our approach allows us to think about building a big foundation model for astronomy, its potential role in artificial intelligence, and its application in astronomy \citep{2023RSOS...1021454S}.

This paper is organized as follows. We provide the motivation behind using a Transformer-based model in \secname\ \ref{sec:transformers} and give an overview of the relevant deep-learning methodology and terminology. \secname\ \ref{sec:implem} describes details of our model implementation. \secname\ \ref{sec:datasets} describes the datasets that we use to train our model. \secname\ \ref{sec:train} describes the training process of our model. \secname\ \ref{sec:result} describes the results of our trained model on various tasks. \secname\ \ref{sec:discussion} discuss our models dependencies on various combination of stellar parameters and observations as well as the embedding learned by our model, the reliability and expectation of our model, and the model's role as a foundation model. Finally, \secname\ \ref{sec:conclusion} gives a conclusion to this paper.

\section{Transformer-based Neural Network}\label{sec:transformers}

Transformer-based neural networks (thereafter Transformers; \citealt{2017arXiv170603762V}) refers to a kind of neural network architecture that utilizes an algorithm called the attention mechanism \citep{2014arXiv1409.0473B}, which allows the model to attend to different parts of an input sequence and transform to a context vector for downstream layers. The attention mechanism is important especially in natural language tasks like neural machine translation as words have different meanings in different contexts; the attention mechanism provides neural pathways to learn how to get the context.

In this section, we discuss the motivation and overview of our framework in \secname\ \ref{subsec:evolution}, then briefly introduce the basic concept of tokens and embedding in \secname\ \ref{subsec:embedding}, and the attention mechanism that is the core component of Transformer-based neural networks in \secname\ \ref{subsec:attention}. The detailed implementation of our model is given in \secname\ \ref{sec:implem}. For in-depth introductions of this terminology and methodology, we refer interested readers to an online open-source interactive book \textit{Dive into Deep Learning}\footnote{\url{https://d2l.ai/}} \citep{2021arXiv210611342Z}.

\subsection{Dense NN to Transformer-based Encoder-Decoder}\label{subsec:evolution}

The main motivation for having a foundation model for stars like the one we implement in this paper (see \figurename\ \ref{fig:model_overview}) is to give the model the ability to perform multiple tasks directly without fine-tuning.
%, which allows a researcher to provide a sequence of observations as a context for stars you can then request for information like unmeasured observations, hence giving the model capability to do multiple tasks directly without fine-tuning since most of the tasks involve relationship between variables. 
To achieve this, we need to build the model using various components like an embedding and the attention mechanism. It is critical to understand the motivation behind and function of each of them. We introduce these components here by imagining that we want to ``evolve" a simple dense neural network (that simply maps \gaia\ XP coefficients to stellar parameters) to the Transformer-based encoder-decoder model that we develop in this work. Dense neural networks are commonly used in astronomy for data-driven machine learning problems or as emulators of complex functions.

Simple dense neural networks fail if one shuffles the order of the input XP coefficients, even though the input then has essentially the same information content. The dense neural network fails, because while the information content is the same after shuffling, there is no way for the model to know which input value corresponds to which coefficient. Unlike in natural or programming languages, the actual ordering of the coefficients does not intrinsically matter, because the information is the same regardless of the order. This is why we need to introduce an embedding, specifically an embedding that includes both the type of observation and its corresponding value. Unlike in natural language processing (NLP), we do not require a positional encoding to be part of the embedding although this may be useful in some contexts (e.g., \citealt{2021arXiv210506178A, 2023A&A...670A..54D, 2023arXiv230615703R, 2023arXiv230806404M}). With such an embedding, a simple dense neural network would be robust to random shuffling of the same inputs (i.e., always providing 110 XP coefficients, but in random order), because the model can learn to approximate pooling operations such as global averaging.

The addition of an embedding layer is not enough when the input content and length are not always the same, for example, when different subsets of inputs are used in random order. This is because it is difficult for neural networks to get the context if the inputs are not always the same, unlike the scenario we discussed in the previous paragraph where the 110 coefficients always exist in the inputs, just in random order. The encoder block of a Transformer using the attention mechanism---specifically self-attention on the input sequence---is needed to train such model. The attention mechanism provides a way for the model to get the ``context'' of the input sequence while handling variable length inputs by masking non-existing inputs using padding tokens.

If we further desire flexibility in what the output node returns rather than having a model that can only predict a fixed set of stellar parameters, we can add the decoder block of a Transformer to perform cross-attention between the context vector obtained from the input sequence and the output request vector. This allows the model to make predictions based on the context of the input sequence for a particular output request. This provides tremendous flexibility to the model and training sequence, because this allows the input node and the output node to be anything in the data set.

\begin{figure*}
\centering
\includegraphics[width= 0.95 \textwidth]{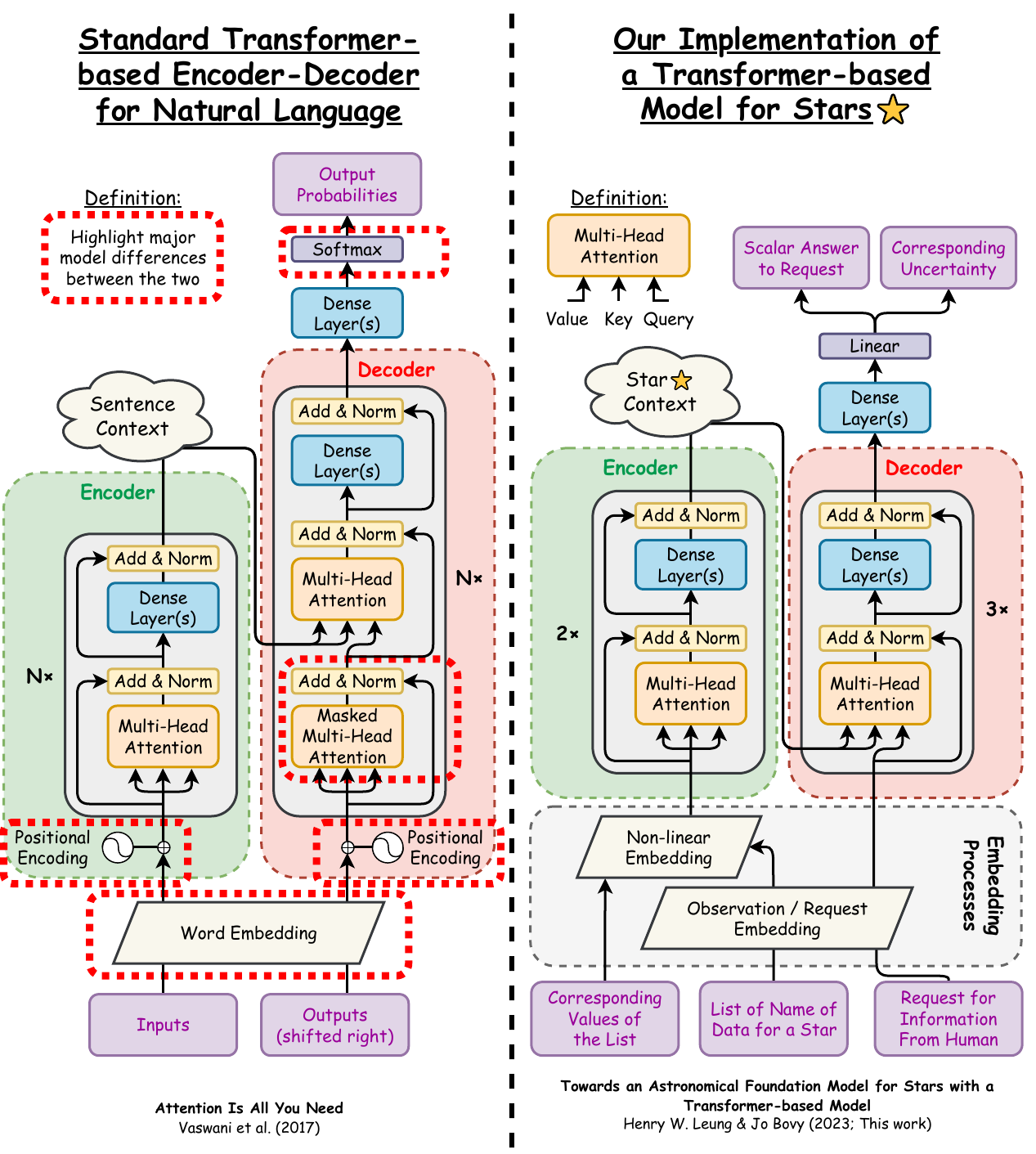}
\caption{Neural network architecture of the model used in this work (right) compared to that of the standard Transformer-based encoder-decoder model (left; \citealt{2017arXiv170603762V}). The encoder in our model has a very similar architecture to the encoder block in the standard Transformer, except that we do not use positional encoding, but rather perform a non-linear embedding of the type of observations and their corresponding values. The decoder blocks in our model have the same architecture as the encoder blocks in the encoder, because we do not need positional encoding for the decoder either in our application, which is different from the standard Transformer-based encoder-decoder where the decoder blocks are different from the encoder blocks, because they need to consider the current output sequence in addition to the context and request.}
\label{fig:model_spec}
\end{figure*}

\subsection{Tokens and Embedding}\label{subsec:embedding}

Tokens and embedding are easiest to understand in the natural language context. A neural network can only accept floating-point numbers as inputs and a process to turn words into floating-point vectors (i.e., a word embedding) is needed. Let's say the goal is to have $N$ unique words to be turned into vectors $\vec{e}_i$ with an embedding dimension $d$,; all the words/vectors then make up the embedding database $\mathbb{E} = \{\vec{e}_1, \vec{e}_2 ... \vec{e}_N\} $ (e.g., \texttt{word2vec} for natural language; \citealt{2013arXiv1301.3781M}). Each word can then be turned into (or ``tokenized'') the integer that correspond to the column index in $\mathbb{E}$ in order to retrieve embedding that correspond to that word. In order to build-up such a database $\mathbb{E}$, each vector $\vec{e}_i$ is part of the neural network optimization problem such that the model can learn the weights that make up a good embedding database (e.g., \citealt{10.5555/944919.944966}). Moreover, there are tokens that carry out special functions, like the padding token which is ignored by the attention mechanism and which is usually used to pad variable-length sentences to all have the same length.

The input astronomical data in our model passes through a similar embedding processing as that which is used in NLP. Our tokenization process is fairly simple, because we simply look up the exact match of the observation type to an indexed database of possible observation types and so we require no heuristics currently in this process as are used in NLP (which needs to deal with, e.g., typos and contractions). This also means that 64 tokens always mean 64 observations with their value (as opposed to NLP, where the number of tokens is generally bigger than the number of individual words).

We have implemented a custom embedding process for our data that we refer to as a ``non-linear embedding''. This works by embedding data of type $x$ using
\begin{equation}\label{eq:embedding}
    y_x = f(w_x \cdot M_x) + w_{b, x}
\end{equation}
where $y_x$ is the final vectorized data for a particular type of data $x$ with value $M_x$ that will be fed into the model. The function $f$ is a typical activation function used in neural networks, $w_x$ are the trainable embedding weights, $M_x$ is the value of the data, and $w_{b, x}$ is a trainable bias weight; all of these are particular to the kind of data $x$. Bias weights are a necessary part of the embedding, because without them, the neural network would receive a vector of zeros for all data with a value of zero and would have no way of knowing which observation has a value of zero. However, a data value of zero has different meanings for different kinds of data.

It is critical to embed the data value information in a non-linear way in the embedding instead of providing it as a scalar value, as we will discuss further in \secname\ \ref{subsec:attention} and \secname\ \ref{subsec:dependences}, because the attention on the data is expected to depend on both the type of observation and its value.

\subsection{Attention Mechanism}\label{subsec:attention}

The attention mechanism is the core algorithm underlying Transformers. Here we discuss a specific implementation of attention called scaled dot-product attention that we use in our model. Unlike recurrent neural networks (RNNs), the attention mechanism can process a whole input sequence all at once and disregard the ordering of input sequences. The attention mechanism was mainly developed to solve the issue of distance-independent dependencies, that is how each element relates to each other regardless of how far apart elements are in an input sequence. Long-distance dependence in the input sequence is important especially in NLP, because words can have multiple meanings and context is needed to allow readers to accurately interpret the meaning of a sentence. That nearby words are more likely important for interpreting a given word is incorporated in Transforms using positional embeddings.

The core algorithm of the scaled dot-product attention, which has no trainable weights, requires three inputs---Queries $Q$, Keys $K$, and Values $V$---that are used to calculate a context vector $C$. First, the output alignment scores $S$ are calculated as the matrix multiplication between $Q$ and transpose of $K$
\begin{equation}\label{eq:alignment}
  S = Q K^T\,,
\end{equation}
which essentially gets the similarity between the query vectors $Q$ and key vectors $K$. The alignment scores $S$ are then converted to attention weights $A$ by
\begin{equation}\label{eq:attention}
  A = \text{Softmax} \bigg(\frac{S}{\sqrt{d_K}}\bigg)\,,
\end{equation}
where $d_K$ refers to an integer constant that is the number of dimensions in $K$, which is usually the embedding dimension, to ensure unit variance of S to improve numerical stability in the gradients used during training when $d_K$ is large. The \texttt{Softmax} function is applied to ensure that the attention weights for each input sequence sum up to one. Finally the context vector $C$ is computed by the matrix multiplication between $A$ and $V$
\begin{equation}\label{eq:context}
  C = \text{Attention}(Q, K, V) = A\, V = \text{Softmax} \bigg(\frac{Q K^T}{\sqrt{d_K}}\bigg)\,V\,,
\end{equation}
where $V$ is the input sequence Values vector.

These equations allow the attention mechanism to effectively act as a database software, where one can execute a query against a database of key--value pairs. In a simple regression setting, the attention weight $A$ effectively allow the model to determine the regions where the regression should be carried out in the input sequence for a regression to a query $Q$, where the keys $K$ are the kind of data and the values $V$ are the input regression floating-point values corresponding to each input key $K$. This is similar to algorithms such as non-parametric kernel regression. Although in theory floating-point values can be passed via $V$, in practice and in this work we encode value information in the key vector $K$ as well and pass it as $K$ and $V$. We discuss this further in \secname\ \ref{subsec:dependences}.

When the Queries $Q$ and the Keys $K$ are the same sequence, the algorithm above is known as \emph{self-attention}, which means attention of one sequence applied to itself. When the Queries $Q$ and the Keys $K$ are different sequences, the algorithm is known as \emph{cross-attention}, because it calculates the attention of one sequence to a different sequence.

In practice, there are trainable weights and multiple attention mechanisms are applied in parallel to the same input sequence (referred to as ``multi-head attention''), which allows the model to attend to different subspaces of the embedding. As part of a neural network, this mechanism constitutes a Multi-Head Attention layer (MHA, called \texttt{MultiheadAttention} in \texttt{PyTorch} \citep{2019arXiv191201703P}). Within the MHA layer, there are $h$ heads where $h$ is a factor of $d_K$ such that each attention head is responsible for a subspace with dimension $d_K/h$ of the embedding space. Within each attention head $i$ in MHA, $Q$, $K$ and $V$ all have their own trainable weights $W_i^Q$, $W_i^K$ and $W_i^V$ to linearly transform $Q$, $K$ and $V$ by matrix multiplication before the attention mechanism is applied within each head $i$. The results from all heads are then concatenated and  are finally linearly transformed through matrix multiplication with another weights matrix $W^O$ to obtain the final output. That is, we do
\begin{equation}\label{eq:weightedattn}
  C = W^O \text{Concat}\big(\text{Attention}(QW_i^Q, KW_i^K, VW_i^V)\big)\,.
\end{equation}

The Attention mechanism also provides a way to handle variable length inputs. That is, one can have missing data in the input sequences $Q$, $K$, and $V$. Any missing parts of the sequence compared to a model's context length can be padding with an arbitrary constant and be ignored in the attention mechanism by setting the attentions to those part to zero.

\section{Model Implementation}\label{sec:implem}

We implement our model using \texttt{PyTorch}\footnote{\url{https://github.com/pytorch/pytorch}} \citep{2019arXiv191201703P} with $\approx 8.8$ millions trainable parameters including those from the embedding space. We have deliberately over-parameterized the model to achieve the best performance; a scaled-down model with $\approx 1$ million trainable parameters should achieve similar results. A high-level overview of the model is given in \figurename\ \ref{fig:model_overview} and the detailed architecture is presented in \figurename\ \ref{fig:model_spec}; our model mimics a typical Transformer-based encoder-decoder model (as opposed to other commonly-used architectures like Vision Transformer \citep{2020arXiv201011929D}, which is a Transformer-based encoder model). 

Our Transformer-based model shares a common limitation of Transformers, which is that there is a limit to how many tokens we can process at once; this limit is known as the ``context window''. Our model has a context window set to 64, that is, our model can only handle up to 64 tokens at once and we did not employ any methodology or mechanism to handle longer contexts. This is different from more traditional neural networks, where there are dedicated input neurons for each data type. We do not have that ``luxury'' here. But this limitation is actually important to demonstrate the model's performance and generality, because once we train on a large variety of data, it is infeasible to have dedicated neurons for every possible type of data, so the context window will eventually be much smaller than the possible types of data. For the embedding space, we use a dimension of 128. The computational cost of each scaled dot-product attention mechanism is $\mathcal{O}(n^2 d_K)$ where $n$ is the context length and $d_K$ is the embedding dimension (note that the shape of the $Q$, $K$, and $V$ matrices is $n \times d_K$; the $\mathcal{O}(n^2 d_K)$ scaling comes from the scaling of the involved matrix multiplications). Thus, the attention mechanism in each layer scales quadratically with the length of the context window in terms of required computational power. This is why it is expensive to increase the size of the context window.

Our encoder has two Transformer encoder blocks doing self-attention on the input sequence with intermediate dense layers of 1024 and 512 neurons, respectively, for the first and second block. Our decoder has three Transformer decoder blocks doing cross-attention between the final output from the encoder, which represents the context of a star, and the information request vector. These blocks have 3096, 1548, and 774 neurons in their intermediate dense layers, respectively. All of the Transformer blocks in the encoder and decoder have 16 attention heads, that is, each attention head is responsible for 8 embedding dimensions and then we concatenate all the results from each head as the results. To map from the output of the final decoder block to the final output, we use a three-layer dense neural network with 3096, 1548 and 774 neurons respectively. For the activation function of these dense layers (including those in the Transformer encoder/decoder blocks), we use Gaussian Error Linear Units (GeLU; \citealt{2016arXiv160608415H}) globally when appropriate. A dropout \citep{10.5555/2627435.2670313} rate of 0.1 is applied globally for the dense layers, except in the embedding space, during training, but is disabled during testing. There are \texttt{Add \& Norm} operations in all Transformer blocks. This operation is composed of two operations, \texttt{Add} and \texttt{Norm}. The \texttt{Add} operation simply element-wise adds two inputs together, which usually are the input and the output of the previous layer to allow residual connections \citep{2015arXiv151203385H}. The \texttt{Norm} operation (which is called \texttt{LayerNorm} in \texttt{PyTorch}) performs normalization of the layer \citep{2016arXiv160706450L}, by tranforming the input $x$ to $y$ given by
\begin{equation}\label{eq:layernorm}
y = \frac{x-\text{Mean}[x]}{\sqrt{\text{Var}[x]+\epsilon}} \times \gamma + \beta\,,
\end{equation}
where the gain $\gamma$ and the bias $\beta$ are trainable parameters initialized as ones and zeros, respectively, at the beginning of training, thus simply standardizing the inputs at the start of training. The parameter $\epsilon$ is a small constant to ensure numerical stability.

\begin{figure*}
\begin{center}
\includegraphics[width=\textwidth]{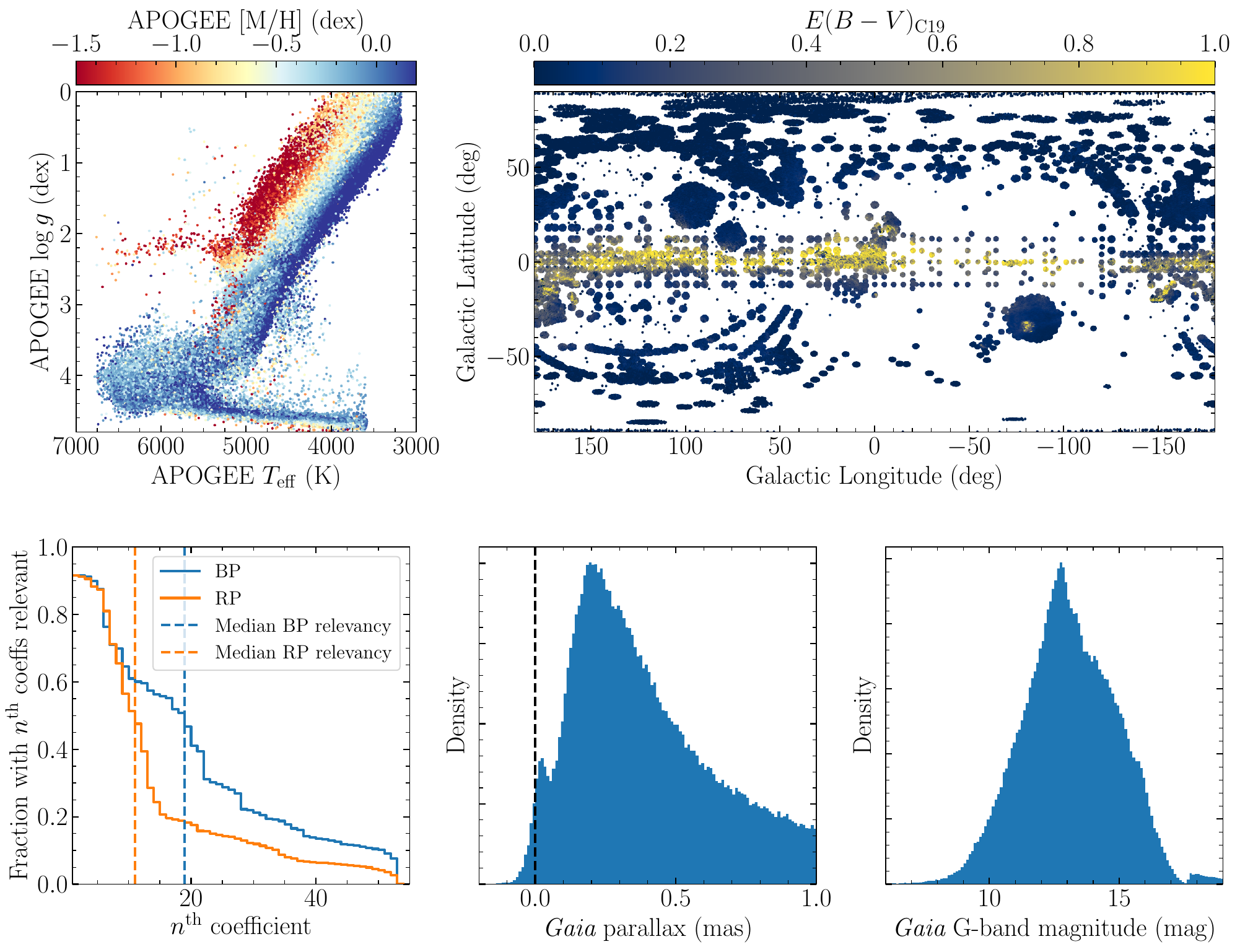}
\caption{Properties of the training set.  The top left panel displays a Kiel diagram of APOGEE \teff\ and \logg\ colored by \xh{M} while the top right panel shows the on-sky distribution of the sample colored by the reddening $E(B-V)$ from the \texttt{Combined19} extinction map. The bottom left panel shows the fraction of stars for which their $n^\mathrm{th}$ BP and RP  XP coefficients are relevant, the bottom middle panel displays the distribution of \gaia\ parallaxes (with some negative parallaxes), and the bottom right panel shows distribution of \gaia\ apparent $G$-band magnitude, which together with the parallax is used to calculate the stellar luminosity. The bottom-left relevancy fraction starts at $\approx 90\%$, because $\approx 10\%$ of stars in the training set do not have XP spectra.}
\label{fig:training_set}
\end{center}
\end{figure*}

The ``unit vector'' $w_x$ in \eqnname\ \ref{eq:embedding} for the non-linear embedding is also used as the request ``token'' given to the decoder (shown in \figurename\ \ref{fig:model_overview} and \figurename\ \ref{fig:model_spec}). We train the decoder such that its output is a scalar value answer corresponding to the requested information vector. This is similar to what happens in LLMs where the next words are determined based on the previous generated words (analog to the information request here) as well as the original user inputs (analog to the data input by users here). As long as we keep the request ``token'' consistent with the input tokens by reusing the ``unit vector'' $w_x$, reuse the knowledge on the embedding learned by the encoder (see further discussion in \secname\ \ref{subsec:new_request} and \figurename\ \ref{fig:new_requests}).

\section{Datasets}\label{sec:datasets}

To train our model, we construct a small, relatively low-dimensional data set that is perfect for fast proof-of-concept prototyping. The dataset we use is composed of data from APOGEE, \gaia, and dust-map data. In theory, our model is flexible enough to train on a highly heterogeneous dataset, that is a dataset where there are lots of missing data. In practice, training on the whole \gaia\ data set, for example, would take a large amount of computational power. That is why we choose to only train on stars observed in APOGEE, where we can still construct a heterogeneous dataset to demonstrate the flexibility of our model, because some stars in APOGEE do not have good observations from APOGEE while having good \gaia\ observation, while some other stars with good APOGEE observations are too dim for \gaia, and yet other stars are so dim that they do not have good \gaia\ or APOGEE observations. The number of stars in our training set satisfying a number of criteria is given in \tablename\ \ref{tab:training}. We discuss the APOGEE data in \secname\ \ref{subsec:apogee}, \gaia\ data in \secname\ \ref{subsec:gaia}, and the data we use on interstellar extinction in \secname\ \ref{subsec:extinction}. The training set has a total of $397,718$ stars while the test set contains $44,080$ stars, with stars ranging from dwarfs to bright giants in both data sets. 

\begin{table} 
\caption{The number of stars in the training set that satisfies certain criteria. A star does not need to have both XP spectra and stellar parameters to be included. A star can have only XP spectra or can have only stellar parameters to be included.} 
\label{tab:training} 
\begin{tabular}{lcc} 
\hline Criteria & Number of Stars\\ 
\hline Stars in the training set & 396,718\\ \hline 
Stars with \gaia\ XP spectra & 363,100 \\ 
Stars with APOGEE \teff & 263,809\\ 
Stars with APOGEE \teff, \logg\ and \xh{M} & 261,182 \\
Stars with \gaia\ XP spectra without APOGEE parameters & 132,657 \\ 
Stars with APOGEE parameters without \gaia\ XP spectra & 33,570 \\
Stars without \gaia\ $G$-band luminosity & 12,065 \\
Stars without \texttt{Combined19} $E(B-V)_\mathrm{C19}$ & 11,947 \\
Stars without \tmass\ colors & 2,031 \\
\hline 
\end{tabular} 
\end{table}

\subsection{APOGEE DR17}\label{subsec:apogee}

We use precise stellar parameter labels from the Apache Point Observatory Galactic Evolution Experiment data release 17 (APOGEE DR17; \citealt{2017AJ....154...28B, 2022ApJS..259...35A}). In particular, we use surface temperature \teff, surface gravity \logg, and the overall metallicity \xh{M}. APOGEE is a high-resolution ($R\sim22,000$), high signal-to-noise ($>100$ per pixel typically) panoptic spectroscopic survey in the near infrared H-band wavelength region of $1.5-1.7\um$ \citep{2017AJ....154...94M, 2019PASP..131e5001W}. The stellar parameter labels are derived from APOGEE spectra using the APOGEE Stellar Parameter and Chemical Abundances Pipeline (ASPCAP; \citealt{2016AJ....151..144G}). We do not use the APOGEE spectra directly here for the training of our model.

In order for a star in our data set to have usable stellar labels, we require that the star does not have any bit set in \texttt{STARFLAG} and does not have bit 19 (\texttt{M\_H\_BAD}) and bit 23 (\texttt{STAR\_BAD}) set in \texttt{ASPCAPFLAG}, to avoid the issue of obvious binary stars and of problematic stellar parameters derived from ASPCAP. Stars with bad APOGEE stellar labels are not excluded as long as they survive cuts made to other survey discussed below andwe simply set bad APOGEE labels to \texttt{NaN}.

In addition to using data from APOGEE itself, we also adapt \tmass\ \citep{2006AJ....131.1163S} photometry in the $J$, $H$, $K_s$ bands and we calculate the \tmass\ colors \jh\ and \jk\ and their uncertainty. The \tmass\ data are taken directly from the APOGEE data file rather than performing our own cross match.

\subsection{\gaia\ DR3}\label{subsec:gaia}

The DR3 release from \gaia\ \citep{2016A&A...595A...1G, 2023A&A...674A...1G} provides an unprecedented number of low-resolution spectra along with precise astrometry. The \gaia\ XP spectra \citep{2021A&A...652A..86C, 2023A&A...674A...2D, 2023A&A...674A...3M} are low-resolution ($R\sim30$ to $100$), optical to near-infrared  ($330$ to $1050\,\mathrm{nm}$) spectra obtained from Blue Photometer (BP) and Red Photometer (RP) aboard the \gaia\ spacecraft. Broad blue and red photometry \bp\ and \rp\ is also obtained from these spectra \citep{2021A&A...649A...3R}. Due to the large amount of data in various formats, we have developed a \texttt{Python} package called \texttt{MyGaiaDB}\footnote{\url{https://github.com/henrysky/MyGaiaDB}} to manage \gaia\ and ancillary photometric and spectroscopic data and to be able to query these using the SQL language on local machines.

For a star be considered to have a valid parallax that we use, a star must have Renormalised Unit Weight Error $\texttt{ruwe}<1.4$ and parallax uncertainty $\sigma_\plx < 0.1\,\mas$ to ensure a good astrometric solution. We adopt the Gaia parallax zero-point correction $\plx_\mathrm{offset}$ from \citep{2021A&A...649A...4L} as implemented in the \texttt{gaiadr3-zeropoint} \footnote{\url{https://gitlab.com/icc-ub/public/gaiadr3\_zeropoint}} \texttt{Python} package and apply the additional correction from \citet{2023MNRAS.519..948L}. The formula for the additional correction is as follows:
\begin{equation}\label{eq:plx_offset}
  \plx_\mathrm{offset} = \begin{cases}
  Z_\mathrm{gdr3} & \mbox{ for $ G\leq13 $}\\
  Z_\mathrm{gdr3} + 5\ \uas\ (G-13)  & \mbox{ for $ 13<G\leq17 $}\\
  Z_\mathrm{gdr3} + 20\ \uas & \mbox{ for $ G>17 $}
  \end{cases}\,,
\end{equation}
where $Z_\mathrm{gdr3} (p_\mathrm{solved}, G, \nu_\mathrm{eff}, \hat{\nu}_\mathrm{eff}, \beta)$ is the official \gaia\ zero-point correction function \citep{2021A&A...649A...4L} that depends on $p_\mathrm{solved}$, which is the number \texttt{astrometric\_params\_solved} of astrometric parameters solved for, $G$ is the apparent \gaia\ $G$-band magnitude \texttt{phot\_g\_mean\_mag}, $\nu_\mathrm{eff}$ is the effective wavenumber \texttt{nu\_eff\_used\_in\_astrometry}, $\hat{\nu}_\mathrm{eff}$ is the astrometrically estimated effective wavenumber \texttt{pseudocolour}, and $\beta$ is the ecliptic latitude \texttt{ect\_lat}. The final corrected parallax is then $\plx=\plx_\mathrm{gdr3}-\plx_\mathrm{offset}$.

Unlike usual stellar spectra, the \gaia\ XP spectra were released as a set of 110 coefficients of an orthogonal basis function expansion of the spectrum, where lower-order coefficients explain large-scale features of the spectra (thus, information like \teff) and higher-order coefficients explain small-scale features including the noise. The data from each photometer has its own expansion, so there are 55 BP coefficients and 55 RP coefficients. The spectra as flux vs. wavelength are only directly available for a small subset of stars. Therefore, we use the coefficients directly and we simply treat each coefficient as a type of data. We normalize the XP coefficients by dividing each coefficient by the \gaia\ $G$-band apparent flux $F_G$, given by\begin{equation}\label{eq:gaia_flux}
2.5\,\log_{10} F_G = G_{0, \mathrm{vega}} - G\,,
\end{equation}
where the \gaia\ DR3 $G$-band photometric zero-point $G_{0, \mathrm{vega}} = 25.6873668671$ mag \citep{2021A&A...649A...3R}.

\gaia\ also provides a relevancy parameter for the coefficients of each photometer for a given star, such that the relevancy $r_\mathrm{BP}$ for BP and $r_\mathrm{RP}$ for RP means that only BP coefficients up to the ${r_\mathrm{BP}}^\mathrm{th}$ and up to the ${r_\mathrm{RP}}^\mathrm{th}$ for RP coefficients are relevant to describe the XP spectrum of a star. In other words, only the first $r_\mathrm{BP}$ BP coefficients and only the first $r_\mathrm{RP}$ RP coefficients are significant compared to thresholds adopted in the \gaia\ data pipeline. During the training process, we set coefficients deemed irrelevant to \texttt{NaN} except for the last two BP and RP coefficients. For the last two BP and RP coefficients, we copy the relevancy from the third last BP and RP coefficients to the last two BP and RP coefficients, because \gaia\ never indicates that any of the last two BP and RP are relevant. Therefore, as long as \gaia\ says that the third last BP or RP coefficient is relevant, we set the same relevancy to the last two BP and RP coefficients, respectively.

For a star to be considered to have a valid XP spectrum (a star can still be selected without XP spectrum, but with good observations from other surveys), the star need to have $6.0<G<17.5$ and $0<\bprp<4$, as well as having the BP and RP flux excess factor less than $2.0$ to avoid problematic stars with erroneous fluxes estimation.

To determine a stellar luminosity proxy, we adopt the luminosity--parallax relation from \cite{2018AJ....156..145A}
\begin{equation} \label{eq:fakemag}
L_\mathrm{pseudo} = \plx 10^{\frac{1}{5}m_\mathrm{apparent}} = 10^{\frac{1}{5}M_\mathrm{absolute}+2}\ ,
\end{equation}
where $L_\mathrm{pseudo}$ is an alternative scaling of luminosity, a pseudo-luminosity. Previously, we have adopted a similar scaling to obtain machine-learned spectro-photometric distances \citep{2019MNRAS.489.2079L} in order to preserve the Gaussianity of the parallax error distribution in the pseudo-luminosity. Unlike the usual luminosity (or pseudo-luminosity) that has been used in the past, which is the true luminosity calculated from the extinction-corrected apparent magnitude and the distance, we simply calculate the pseudo-luminosity using the measured apparent magnitude without applying an extinction correction. Thus, the pseudo-luminosity $L_\mathrm{pseudo}$ is contaminated by extinction, but the advantage of this approach is that any pseudo-luminosity we infer from the trained model can be directly used to convert to apparent magnitude to distance using the \gaia\ measured $G$-band apparent magnitude without having to explicitly consider the $A_G$ extinction.

\subsection{Dust Map}\label{subsec:extinction}

We obtain information on the interstellar extinction to stars in our sample from the  \texttt{Combined19} (thereafter C19) three-dimensional extinction map in the \texttt{mwdust}\footnote{\url{https://github.com/jobovy/mwdust}} \texttt{Python} package \citep{2016ApJ...818..130B}. This map combined data from the extinction maps of \texttt{Drimmel03} \citep{2003A&A...409..205D}, \texttt{Marshall06} \citet{2006A&A...453..635M}, and \texttt{Green19} \citep{2019ApJ...887...93G}. \texttt{Green19} is a three-dimensional extinction map based on \gaia, \tmass\ and Pan-STARRS 1 data that covers the sky with declination above $-30^\circ$ and out to a distance of a few kpc. The \texttt{Marshall06} map is a three-dimensional extinction map based on \tmass\ that covers the sky of $-100^\circ \le l \le 100^\circ$ and $-10^\circ \le b \le 10^\circ$, mainly using giants; this map performs better than \texttt{Green19} when available \citep{2016ApJ...818..130B} and we therefore use it wherever it is available. The \texttt{Drimmel03} map is a three-dimensional extinction map based on COBE/DIRBE \citep{1998ApJ...508...25H} and we use it to fill in the blank for the area not covered by \texttt{Green19} or \texttt{Marshall06}, which is mainly the part of the sky surrounding the south celestial pole.

The reddening $E(B-V)_\mathrm{C19}$ obtained from C19, which we use to train the model, is on the scale of the $E(B-V)_\mathrm{SFD}$ of the SFD extinction map \citep{1998ApJ...500..525S}. This scale can be converted to true reddening $E(B-V)_\mathrm{true}$ using
\begin{equation} \label{eq:sfd}
E(B-V)_\mathrm{true} = 0.884 \times E(B-V)_\mathrm{SFD} = 0.884 \times E(B-V)_\mathrm{C19}\,,
\end{equation}
where the factor of 0.884 comes from the measurement of \citet{2011ApJ...737..103S}. Because we train on $E(B-V)_\mathrm{C19}$, any $E(B-V)$ obtained from our trained model is on the SFD scale as well and needs to be converted to true $E(B-V)$ using the equation above. We assume that the uncertainty in $E(B-V)_\mathrm{C19}$ is solely coming from the uncertainty in the parallax used to obtain the distance at which the extinction map is evaluated. We set $E(B-V)$ to \texttt{NaN} for (a) negative values arising from numerical instability in the extinction map evaluation, (b) values over 10 mag, and (c) stars with unavailable distances. For stars with negative parallaxes but otherwise with good \gaia\ astrometry satisfying $\texttt{ruwe}<1.4$ and $\sigma_\plx < 0.1\ \mas$, we set the distance to 99 kpc for the purpose of obtaining the extinction such that we obtain the extinction at infinity for these distant stars. For each stars, we then draw $1,000$ samples from the zero-point corrected parallax uncertainty distribution and obtain the distribution of $\ln E(B-V)$. $E(B-V)$ values that are exactly equal to zero are set to a small number $e^{-7}$ before taking the logarithm to prevent numerical issues, but also to prevent the model from predicting negative $E(B-V)$. Then we simply calculate the median and robust standard deviation as the final logarithmic $E(B-V)$ and its uncertainty that goes into our model.

\begin{figure*}
\centering
\begin{tikzpicture}
\node (n1) [draw=spec2paramcolor, line width=\figboxlw,rounded corners=\figboxcorner, inner xsep=\figboxsep, inner ysep=\figboxsep]
{\includegraphics[width=\textwidth]{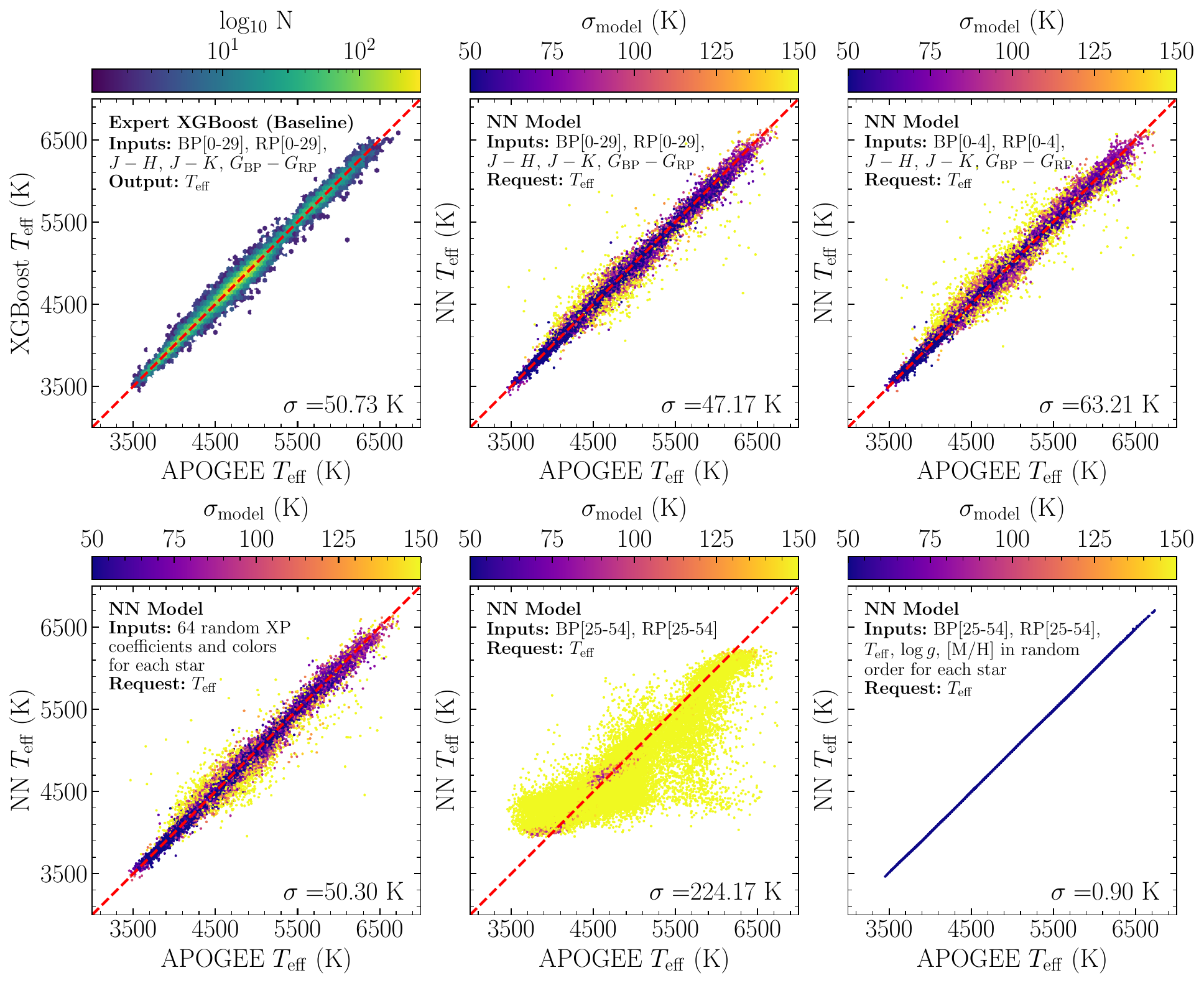}};
\node[font=\fontsize{\figheadfontsize}{0}\selectfont, text=spec2paramcolor] at (-4.5, 7.7) {Task: Stellar Spectra to Stellar Parameters};
\end{tikzpicture}
\caption{Results for predicting \teff\ from different combinations of XP spectra and photometry. The top-left panel shows an expertly trained \texttt{XGBoost} model as a baseline. The top-middle panel shows the performance of our Transformer-based model given the first 30 BP and RP coefficients that describe the XP spectra as well as the \jh, \jk\ and \bprp\ colors. These are the same inputs as the \texttt{XGBoost} model, showing that our general model that is not specifically tuned to these inputs slightly outperforms a dedicated,  expertly-trained \texttt{XGBoost} model. The top-right panel is similar, but only using the first 5  BP and RP coefficients instead of the first 30 BP and RP coefficients. The bottom-left panel displays a task where the inputs are 64 values randomly selected among that 110 BP and RP coefficients and colors for each star and given to the model in random order. The number of 64 is chosen so as to fill the whole context window of the model. In the bottom-middle panel, we provide the model with the most uninformative BP and RP coefficients and also do not give any color, to see if the model correctly returns large uncertainties in this case where the predictions are poor. The bottom-right panel is similar, except that we now mix in \teff, \logg\ and\xh{M} in random order for each star. In this case, the model successfully answers the information request to very high precision. Similar plots for \logg\ and \xh{M}\ are available in \figurename\ \ref{fig:teff_logg_mh} in \appenname\ \ref{sec:performance}.}
\label{fig:teff}
\end{figure*}

\begin{figure*}
\centering
\begin{tikzpicture}
\node (n1) [draw=spec2paramcolor, line width=\figboxlw,rounded corners=\figboxcorner, inner xsep=\figboxsep, inner ysep=\figboxsep]
{\includegraphics[width=\textwidth]{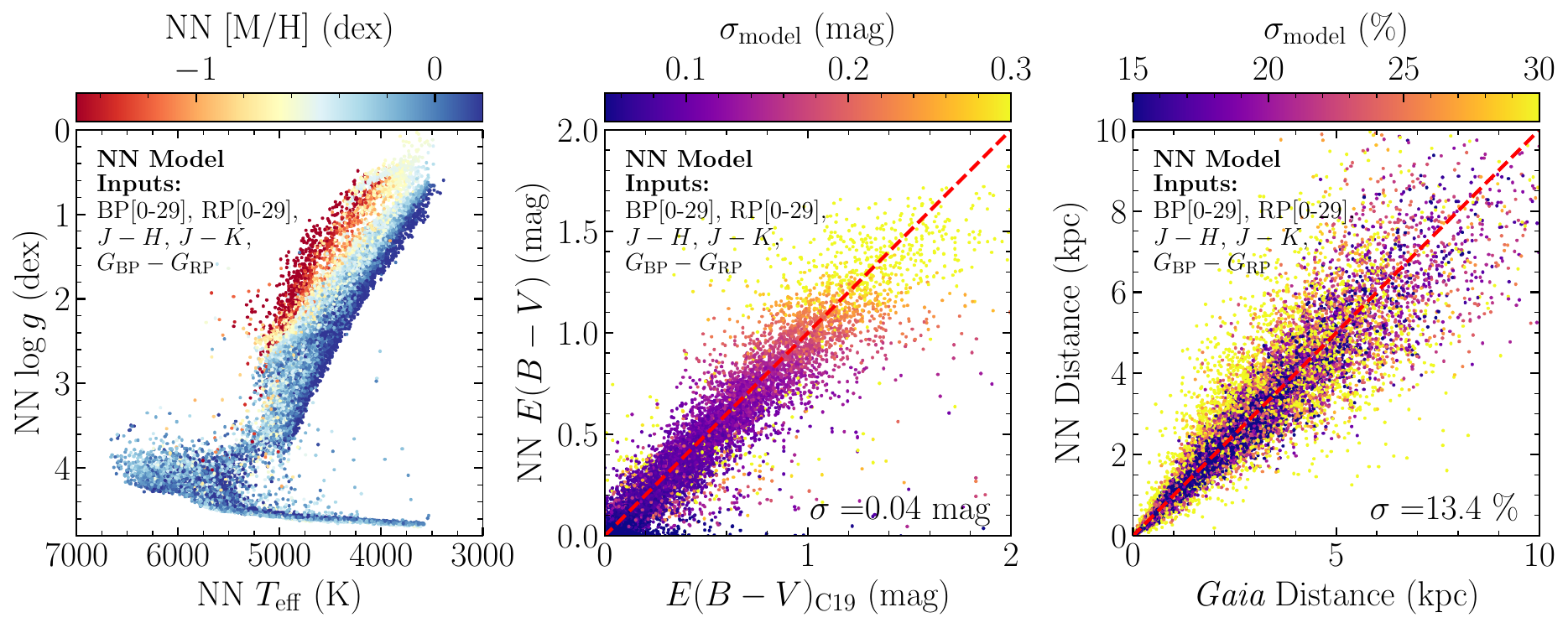}};
\node[font=\fontsize{\figheadfontsize}{0}\selectfont, text=spec2paramcolor] at (-4.5, 4.0) {Task: Stellar Spectra to Stellar Parameters};
\end{tikzpicture}
\caption{Additional performance of our model for the same set of inputs as in the top-middle panel of \figurename\ \ref{fig:teff}, but requesting additional information. The left panel in this figure shows our model's $\teff-\logg$ distribution color-coded by \xh{M} without cuts on the model uncertainty (the performance of predicting \logg\ and \xh{M} can be seen in \figurename\ \ref{fig:teff_logg_mh} in \appenname\ \ref{sec:performance}), where one can clearly see features such as the metallicity--color gradient for giants and red-clump stars. The middle panel displays the model's performance on $E(B-V)$ color-coded by the model prediction uncertainty on $E(B-V)$. The right panel shows the recovery of the spectro-photometric distances obtained using our model color-coded by the model prediction uncertainty.}
\label{fig:disrminate}
\end{figure*}

\section{Training}\label{sec:train}

We train our model with 118 unique kinds of tokens (that is 118 types of data; 110 \gaia\ XP coefficients, \bprp, \jh, \jk, \teff, \logg, \xh{M}, logarithmic $E(B-V)_\mathrm{C19}$, and \gaia\ $G$-band pseudo-luminosity as described in \secname\ \ref{sec:datasets}). The entire training run involves $\approx 16.4$ millions tokens that are not \texttt{NaN} from $\approx397$k stars in the training set. Data that are \texttt{NaN} are converted to a padding token (the same idea used to mask empty spaces in sentences in NLP), which are masked the in attention layers by always having no attention paid to those padding tokens. In order to train the model to obtain general knowledge of stars and the data set, we train without a clear learning objective in terms of input and output combinations, but rather we pick a random set of data for each star as the input and a random data point as the output which may or may not be already included in the input even for stars in the same batch during training. This training procedure necessitates being able to interact with the decoder, because we do not have a fixed output target during training (or testing). This way of training the model is similar to LLM pre-training, where the goal is to predict the next words given the starting words of sentences for OpenAI's GPT models or to predict the masked words in sentences for Google's BERT. But here we do not care about the relative ordering of the input data, but simply with learning the general relationships among data types.

Because we employ a context window length of $n=64$ as discussed in \secname\ \ref{sec:implem}, we randomly choose 5 to 64 elements from all 118 unique tokens for each star as inputs. For stars with less than 64 available non-\texttt{NaN} elements or even a handful of stars with less than 5, \texttt{NaN} elements are selected, but we always prioritize non-\texttt{NaN} elements for selection. Even if there are many stars with enough data to fill the context window, it is important to select only a few data elements sometimes to break co-adaptation between data points. An example is if C is predictable from A and B but A has more predictive power than B to C. If A and B always co-exist in the input sequence, then the model might learn A is related to C but not B. For the output node at each training epoch, we always choose one data type selected from among all non-\texttt{NaN} elements, but not including \bprp\ or \jk, because we use these for testing the model's generalization capabilities (see \secname\ \ref{subsec:new_request}).

To get a sense of how many data type combinations the model sees during training, we note that 64 unique samples (ignoring the ordering) from among 118 possible tokens comprise $\approx 1.6 \times 10^{34}$ combinations, which is 25 orders of magnitude larger than the $\approx 1.6\times 10^{9}$ used during training. It is therefore very unlikely that the model ever sees the same combination twice during training or that it sees the exact combinations on which we test the model below.

As the learning-rate scheduler for the \texttt{AdamW} \citep{2014arXiv1412.6980K, 2017arXiv171105101L} optimizer, we adapt Cosine Annealing with Warm Restarts \citep{2016arXiv160803983L}, which is implemented as \texttt{CosineAnnealingWarmRestarts} in \texttt{PyTorch}. Cosine Annealing is a type of cyclical learning rate schedule that has cycles with a cosine-shaped learning rate. We have set the initial learning rate of each cycle to $1\times10^{-4}$ and the final learning rate to $1\times10^{-10}$ with no early stopping being applied. The training process uses eight of these cycles where each cycle spans 512 epochs (thus, 4096 epochs in total for the training process). The adopted batch-size is 1024 for the best balance between computational performance (a higher batch size requires higher video-card memory) and model performance on the validation set (which is independent from the test set) during training.

As the objective function $J$, we use a robust version of the mean-squared loss that takes uncertainty in the training data into account, that is,
\begin{equation} \label{eq:robust_mse}
   J(y, \hat{y}) =\frac{(\hat{y}-y)^2}{2\,e^s} + \frac{s}{2}\,,
\end{equation}
where $y$ and $\hat{y}$ represent the ground truth and predictions, respectively and $s = \ln \big(\sigma^2_\mathrm{known} + \sigma^2_\mathrm{predictive}\big)$, with $\sigma^2_\mathrm{known}$ the uncertainty of the ground-truth data and $\sigma^2_\mathrm{predictive}$ the model's predictive uncertainty. This is the same loss function as we have used in many of our previous works (e.g., \citealt{2019MNRAS.483.3255L,2019MNRAS.489.2079L}). Further technical details of the training process are given in \appenname\ \ref{sec:technical_dl}.

\section{Results}\label{sec:result}

To illustrate the general capabilities of our trained model, we apply our model to multiple tasks that is was not specifically trained or even fine-tuned to do. During the inference process all weights of the model are used, so the users of the model do not need to specifically programme a routine for each task. For example, models like conditional autoencoders can do discriminative and generative tasks, but only the encoder participates in the discriminative task and only the decoder in the generative task. To the best of our knowledge, this is the first model in astronomy that has such multi-tasking capabilities with all weights participating. In the following subsections, we will show the result of applying our model to map stellar spectra to stellar parameters in \secname\ \ref{subsec:spec2labels}, to determine stellar spectra for given stellar parameters in \secname\ \ref{subsec:label2spec}, to infer a portion of a stellar spectrum based on another portion in \secname\ \ref{subsec:spec2spec}, to find relations between stellar parameters in \secname\ \ref{subsec:label2label}, and to recover the interstellar extinction curve in \secname\ \ref{subsec:emprical}. 

Unless otherwise specified, a scatter $\sigma$ refers to a robust measurement of the scatter that make use of the median absolute deviation (MAD): $\sigma = 1.4826$ MAD. The 1.4826 factor is applied such that for a Gaussian distribution, $\sigma$ equals the Gaussian standard deviation. All results, unless specified otherwise, are obtained either by applying the model to stars in the test set, which is independent from and identically distributed to the training set, or by using the model on data not corresponding to real stars.

\subsection{Spectra to Stellar Parameters}\label{subsec:spec2labels}

Deep learning provides a way to quickly infer stellar parameters from stellar spectra directly. Here, as the first task we test our model on, we will check if the model can infer stellar parameters and their uncertainties reasonably for different sets of inputs. We again emphasize that unlike in traditional machine-learning methods for inferring stellar parameters from observed spectra, we are using a \textit{single} model where all weights participate in the inference and our model has never been specifically trained or even fine-tuned to predict the requested labels from the specific set of inputs that we test on. 

Results for predicting \teff\ are shown in \figurename\ \ref{fig:teff}. An extended version of this figure that also includes results for \logg\ and \xh{M} is given in \figurename\ \ref{fig:teff_logg_mh} in \appenname\ \ref{sec:performance}. To put our results in this section into context, we compare to a baseline model using traditional machine-learning methods. Specifically, we train dedicated \texttt{XGBoost} models for each specific set of inputs and outputs as shown \figurename\ \ref{fig:teff} to give an idea of how our \emph{single} NN model is doing on a variety of input combinations compared to more traditional methods. For the case where we input the first 30 BP and 30 RP coefficients as well as colors, our model performs slightly better than the \texttt{XGBoost} model. The results we obtain using our model are also similar to works such as \citet{2022ApJ...941...45R}, who specifically train an \texttt{XGBoost} model on all XP coefficients and on \tmass\ and WISE photometry. As shown in the top-right panel of \figurename\ \ref{fig:teff}, our model is robust even when only a few XP coefficients are provided as inputs. The bottom-left panel of \figurename\ \ref{fig:teff} shows that our model performs well even for a random subset of XP coefficients and colors. The model's predictive uncertainty also make sense, in that less accurate predictions compared to the ground-truth are assigned high uncertainty by the model. 

Two particularly stringent tests of our model are given in the bottom-middle and bottom-right panels of \figurename\ \ref{fig:teff}. In the middle panel, we give the model the 30 least informative BP and RP coefficients and in this case we find correctly that the predictions are inaccurate compared to the ground truth but also that the model knows that it is very uncertain about those predictions. In the right panel, we randomly mix in \teff, \logg\ and \xh{M} among these least informative XP coefficients on a star by star basis. In this case, the model returns almost perfect predictions with very low uncertainty, because the requested information is already present in the inputs. But we have to emphasize that the model has not been explicitly instructed to learn the identity mapping if the information request already exists in the input sequence. This behavior therefore only occurs within the boundaries of the training set's stellar parameter space. For example, if we give $\teff = 10,000$ K and request \teff, the model returns a \teff\ that is far from the input \teff\ with large uncertainty.

Overall, our model performs well and about as well as can be expected given the data (as shown by comparing to the expertly-trained \texttt{XGBoost} models). \figurename\ \ref{fig:disrminate} shows additional performance tests of our model, where the inputs are the first 30 BP and RP coefficients and the \jh, \jk, and \bprp\ colors along with one padding token, which we believe to be the best combination to fill up the 64 token context window for this task. The left panel displays the Kiel diagram of the test set without any cuts on the model uncertainty. The distribution is reasonable with expected trends and features such as the red clump. The middle panel shows that the model predicts $E(B-V)$ to an accuracy of $0.04$ mag on the scale of $E(B-V)_\mathrm{C19}$, while the right panel demonstrates that we obtain spectro-photometric distances to an accuracy of $13.4\%$. Because our model can use any combination of inputs given to it, we can determine these parameters based on, e.g., XP coefficients and \tmass\ colors or just XP coefficients when \tmass\ colors are unavailable, without having to train separate models for these different input options.

\begin{figure*}
\centering
\begin{tikzpicture}
\node (n1) [draw=param2speccolor, line width=\figboxlw,rounded corners=\figboxcorner, inner xsep=\figboxsep, inner ysep=\figboxsep]
{\includegraphics[width=\textwidth]{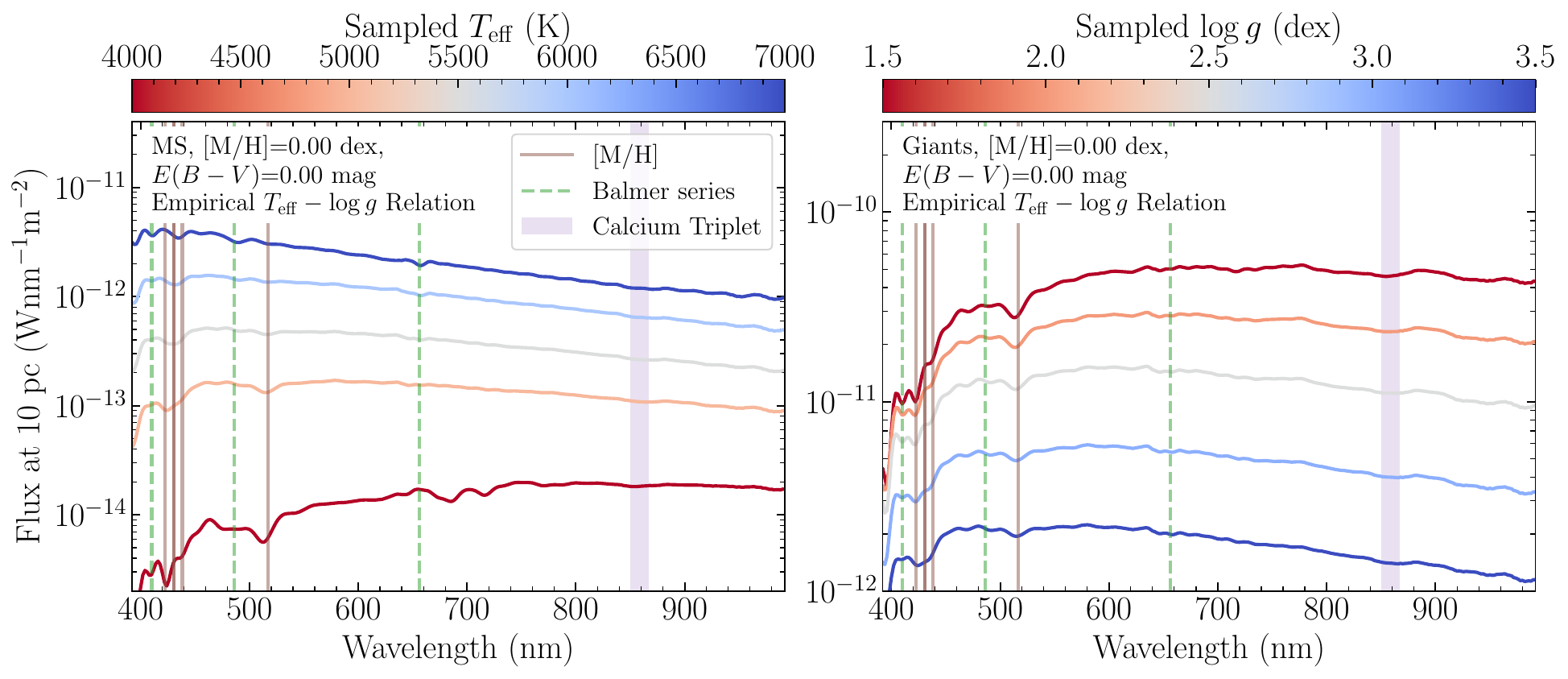}};
\node[font=\fontsize{\figheadfontsize}{0}\selectfont, text=param2speccolor] at (-4.5, 4.3) {Task: Stellar Parameters to Stellar Spectra};
\end{tikzpicture}
\caption{Predicting stellar spectra based on stellar parameters using our model. The left panel shows our model's prediction of \gaia\ XP spectra for dwarf-type stars with solar metallicity and zero extinction for \teff\ sampled between $4,000\,\mathrm{K}$ to $7,000\,\mathrm{K}$ for every $1,000\,\mathrm{K}$ with \logg\ given by the empirical $\teff-\logg$ relation given by \eqnname\ \ref{eq:ms_teffogg}. The right panel gives our model's prediction of \gaia\ XP spectra for giant-type stars of solar metallicity and zero extinction with \logg\ sampled between $1.5$ and $3.5$ for every 0.5 dex and \teff\ obtained from the empirical $\teff-\logg$ relation from \eqnname\ \ref{eq:rgb_tefflogg}. Both panels display the locations of a few \xh{M} Fraunhofer lines, of the Balmer series, and of the Calcium Triplet. The spectra are the predicted flux densities at 10 pc, which are directly predicted by our model as it can return the absolute magnitude, so both the shape and the flux-density level is predicted by the model. The model correctly predicts that hot main-sequence stars are brighter than cool main-sequence stars, while low-\logg\ giants are predicted to be brighter than higher \logg\ giants.}
\label{fig:spec_recon_logg}
\end{figure*}

\begin{figure*}
\centering
\begin{tikzpicture}
\node (n1) [draw=param2speccolor, line width=\figboxlw,rounded corners=\figboxcorner, inner xsep=\figboxsep, inner ysep=\figboxsep]
{\includegraphics[width=\textwidth]{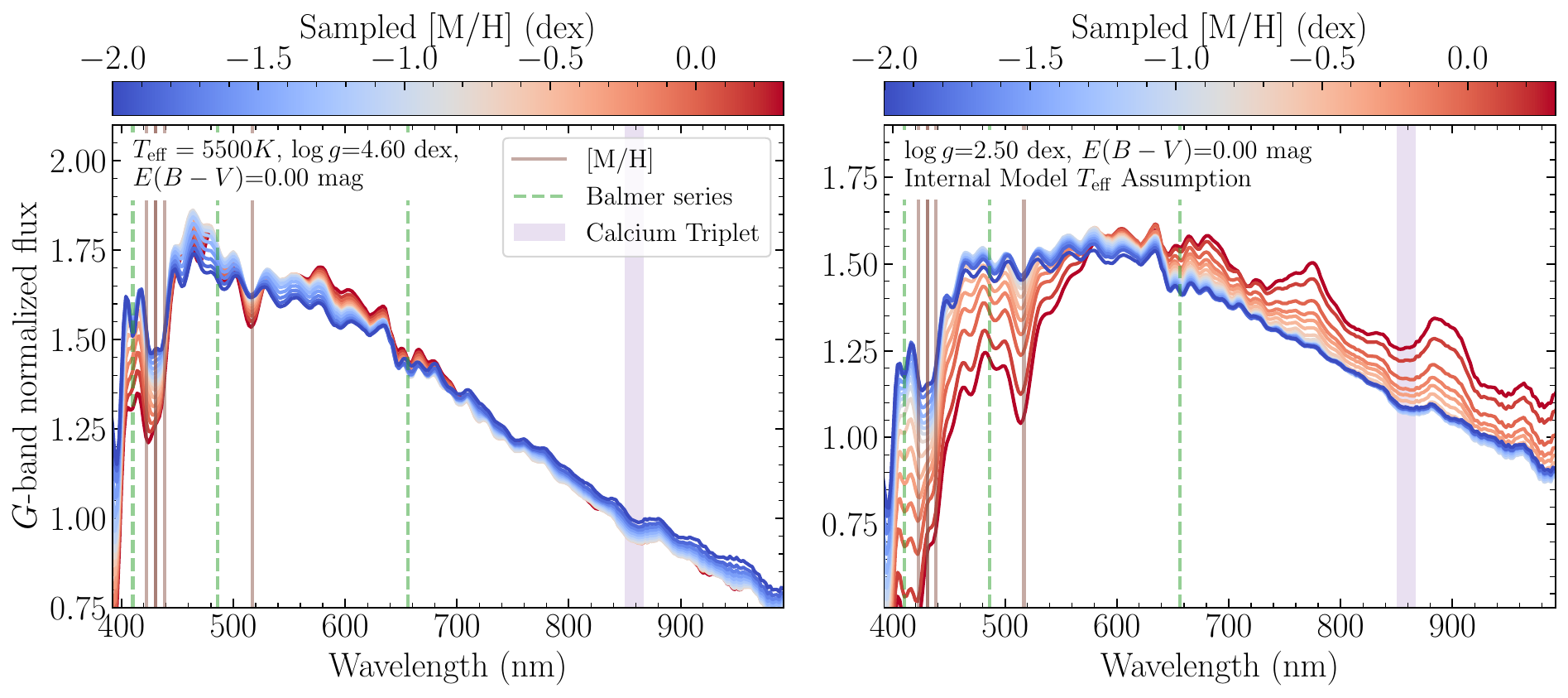}};
\node[font=\fontsize{\figheadfontsize}{0}\selectfont, text=param2speccolor] at (-4.5, 4.4) {Task: Stellar Parameters to Stellar Spectra};
\end{tikzpicture}
\caption{Spectral shape and absorption lines of spectra predicted by our model. The spectra shown in this figure are similar to those in \figurename\ \ref{fig:spec_recon_teff_mh}, except that we use the $G$-band normalized flux density (that is, the flux density divided by the $G$-band flux) instead of the flux density at 10 pc, because we want to focus on the shape of the predicted spectra. The left panel shows our model's predicted normalized \gaia\ XP spectra for main-sequence stars with $\teff=5,500$ K, $\logg=4.60$ dex, zero extinction, and \xh{M} sampled between $-2.0\,\mathrm{dex}$ and $0.3\,\mathrm{dex}$. The right panel panel shows predicted normalized \gaia\ XP spectra for giant-type stars with $\logg=2.50$ dex, zero extinction, and \xh{M} sampled between $-2.0\,\mathrm{dex}$ to $0.3\,\mathrm{dex}$. In the right panel, we do not provide \teff\ as an input, so the model has to internally assume a \teff\ from a $\teff-\logg-\xh{M}$ relation that it has learned during training. For giants, this means that the model must know that metal-rich giants are cooler than metal-poor giants at the same \logg, as can be seen in training data in \figurename\ \ref{fig:training_set}. In both panels, we see that our model correctly produces deeper metal absorption lines, especially at $516.891\,\mathrm{nm}$, for high \xh{M}\ stars.}
\label{fig:spec_recon_teff_mh}
\end{figure*}

\begin{figure*}
\centering
\begin{tikzpicture}
\node (n1) [draw=param2speccolor, line width=\figboxlw,rounded corners=\figboxcorner, inner xsep=\figboxsep, inner ysep=\figboxsep]
{\includegraphics[width=\textwidth]{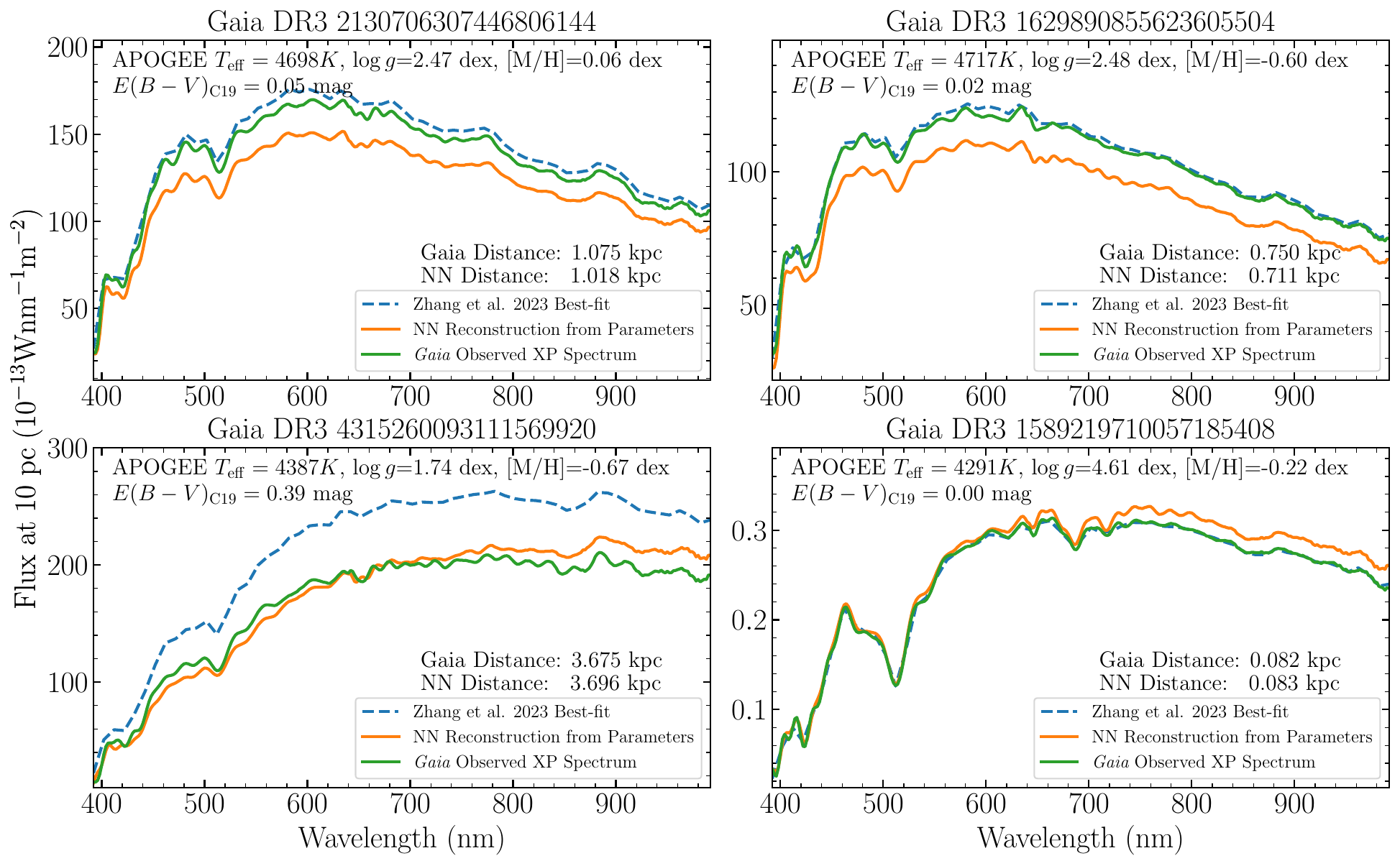}};
\node[font=\fontsize{\figheadfontsize}{0}\selectfont, text=param2speccolor] at (-4.5, 5.95) {Task: Stellar Parameters to Stellar Spectra};
\end{tikzpicture}
\caption{Model reconstruction of the \gaia\ XP spectra for stellar parameters sampled from four real-world APOGEE stars. We predict these spectra by giving the model the observed APOGEE \teff, \logg, \xh{M}, and the reddening $E(B-V)_\mathrm{C19}$ obtained from the extinction map at the distance given by the inverse \gaia\ parallax. We then request all \gaia\ XP coefficients as well as the expected stellar luminosity contaminated with extinction from the model, to obtain the predicted flux density at $10$ pc. The top-left panel has a giant-type star with near-solar metallicity and almost no extinction, while the top-right panel displays a giant with similar extinction, \teff, and\logg as the top left star, but with much lower metallicity. The bottom-left panel displays a lower metallicity bright giant with some extinction. Finally, the bottom-right panel has a dwarf-type star with no extinction. In all cases, our model accurately predicts the shape of the spectra and the main noticeable difference is the slight amplitude offset that results from a small error in the predicted distance. Best-fit spectra from \citet{2023MNRAS.524.1855Z} are also shown as dashed lines.}
\label{fig:spec_comparison}
\end{figure*}

\subsection{Stellar Parameters to Spectra}\label{subsec:label2spec}

Given the flexibility in the inputs and outputs of our model, we can invert the machine-learning problem from \secname\ \ref{subsec:spec2labels} without needing to change the model. That is, we can simply provide different sets of input and of output requests without needing to consider which parts of the model need to be involved, because we again use all the same weights inside the model as before. Here we explore the performance of our model when predicting optical stellar spectra for given sets of stellar parameters. Because we are working with XP coefficients, we use the \texttt{GaiaXPy}\footnote{\url{https://github.com/gaia-dpci/GaiaXPy}} \texttt{Python} package to convert XP coefficients to physical stellar spectra of flux density versus wavelength to visualize the results.

In some of our results, we need to adopt approximate $\teff-\logg$ relations for dwarf-type and giant-type stars used in \citet{2023MNRAS.524.1855Z}. These are
\begin{equation}\label{eq:ms_teffogg}
  {\logg}_\mathrm{dwarf} = \begin{cases}
  4.6 & \mbox{ for $ \teff < 5000 $ K}\\
  4.6-0.0005 (\teff-5000)  & \mbox{ for 5000 K $\leq \teff<$6300 K}\\
  3.95 & \mbox{ for $ \teff \geq$6300 K}
  \end{cases}
\end{equation}
for for solar-metallicity dwarfs and \begin{equation}\label{eq:rgb_tefflogg}
  {\teff}_\mathrm{,\ RGB} = \begin{cases}
  5200-441.86(3.65-\logg) & \mbox{ for $ \logg < 3.65 $ dex}\\
  5900-1400(4.15-\logg) & \mbox{ for $ \logg \geq 3.65 $ dex}
  \end{cases}
\end{equation}
for solar-metallicity giants.

Simulated spectra with a range of stellar parameters are shown in \figurename s \ref{fig:spec_recon_logg} and \figurename\ \ref{fig:spec_recon_teff_mh}. \figurename\ \ref{fig:spec_recon_logg} displays spectral reconstructions of the flux density at a distance of 10 pc. The left panel has spectra for dwarf-type stars along the sequence given by Equation \eqref{eq:ms_teffogg} and shows that the higher the \teff, the higher is the luminosity, but also the clearer the Balmer lines are, and the weaker most metal lines are. The right panel is for giant-type stars along the sequence given by Equation \eqref{eq:rgb_tefflogg} and shows that giants with higher \teff\ and, thus, higher \logg\ have lower luminosity, while a similar \teff\ attenuation effect on the metal lines can be observed as for dwarfs.

\figurename\ \ref{fig:spec_recon_teff_mh} shows predicted spectra of simulated stars with different metallicity \xh{M}, with the flux density now normalized by the $G$-band apparent flux to be able to focus on the effect of changing \xh{M} on the shape of the spectrum and on the appearance of absorption lines. The left panel shows dwarf-type stars for which we can clearly see the metal lines deepening as the metallicity increases. The right panels show giant-type stars, but rather than specifying \teff, we let the model internally assume \teff. That is, we do not use any empirical $\teff-\logg-\xh{M}$ relation to calculate what the \teff\ should be, but instead just do not provide any \teff\ to the model (see the discussion in \secname\ \ref{subsec:label2label} below). As can be seen in the top-left panel of \figurename\ \ref{fig:training_set}, for giants of a fixed \logg, increasing \xh{M} corresponds to decreasing \teff. This effect can be seen in the simulated spectra, where spectra have vastly different shapes due to the changing internally-assumed \teff. Metal lines are clearly deeper as \xh{M} increases.

\figurename\ \ref{fig:spec_comparison} shows a few comparisons of predictions from our model to real-world stellar data to check if the simulated \gaia\ XP spectra predicted using our model based on precise stellar parameters derived from high resolution APOGEE compare well to the actual observe \gaia\ XP spectra for a few example stars. As in \figurename\ \ref{fig:spec_recon_logg}, we normalize all of the real observed spectra to give the flux density if the star were at 10 pc using the \gaia\ $G$-band apparent magnitude as well as the observed parallax. In \figurename\ \ref{fig:spec_comparison}, the top-left panel has a giant-type star with near solar metallicity, while the top-right panel has a star with similar parameters except that its metallicity is lower. In both cases, the reconstructed spectra match the observed spectra overall well and all large- and small-scale features match those seen in the observed spectra. The main difference is a small error in the intrinsic luminosity which causes the overall flux density level to be slightly off. The distance inferred by our model as well as the geometric distance obtained from the \gaia\ parallax is also given in the plots. We see that the inferred distances are actually highly accurate, even if the flux level visually looks to be quite off. The bottom-left panel shows a bright giant with some extinction and the bottom-right panel has a dwarf-type star with no extinction. In both cases, the reconstructions are quite good and the model predicts the luminosity quite well, leading to a highly accurate predicted distance and a good match between the flux density amplitude of the predicted and observed spectra. In all panels, we show best-fit spectra from \citet{2023MNRAS.524.1855Z} for the same stars as reference comparison. One thing to note is that \citet{2023MNRAS.524.1855Z} find the stellar parameters that best fit the observed spectrum while our reconstruction comes purely from ground-truth stellar parameters. This causes the \citet{2023MNRAS.524.1855Z} reconstruction to in general be much closer to \gaia, except in the bottom right panel where \citet{2023MNRAS.524.1855Z}'s spectrum is off due to the difference in zero-point correction used in this work.

Overall, our model is able to predict realistic XP spectra for stars within the wide range of stellar parameters covered by the training set.

\begin{figure*}
\centering
\begin{tikzpicture}
\node (n1) [draw=spec2speccolor, line width=\figboxlw,rounded corners=\figboxcorner, inner xsep=\figboxsep, inner ysep=\figboxsep]
{\includegraphics[width=\textwidth]{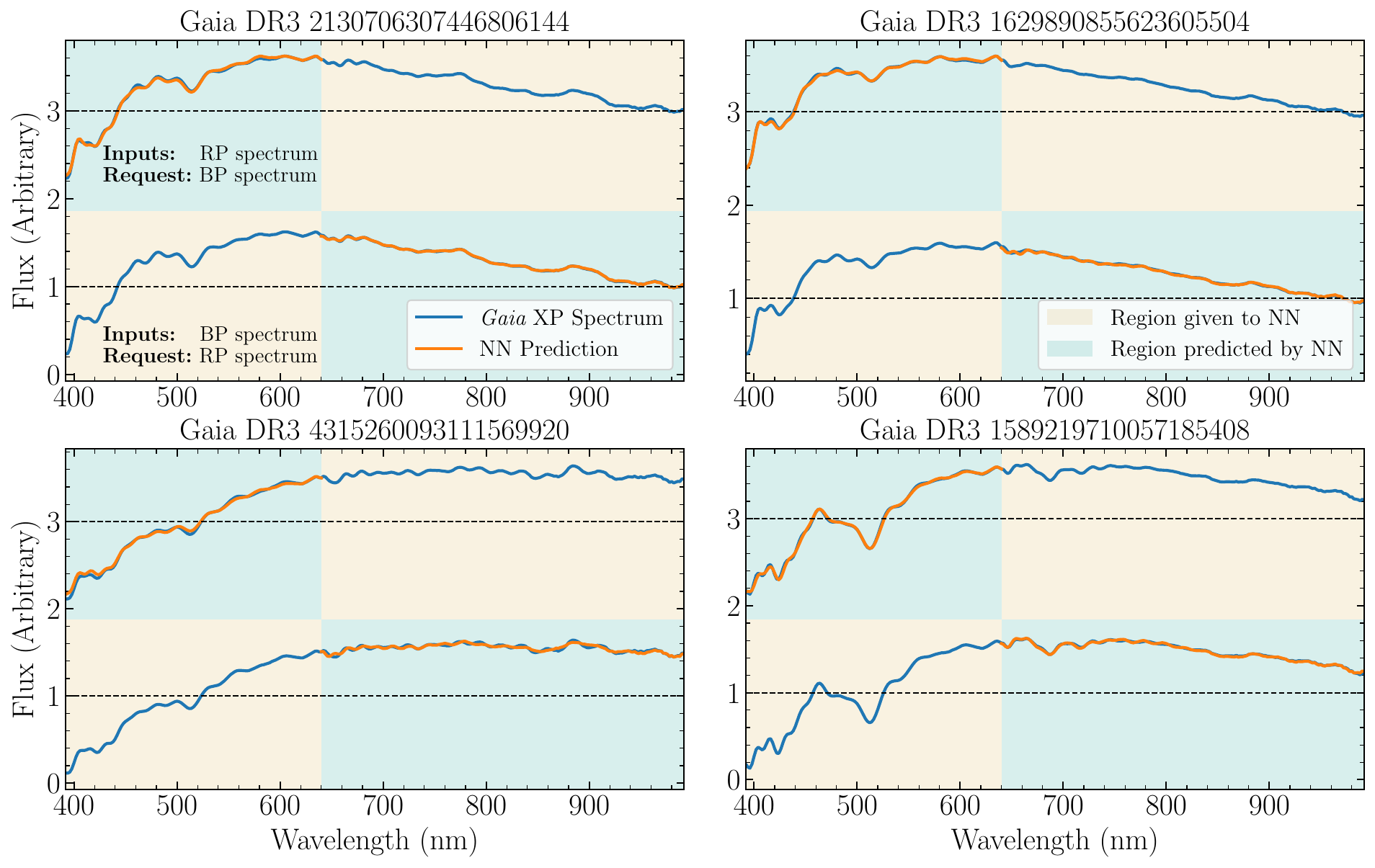}};
\node[font=\fontsize{\figheadfontsize}{0}\selectfont, text=spec2speccolor] at (-4.9, 6.05) {Task: Stellar Spectra to Stellar Spectra};
\end{tikzpicture}
\caption{Reconstructing parts of a spectrum based on other parts. Each of the four panels in this figure shows our model's reconstruction of a portion of the $G$-band normalized \gaia\ XP spectrum when providing the model with another portion of the \gaia\ XP spectrum of the same star (for the same APOGEE stars as in \figurename\ \ref{fig:spec_comparison}). The blue shaded area represents the rough wavelength range given to the model while the beige shaded area is the rough wavelength range predicted by the model. The black dashed lines in each panel are reference lines that allow one to compare the top and bottom spectra in each panel. In all cases, the model successfully reconstructs portions of the \gaia\ XP spectra.}
\label{fig:spec2spec}
\end{figure*}

\subsection{Spectra to Spectra}\label{subsec:spec2spec}

In addition to inputting or outputting stellar parameters, our model also works without involving any stellar parameters and can go directly, e.g., from one portion of a stellar spectrum to another portion, because the input portion of the spectrum provides enough context. In other words, the model has a good internal representation of a star given just a portion of the spectrum without explicitly having to involve the stellar parameters.

\figurename\ \ref{fig:spec2spec} displays $G$-band normalized XP spectra for the same four stars as in \figurename\ \ref{fig:spec_comparison}. In the test in \figurename\ \ref{fig:spec2spec}, we provide the RP portion of the spectrum to the model and then request both the RP and BP portions of the same spectrum, because we need both the BP and RP portions to use \texttt{GaiaXPy} to convert back to physical spectra and we use the RP and BP coefficients returned by the model to create the final predicted XP spectrum. For each star, we repeat this procedure giving the BP portion of the spectrum and requesting the RP portion. The BP and RP portions of a star's spectrum only overlap over a very short wavelength range, so the portion of the spectrum that is given to the model does not contain much information about the portion that we request. In both cases of giving BP or giving RP, the model reconstructs the missing portion of the spectrum well and, in fact, much better than in \figurename\ \ref{fig:spec_comparison}, where the spectra are reconstructed based on the stellar parameters. This is likely because the measurement of the stellar parameters by APOGEE is uncertain, while the reconstruction in \figurename\ \ref{fig:spec2spec} does not involve uncertain stellar parameters, but just certain spectral portions. This is the same reason as why many works now prefer to employ unsupervised or self-supervised training of models that do not include very precise labels. For example, see the papers of \citet{2023MNRAS.521.2745S} and \citet{2023arXiv230706378L} for applications to \gaia\ XP spectra specifically.

\begin{figure*}
\centering
\begin{tikzpicture}
\node (n1) [draw=param2paramcolor, line width=\figboxlw,rounded corners=\figboxcorner, inner xsep=\figboxsep, inner ysep=\figboxsep]
{\includegraphics[width=0.95\textwidth]{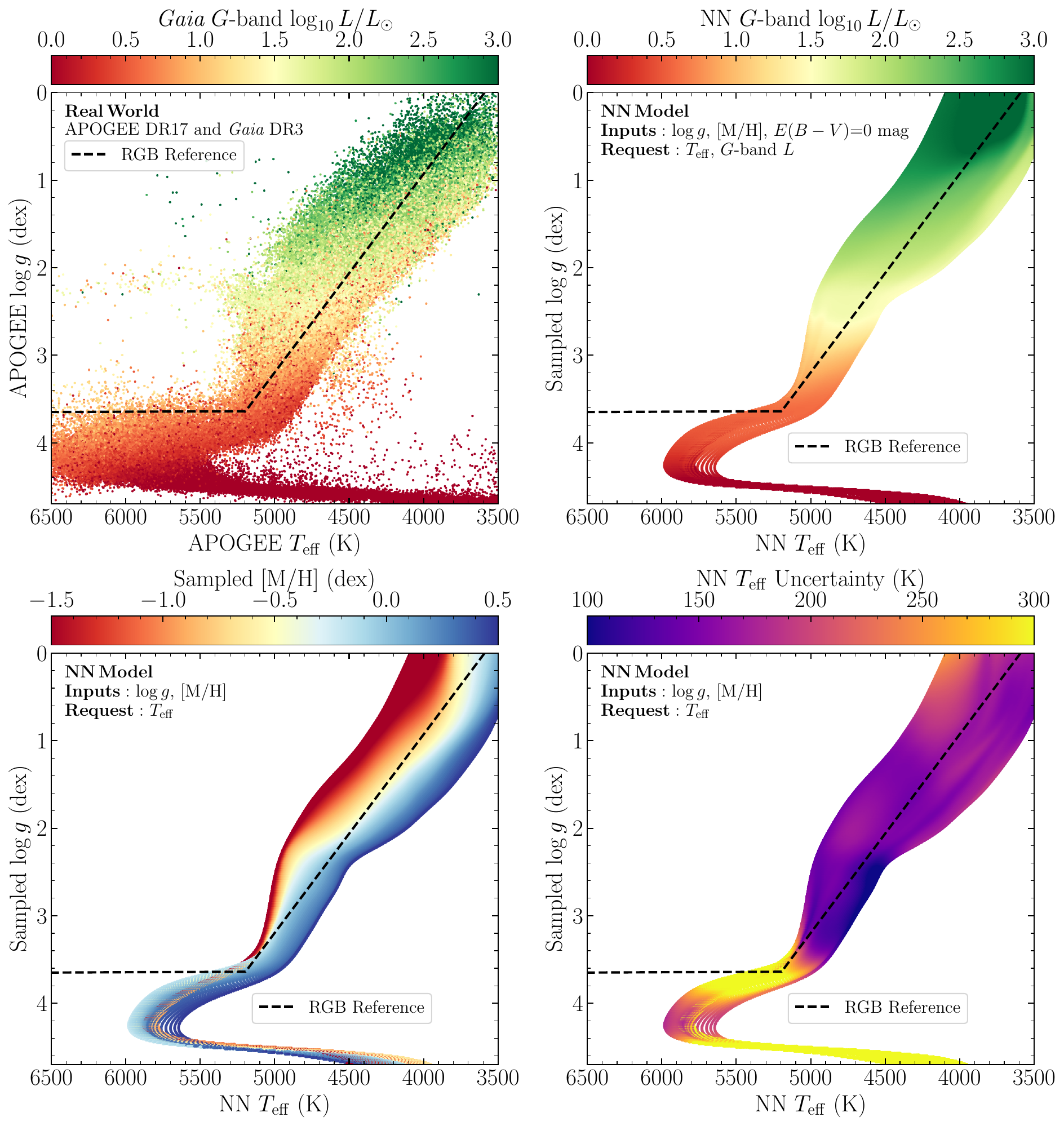}};
\node[font=\fontsize{\figheadfontsize}{0}\selectfont, text=param2paramcolor] at (-
3.7, 9.4) {Task: Stellar Parameters to Stellar Parameters};
\end{tikzpicture}
\caption{Our model's perception of the Kiel diagram. To investigate our model's understanding of the relations among stellar parameters, we sample a grid of evenly-spaced \logg\ from $0\,\mathrm{dex}$ to $5\,\mathrm{dex}$ dex with a spacing of $0.01\,\mathrm{dex}$ and \xh{M}\ from $-2\,\mathrm{dex}$ to $0.5\,\mathrm{dex}$ with a spacing of $0.02\,\mathrm{dex}$ and we request \teff\ and the stellar luminosity. The spacings are simply chosen to create a dense sampling of the diagram.
%In an ideal scenario, the model should have learnt parameters to parameters relationships such that those relationships will minimize the objective function if it is during training as opposed to the model actually learnt the underlying physical laws. 
\textbf{Top left} panel: the ground-truth diagram from APOGEE color-coded by the \gaia\ $G$-band logarithmic luminosity in solar units. \textbf{Top right} panel: our model's Kiel diagram similarly color-coded with our model's stellar luminosity. Neither of the luminosities in the top left and right panels are corrected for extinction. One can clearly see that the model reproduces the patterns seen in the top left panel, reconstructing both the \teff\--\logg\ stellar evolutionary tracks as well as the luminosity along those tracks. \textbf{Bottom left} panel: our model's Kiel diagram color-coded by our the input \xh{M}, which again displays the expected trends for red giants (see \figurename\ \ref{fig:training_set} for the ground truth in the training set). \textbf{Bottom right} panel: our model's Kiel diagram color-coded by the uncertainty in the predicted \teff, demonstrating that while the $\teff-\logg-\xh{M}$ trend is clear and tight in red giants, this is not the case for sub-giants and dwarfs. The bottom left panel shows that tracks with different \xh{M} overlap for these stellar types and the bottom right panel shows that this leads to the \teff\ prediction based on \logg\ and \xh{M} being quite uncertain for sub-giants and dwarfs. The red dashed line is a reference line for red-giant branch stars, which is described in \eqnname\ \eqref{eq:rgb_tefflogg}.}
\label{fig:kiel_perception}
\end{figure*}

\subsection{Stellar Parameters to Stellar Parameters}\label{subsec:label2label}

To assess the quality of the astrophysical knowledge that our model has acquired during training, we can request relations between stellar parameters from our trained model without involving any observational data. This way, we can test if the model has actually learned basic relationships between stellar parameters when it creates an internal ``perception'' of a star. 

\figurename\ \ref{fig:kiel_perception} shows that our model can reconstruct the distribution of stars observable by APOGEE in the Kiel diagram of \logg\ versus \teff\ (like the one in \figurename\ \ref{fig:training_set}). Although one cannot currently simply ask questions such as ``Please give me your concept of the Kiel diagram'', we can still extract this knowledge from the model by, e.g., creating uniformly sampled grids of \logg\ and \xh{M}\ as inputs to the model and then asking for \teff\ or other stellar parameters. If the model has learned the general properties of stars (specially stars within the training set's boundaries), the model should predict \teff\ in a way that recreates the Kiel diagram of the training set. And indeed, we see that the model is clearly able to recreate the Kiel diagram. The top two panels of \figurename\ \ref{fig:kiel_perception} show the ground truth of $\teff-\logg$ color-coded by the logarithmic $G$-band luminosity from \gaia\ as well as from the model. We see that not only can the model reproduce the expected \teff\ trends, it can also reproduce the run of stellar luminosity just from stellar parameters alone. The bottom left panel of \figurename\ \ref{fig:kiel_perception} shows the usual Kiel diagram color-coded by \xh{M} and the expected clear \xh{M} trend can be clearly seen for giant-type stars (see \figurename\ \ref{fig:training_set} for the ground truth as a reference), where metal-rich giants are cooler than metal-poor giants at the same \logg. The bottom right panel of the same figure shows the uncertainty in the predicted \teff, which is small for giant-type stars, because there is a simple relation between $\teff-\logg-\xh{M}$ for giants, thus, making \teff\ highly predictable in this scenario. For dwarf-type stars, this is not the case (i.e., we do not see a nice \xh{M} color gradient for dwarfs in the left panel), so the model's uncertainty on \teff\ is. Thus, our model internally contains information similar to stellar isochrones and we can extract these using this procedure. 

\begin{figure*}
\centering
\begin{tikzpicture}
\node (n1) [draw=empiricalcolor, line width=\figboxlw,rounded corners=\figboxcorner, inner xsep=\figboxsep, inner ysep=\figboxsep]
{\includegraphics[width=\textwidth]{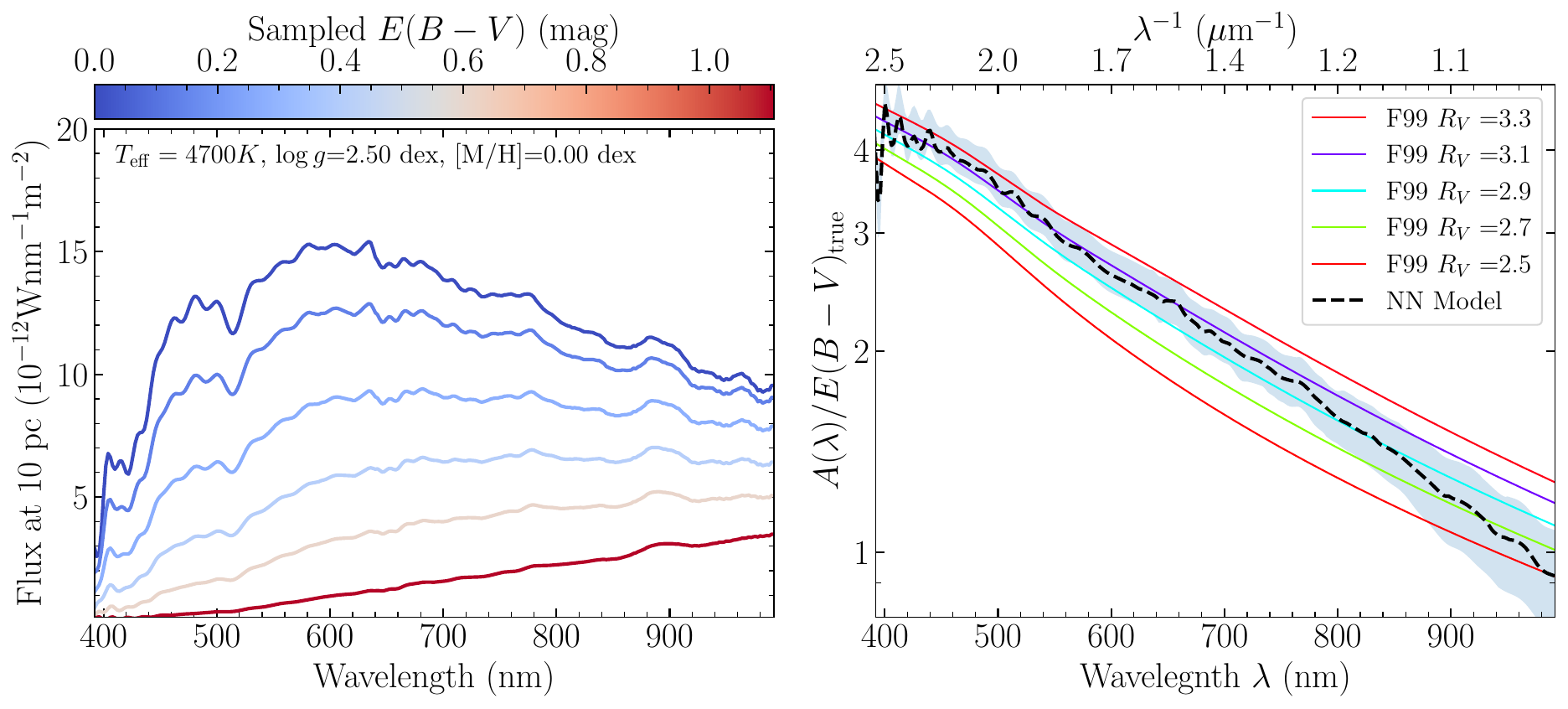}};
\node[font=\fontsize{\figheadfontsize}{0}\selectfont, text=empiricalcolor] at (-5.7, 4.45) {Task: Empirical Law Recovery};
\end{tikzpicture}
\caption{Our model's recovery of the interstellar extinction curve. The left panel demonstrate the effect of extinction on stellar spectra as learned by the model by showing the model's reconstruction of \gaia\ XP spectra for a giant with $\teff=4,700$ K, $\logg=2.50$ dex,  solar metallicity, and with different reddening values $E(B-V)$. By comparing to the zero extinction spectra, we can derive the wavelength dependence of the extinction curve and this is shown in the right panel. In this panel, the dotted black line displays this learned extinction curve $A(\lambda)/E(B-V)$ with the blue shaded region representing the uncertainty on this curve. The colored lines are the extinction curve from \citet{1999PASP..111...63F} for different $R_V$. Our derived extinction curve agrees well with the standard $R_V = 3.1$ curve, except at longer wavelengths, where the lower $R_V$ curves match better.}
\label{fig:spec_recon_extincted}
\end{figure*}

\subsection{Recovery of the Interstellar Extinction Curve}\label{subsec:emprical}

As an example of an empirical law that is not at all included in the model but that we can recover from our model, we predict the ratio of the total-to-selective extinction $R(\lambda)$ as a function of wavelength $\lambda$, defined as

\begin{equation}\label{eq:extinction_curve}
R(\lambda) = \frac{A(\lambda)}{E(B-V)_\mathrm{true}}\,,
\end{equation}
where $A(\lambda)$ is the extinction at wavelength $\lambda$ and $E(B-V)_\mathrm{true}$ is the reddening on the true scale (as opposed to the \texttt{SFD} or \texttt{C19} scale). \figurename\ \ref{fig:spec_recon_extincted} shows the extinction curve that we recover from our model.

Our model does not contain a fixed or pre-defined extinction or reddening law setup  (unlike the work of \citet{2023MNRAS.524.1855Z}), the only information about extinction in our model is the $E(B-V)_\mathrm{C19}$ value when it is available. So to determine the extinction curve, we simulate spectra of fixed stellar parameters with different $E(B-V)$ and we can then derive the model's implied extinction curve by comparing the spectra for different $E(B-V)$.  We first set up a grid of stellar parameters. To simplify the process, we use solar-metallicity giants with \logg\ ranging from $1.0\,\mathrm{dex}$ to $2.5\,\mathrm{dex}$ for every $0.1\,\mathrm{dex}$ with \teff\ determined using \eqnname\ \eqref{eq:rgb_tefflogg}, and using $E(B-V)$ of 0.0, 0.4, 0.8 and 1.2 mag, and we then request the \gaia\ XP spectra and the luminosity at 10 pc so we can obtain the absolute magnitude. The emulated spectra with $E(B-V)=0$ act as the reference extinction-free spectra $F_0(\lambda)$, while those spectra $F_E(\lambda)$ with $E(B-V)_{\mathrm{C19}}>0$ are extinguished. The extinction curve $R(\lambda)$ is then calculated as follow:
\begin{equation}\label{eq:extinction_spec}
R(\lambda) = \frac{A(\lambda)}{E(B-V)_\mathrm{true}}=\ln{\bigg(\frac{F_0(\lambda)}{F_E(\lambda)}\bigg)} \frac{1}{0.884\,E(B-V)_{\mathrm{C19}}}\,,
\end{equation}
and we do this calculation for the entire grid of stellar parameters and extinctions. We then take the median of the derived extinction curves, which is shown as the black dashed line in \figurename\ \ref{fig:spec_recon_extincted}, with the standard deviation among the grid points given as the blue curve. It is clear that the black dashed line mostly agrees with the mean extinction curve\footnote{Calculated using the \texttt{extinction} \texttt{Python} package  \citep{barbary_kyle_2016_804967}.} of \citet{1999PASP..111...63F} with $R_V\approx 3.1$, although at the red end of the spectrum, the model extinction curve prefers a lower $R_V\approx2.5$.

\begin{figure}
\centering
\includegraphics[width=0.475\textwidth]{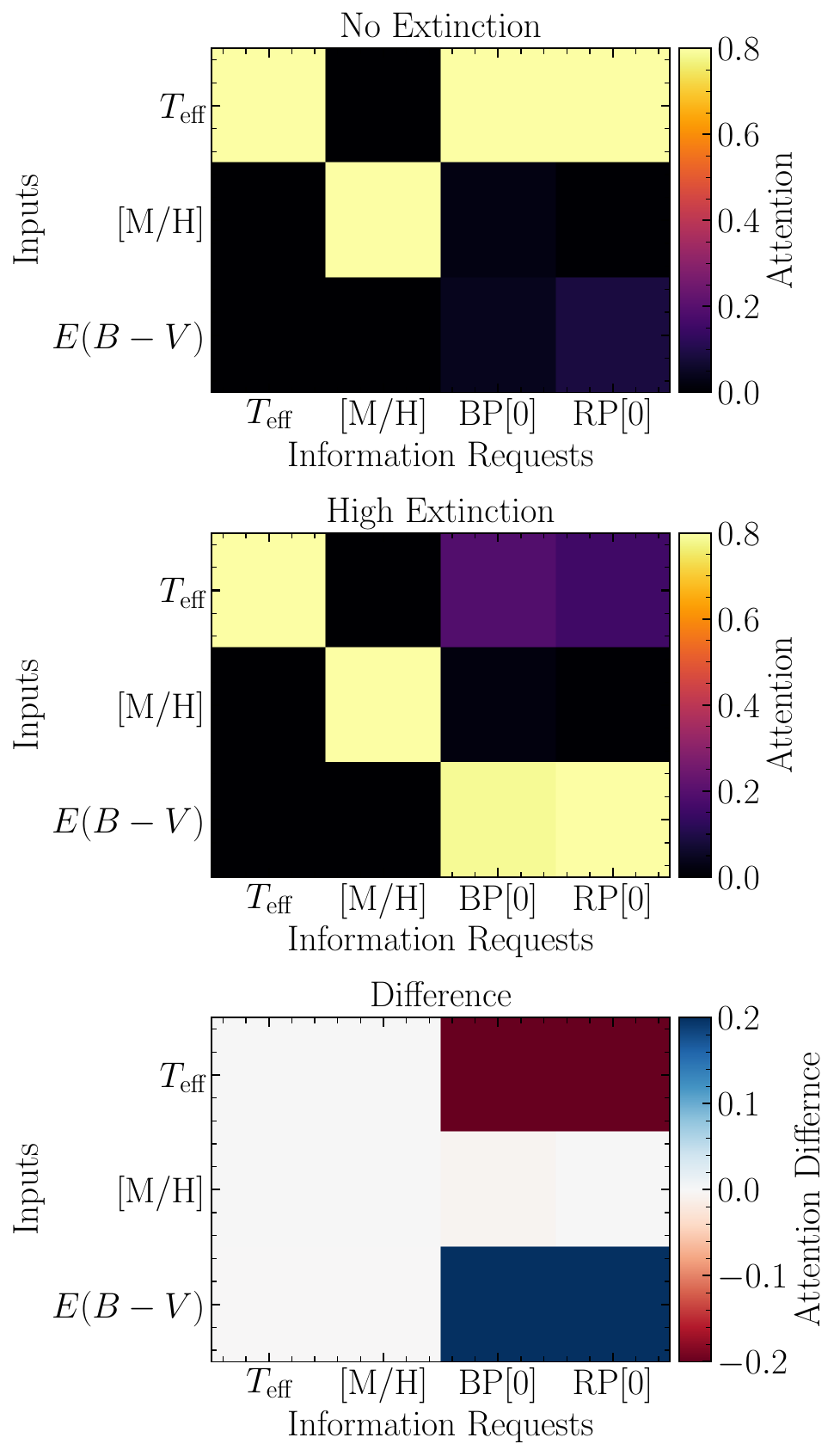}
\caption{Average cross-attention weights $A$ (see \eqnname\ \ref{eq:attention}) across all attention heads between an input sequence and a request vector. That is, for a request for information from the model, the cross-attention between the request and the inputs shows which parts of the input the model has learned are relevant. The attention weights are calculated in such a way as to make each column sum up to one. In all three panels, the inputs are \teff, \xh{M}, and $E(B-V)$ and we request \teff, \xh{M}, or the first BP or RP coefficient. The top panel considers a star with $E(B-V)=0.0$, while the middle panel uses the same star, but with high $E(B-V)$. The bottom panel gives the difference in the average attention score between the two. We see that in both $E(B-V)$ cases, the model pays attention to \teff\ and \xh{M} when determining \teff\ and \xh{M}, respectively, but that to determine the BP/RP coefficients, the model uses \teff\ in the $E(B-V)=0$ case, but relies on $E(B-V)$ in the high extinction case. This trend in attention is not possible without encoding the value of the observation in the embedding space.}
\label{fig:attention}
\end{figure}

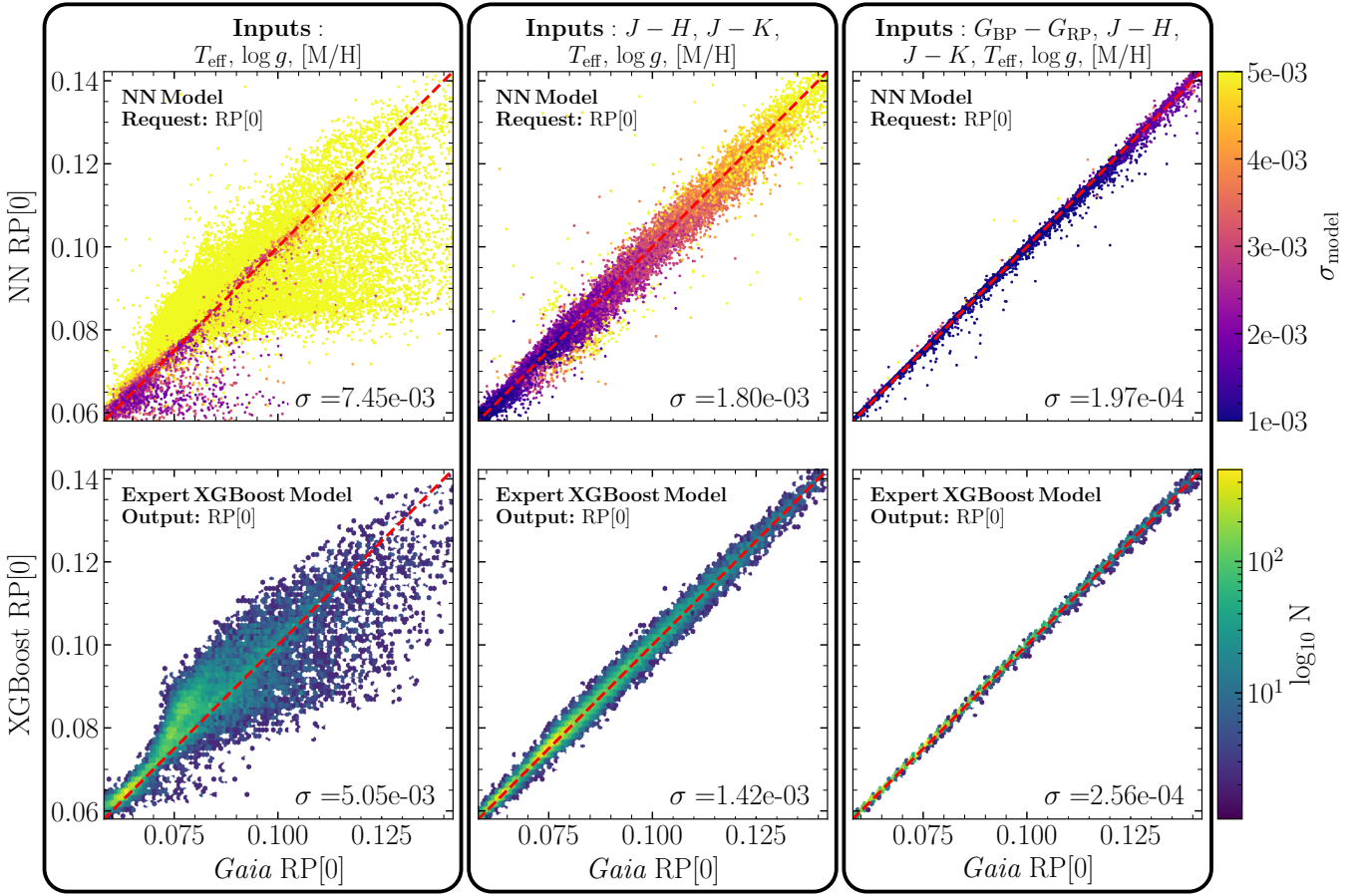
\begin{figure*}
\centering

\begin{tikzpicture}
\node[inner sep=0] at (0,0)
{\includegraphics[width=\textwidth]{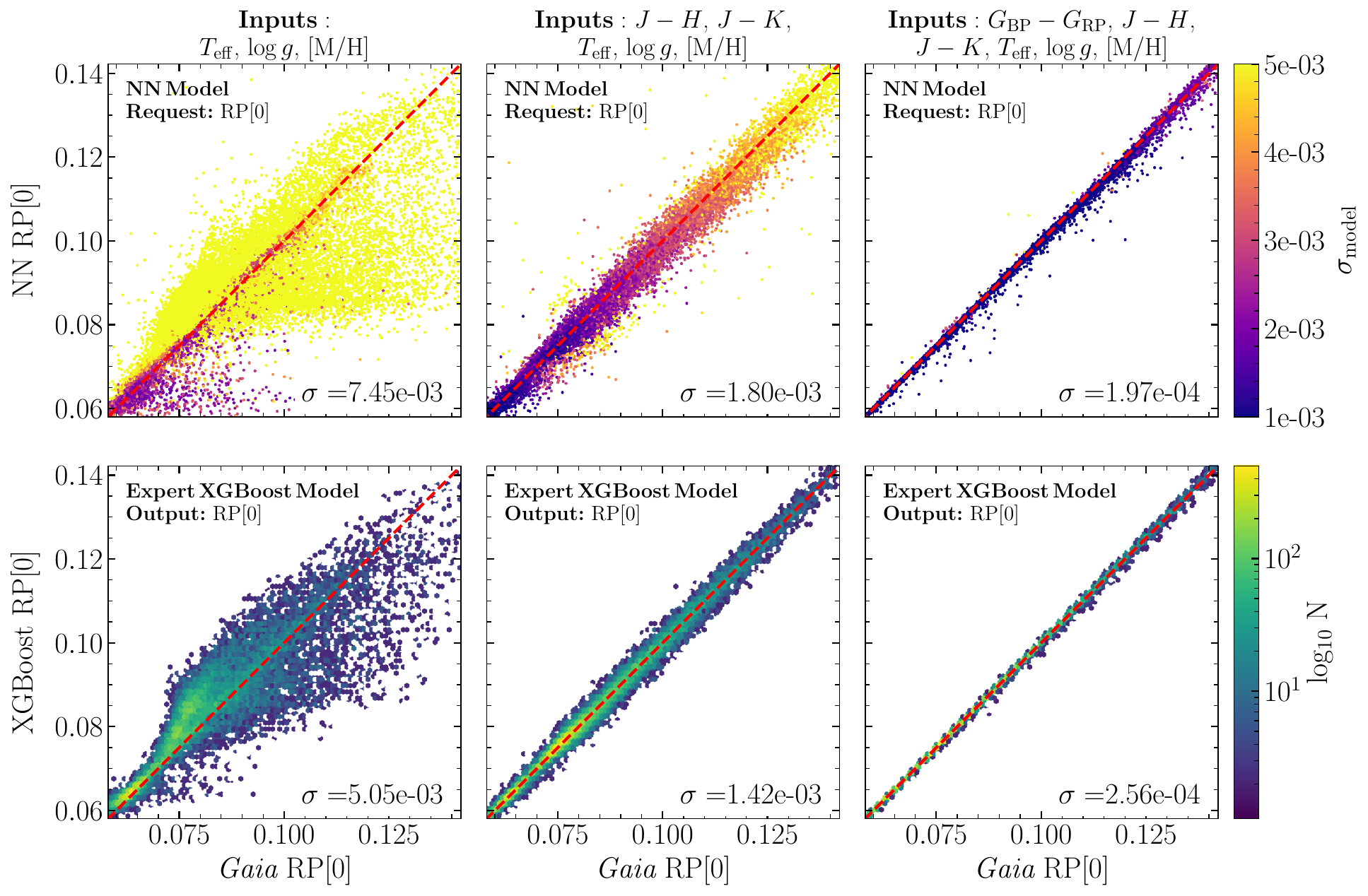}};
\draw[black,line width=\figboxlw,rounded corners=\figboxcorner] (-8.3, -5.8) rectangle (-2.8, 5.9);
\draw[black,line width=\figboxlw,rounded corners=\figboxcorner] (-2.7, -5.8) rectangle (2.15, 5.9);
\draw[black,line width=\figboxlw,rounded corners=\figboxcorner] (2.25, -5.8) rectangle (7.1, 5.9);
\end{tikzpicture}
\caption{Assessing our model by comparing to expert \texttt{XGBoost} models. The top panels show our \textit{single} model's prediction for the first RP coefficient color-coded by the model uncertainty for different combinations of inputs, while the lower panels show the predictions from separately trained \texttt{XGBoost} models for the same combinations of inputs. Both models struggle when only provided with \teff, \logg, and \xh{M} (left panels), but return good predictions if \jh\ and \jk\ are provided as well (middle panels), because these contain reddening information. When \bprp\ is provided as well, both models perform well as expected (right panels). In all cases, our model has similar performance as expertly trained models.}
\label{fig:dependence}
\end{figure*}

\begin{figure}
\centering
\includegraphics[width=0.475\textwidth]{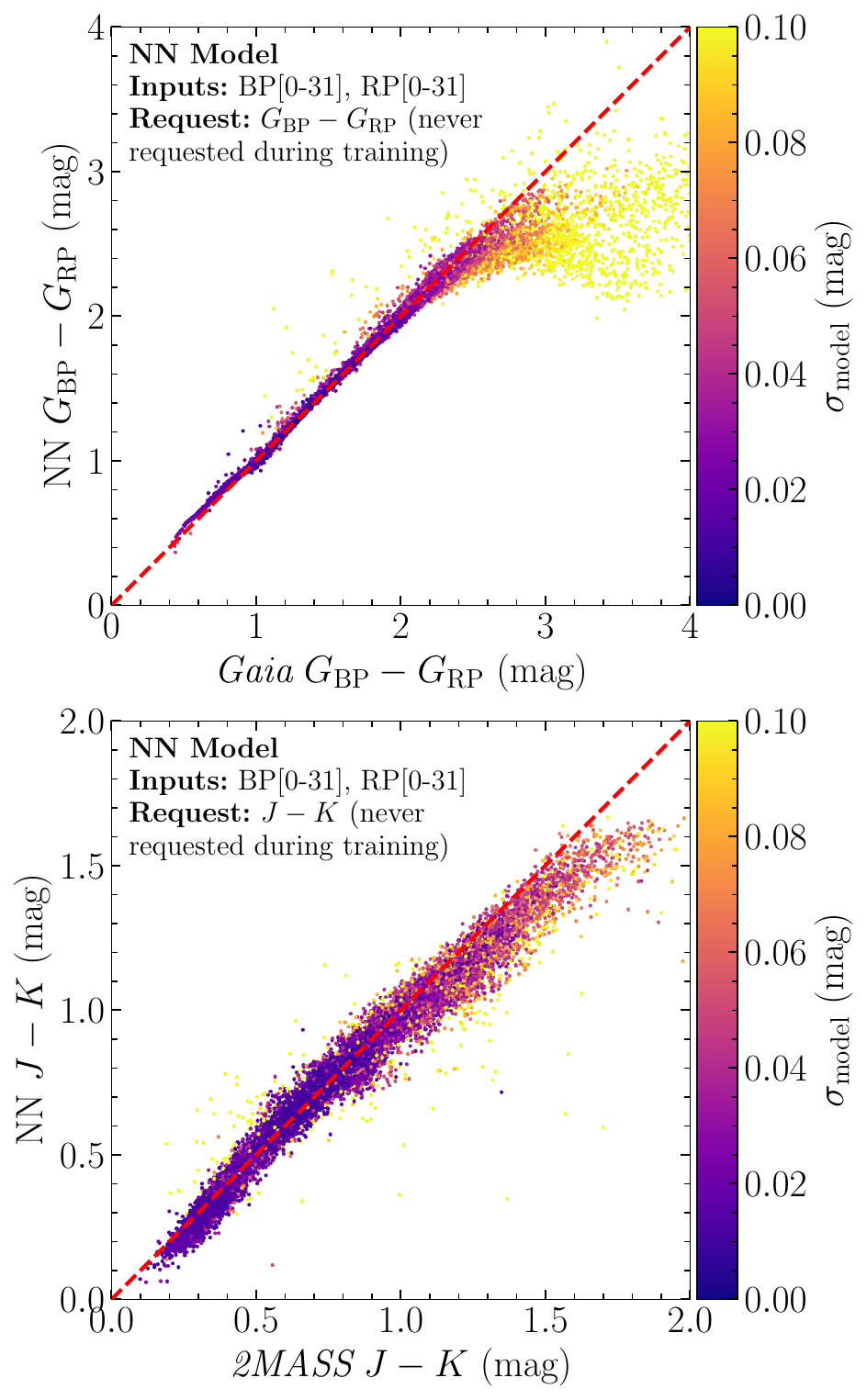}
\caption{Model performance for outputs never requested during training. The two panels of this figure show our model's prediction for the \bprp\ and \jk\ colors based on input XP spectra only. These two colors were never requested from the decoder during the training although they are used in the encoder to learn the embedding (i.e., vector representation) of these two types of data. We can clearly see that the trained model is still able to predict the two colors when requested. This demonstrates that the unit vector embedding as discussed in \secname\ \ref{subsec:embedding} and \eqnname\ \eqref{eq:embedding} is a meaningful embedding learned by the encoder that can be understood by the decoder reasonably well.}
\label{fig:new_requests}
\end{figure}

\section{Discussion}\label{sec:discussion}

In this paper, we have trained and tested a single model to perform a variety of tasks. Ultimately, the goal of this type of modeling is to create a foundation model for stars or for more general parts of astronomy (a review of neural networks and the role of foundation model in astronomy can be found in \citealt{2023RSOS...1021454S}). Our model is only a step towards this, which is why we refer to it as a ``proto-foundation model''. In this section, we discuss various aspects of our model in the context of foundation models: how our model processes information in \secname~\ref{subsec:dependences}, how the embedding that we create is meaningful and generalizable in \secname~\ref{subsec:new_request}, the reliability of our model and of future foundation models built using the same technology in \secname~\ref{subsec:reliability}, and a comparison of our model to other foundation models in astronomy in \secname~\ref{subsec:foundation}. 

\subsection{Parameters Dependencies}\label{subsec:dependences}

One quick way to check what the model has learned is to investigate the dependencies among parameters inside the model. We provide two examples of such investigations: one looking at the attention scores from the attention layers in the model and one comparing our single model's performance to separately-trained \texttt{XGBoost} models for different combinations of input parameters.

We can learn about how our model processes inputs and output information requests by looking at the attention paid to them by the model. As an example, we consider the cross-attention between input sequences and information requests. \figurename\ \ref{fig:attention} shows how the cross-attention changes for two cases, one with no extinction and one with high extinction. The first feature that can be seen in this figure is that in both cases, requesting \teff\ and \xh{M}\ solely pays attention to \teff\ and \xh{M} and it is therefore not surprising that the model can predict very accurate parameters when the parameters are given as inputs (see \figurename\ \ref{fig:teff}). That is, the model learns to return the inputs when they are requested, at least within the boundaries of the training set (see the discussion in \secname~\ref{subsec:spec2labels}). When requesting the first BP and RP coefficients, which both describe the large-scale features of the \gaia\ XP spectra, the model pays attention to \teff\ in the case of no extinction, but to $E(B-V)$ in case of high extinction. This is the expected behavior, because in the case of no extinction, the spectral shape is largely determined by \teff, while the large-scale shape is determined by the extinction when it is large (see \figurename~\ref{fig:spec_recon_extincted}). This also demonstrate the important of having the value included in the embedding in our application rather than just providing the value as $V$ in the attention layer that we have introduced in \secname\ \ref{subsec:attention}. Without embedding the value as well, the alignment scores and attention from Equations \eqref{eq:alignment} and \eqref{eq:attention} would not be able to depend on the value (of $E(B-V)$ in this case).

Another way to check whether our model correctly deals with different combinations of inputs is to compare its performance to separate expertly-trained \texttt{XGBoost} models. Examples of this are shown in \figurename\ \ref{fig:dependence}. The example task here is to predict the first RP coefficient provided a combination of stellar parameters and colors. We see that in all the example combinations of inputs, our model behaves very similarly to the \texttt{XGBoost} models even though our model is trained very differently from the \texttt{XGBoost} models, which are trained on each specific task (hence the name of expert model) while our model is trained in a general, self-supervised manner. Moreover, our model also produces reasonable uncertainties in all cases. In the case of only supplying stellar parameters, we cannot predict the first RP coefficient accurately, because of the important role of interstellar reddening and our model correctly indicates that its predictions have large uncertainties. When adding \jh\ and \jk\ to the inputs, which provide a handle on the reddening, both our model and the \texttt{XGBoost} model provide fairly good predictions and the uncertainty from our model is lower. When we further add the \bprp\ color, which is calculated from the XP spectrum, both models are able to give highly accurate predictions.

\subsection{The Meaningfulness of the Embedding}\label{subsec:new_request}

To assess the meaningfulness of the embedding learned by our model, we have withheld two labels \bprp\ and \jk\ from the training process as discussed in \secname\ \ref{sec:train}. This allows us to see if the decoder can answer information requests for outputs it was never trained on. If the embedding is not meaningful---i.e., it contains no knowledge of the fact that there are intrinsic properties like \teff\ and properties only seen from Earth like extinction---the encoder will understand the label embedding, but the decoder will not. In other words, a good embedding is one that is meaningful with respect to the property of stars, not meaningful with respect to the encoder only. We chose to withhold the two labels \bprp\ and \jk, because \bprp\ can be inferred from \gaia\ XP spectra directly, while \jk\ requires the model to extrapolate. \figurename\ \ref{fig:new_requests} demonstrates that the decoder in our model is indeed able to interpret the information requests for \bprp\ and \jk, producing precise values for all but the reddest stars.

In an earlier prototype of the current model with only 55k trainable parameters, we observed similar behavior. This prototype was trained on only \teff, \xh{M}, and the XP spectra and we only requested \teff, \xh{M}, and the first two coefficients during training. We found that the model still managed to infer tight sequence of the next few coefficients, similar to what can be seen in \figurename\ \ref{fig:new_requests}. We emphasize that the ability to predict a tight sequence is important rather than an accurate sequence, because the decoder was never trained on the requested outputs and it has therefore little way of knowing what the correct scale of the outputs is. A small amount of fine-tuning would allow the tight sequence to be corrected to an accurate sequence as well. Thus, because our model learns meaningful embeddings of the properties of stars, it can easily be fine-tuned to predict other types of data or parameters not used during training.

\subsection{Reliability of our Model and Reasonable Expectations}\label{subsec:reliability}

Traditional machine-learning methods in astronomy are trained to perform a single task, such as determine stellar parameters from stellar spectra. As such, their reliability in different parts of parameter space can be relatively straightforwardly be determined by assessing its performance using a test set. The type of model that we have introduced in this work is far more flexible both in terms of the large number of trainable parameters and in terms of input--output combinations that are possible. However, this flexibility means that it is difficult to test the model, because we cannot realistically assess the model's performance  and uncertainty estimation on all possible combinations of inputs and outputs. The type of flexible foundation models that we use here will eventually have orders of magnitude more diverse capabilities than the proto-foundation model that we have trained in this work. Thus, while in this work we have still been able to assess the model's performance on many of the plausible tasks that our model could be used for, in the future this will not be possible.

Natural-language-focused generative models like chat-bots suffer from the problem of ``hallucination'' \citep{2020arXiv200500661M} in that they can make up non-factual information that sounds convincing, because there is no intrinsic mechanism inside the model to care about whether the model's outputs are facts or not. Our model is similarly trained to produce plausible outputs given inputs and this means that the model may make silent assumptions during inference gleaned from the training data if not given with detailed enough input data to make a clear prediction. For example, this happens in our model when the request for information cannot be conclusively inferred from the data provided, in which case the model output will be strongly biased towards the distribution of the training set without using true physics-driven assumptions. An example of this is the following. If you request \logg\ by just giving the model $\teff=3,700\,\mathrm{K}$ and $\xh{M}=0.0\,\mathrm{dex}$, the model returns  $\logg=3.8\pm2.5\,\mathrm{dex}$. This correctly has a large uncertainty, because  there can be giant-type stars with low \logg\ as well as dwarf-type stars with high \logg\ that share the same \teff. But if in addition you provide $E(B-V)=0$ in the input data and request \logg, the model will confidently say that this star has high \logg, because the model essentially thinks it must be a dwarf-type star as small reddening values are always associated with dwarf-type stars in the training set (because generally stars must be close to us in order to be observed with low reddening).

%We should have reasonable expectation of the ability of our model since it is still quite limited. 
In the near term, we can scale our model up to train on more stars and on a wider variety of data from different surveys, which might lead to emergent abilities of the model \citep{2022arXiv220607682W} (that is, the ability to perform tasks that were not possible when training on smaller data sets). Nevertheless, this would still be very far from building intelligent models for astronomy. One of the challenges involved in building such intelligent models might be a version of the alignment problem, which usually refers to the problem of aligning an artificial intelligence systems with human values. In science we might want intelligent models to have scientific values such as Occam's razor and simple Transformer-based models trained with a self-supervised training procedure like we use in this work or physics-inspired networks do not solve the general, fundamental issue (even if they may be able to partially solve the problem). The solutions to some interesting problems will require future models to learn and align with basic scientific values as they gain the ability to create and manipulate variables (i.e., have sub-goals) to explain the (ideally simple and physics-driven) relationships inherent in the data. For example, in this work, we were able to retrieve knowledge of the extinction curve (even while giving the model extinction information with the wrong scaling to the true quantity, see discussion in \secname\ \ref{subsec:new_request} and \figurename\ \ref{fig:new_requests}) without giving the model any explicitly understanding of what an extinction curve is, but simply from manipulating the input stellar parameters and output stellar spectra. Future models, however, might autonomously learn internal representations of the extinction curve that they can more directly use in making predictions. However, the use of such types of intelligent models, even if specifically trained for astronomy, should be treated with caution and they should be thoroughly evaluated when applied to a specific task that they were not specifically trained to do (and they can be fine-tuned if it turns out they perform less well than desired).

\subsection{Foundation Models}\label{subsec:foundation}

There has been much recent interest in foundation models in astronomy, for example, the works of \citet{2021ApJ...911L..33H}, \citet{2022ApJ...932..107S}, \citet{2022mla..confE..29W}, \citet{2023arXiv230615703R}, and \citet{2023arXiv230516127S}. In general, a ``foundation model'' refers to a model trained on a large amount of data that is constructed in such a way that it can be adapted by minimal fine-tuning to a wide range of downstream tasks (hence the model is pre-trained for them from the perspective of the downstream tasks). For example, \citet{2021ApJ...911L..33H}, \citet{2022mla..confE..29W} and \citet{2023arXiv230516127S} adopt a commonly-used convolutional neural network architecture such as \texttt{ResNet-N} \citep{2015arXiv151203385H}, but train on unlabelled images using contrastive objectives to learn useful representation of the data (e.g., \citealt{2019arXiv191105722H}) in a self-supervised, task-independent manner. This makes the model useful for downstream tasks. Pre-training models in this way has been shown to outperform simple supervised learning (e.g., \citealt{2020arXiv200411362K}).

Our model has a blurry boundary between data and labels because of the flexibility in the inputs and outputs, which can be anything that can also be an input. Our model highly mimics standard LLM architectures for which lots of work has been done recently to adapt these models to downstream tasks using fine-tuning methods such as Low-Rank Adaptation (LoRA; \citet{2021arXiv210609685H}) or using the encoder of our model for tasks such as encoding multi-domain data to a common embedding space as in \citet{2021arXiv210300020R}. Ultimately depending on the type of astronomical data included, a combination of all of these techniques would be necessary to create a big universal foundation model that works with multi-domain data from low to very high dimension in astronomy (i.e., we might need a pool of experts to deal with the different modalities of the data). As an example, we present an application of using our model as a foundation model for the purpose of searching for stellar spectroscopy -- stellar parameter pairs in a data set using a contrastive objective function in \appenname\ \ref{sec:foundationdemo}.

Much work remains to be done to turn the proto-foundation model introduced in this paper into a proper foundation model with wide applicability to problems in astronomy. One possible improvement to the model might be to have a proper tokenizer. In NLP, tokenization takes care of the fact that there are many uniquely spelled words and it is difficult to create separate embedding for all of those words (e.g., \citealt{10.1145/1461518.1461544}). The solution is to have a tokenizer (e.g., \citealt{2018arXiv180205365P}), which breaks up words like ``non-existent" into ``non existent'', which can be tokenized into two tokens. Similarly, not commonly used words like ``non-star'' or ``non-galaxy'' can be understood as ``not star'' and ``not galaxy''. In the context of an astronomical foundation model, we could have a tokenizer that understands, for example, different combinations of colors without having each of the combinations be a unique token, while taking advantage of the observations and the ``non-linear'' embedding already learned by our model.

\section{Conclusion}\label{sec:conclusion}

In this paper, we have introduced a novel framework of utilizing a Transformer-based encoder-decoder model towards building and training a foundation model for stars in astronomy. To illustrate this framework, we have trained a proto-foundation model on only a small subset of stars selected from only a few surveys due to the high computational cost involved in training these models\footnote{For example, the estimated cost of training OpenAI's GPT-4 is \href{https://www.wired.com/story/openai-ceo-sam-altman-the-age-of-giant-ai-models-is-already-over/}{\$100 million}.}. But using this limited set of data, we believe we have successfully demonstrated that this framework can be trained to build a foundation model for stars. Our pre-trained model can be used for multiple tasks and can, furthermore, be fine-tuned for other downstream tasks. While our model is competitive with traditional machine-learning models run on the same data and is, therefore, already useful in its own right, the main purpose of this paper is to show that implementing versatile, unified, cross-survey, cross-domain artificial intelligence in astronomy is well within reach and with this work, we hope to accelerate the development of such models. Our Transformer-based model adopts and adapts ideas and technology used in LLMs and inherits many of their advantages, but also disadvantages (such as possible ``hallucinations''). That our model so closely tracks LLMs means that progress in the ideas and technology behind LLMs---an area of high activity currently in academic research and industry---can be easily translated and used by our model.

Our specific application in this paper was to train a Transform-based encoder-decoder model on data from APOGEE, \gaia, \tmass\ and extinction data in a self-supervised manner where all data can be in the input and output nodes and our model additionally provides a predictive uncertainty that represents the model's confidence in its own prediction. We use a single, non-fine-tuned training process to train the model, yet find that the model can perform multiple tasks of interest to astronomers, including but not limited to (i) mapping from stellar spectra to stellar parameters, (ii) from stellar parameters to stellar spectra, (iii) from one portion of a stellar spectrum to another portion, (iv) from stellar parameters to other stellar parameters. We also demonstrated that we can recover the interstellar extinction curve from the trained model, as the model internally clearly learns about extinction. For the common application of mapping from stellar spectra to stellar parameters, our model's prediction accuracy is comparable (or even slightly better) in the same setting as fine-tuned \texttt{XGBoost} models with an accuracy of 47 K in \teff, 0.11 dex in \logg, and 0.07 dex in \xh{M} compared to the APOGEE ground truth. For generative tasks such as predicting stellar spectra from stellar parameters or mapping a portion of the spectrum to another portion, our model is able to predict spectra that are very close to those of real-world stars with the same stellar parameters. And even without explicitly telling the model about the existence of an interstellar extinction curve---we only tell the model about $E(B-V)$---, the model still managed to learn the extinction curve on its own.

While we have focused on a relatively small proof-of-concept setting, training more general models on large cross-survey, multi-domain astronomical data sets with a huge amount of missing data due to the different footprints of the surveys, etc., is in principle straightforward, albeit computationally costly. For example, already in our training set, there are stars that only have spectra and stars that only have stellar labels and the model is able to use all of these for training. Of course, the astronomy community's wide past experience with applying machine learning to astronomical data will be crucial in the construction of these larger foundation models.

In this paper, we have focused on implementing the most basic version of a foundation model in astronomy. There are still many important aspects to add to the model and many open questions as to how more general models should be constructed. These include questions such as how to incorporate data from a wide range of spectral resolutions (broadband photometry to high-resolution spectra), how to make use of theoretical models (e.g., stellar isochrones and theoretical photometry and spectra), how to add time-domain observations (e.g., asteroseismology), and whether it is useful and desirable to create even larger foundation models that incorporate not only stars, but also galaxies over a wide range of redshifts, because the observations of stars and galaxies have many commonalities.

\section*{Acknowledgements}

We thank the anonymous referee, Joshua S. Speagle, and Alexander Laroche for helpful comments. HL and JB acknowledge financial support from the Natural Sciences and Engineering Research Council of Canada (NSERC; funding reference number RGPIN-2020-04712). 

Funding for the Sloan Digital Sky Survey IV has been provided by the Alfred P. Sloan Foundation, the U.S. Department of Energy Office of Science, and the Participating Institutions. SDSS-IV acknowledges support and resources from the Center for High Performance Computing at the University of Utah. The SDSS website is www.sdss.org.

This work has made use of data from the European Space Agency (ESA) mission
{\it Gaia} (\url{https://www.cosmos.esa.int/gaia}), processed by the {\it Gaia}
Data Processing and Analysis Consortium (DPAC,
\url{https://www.cosmos.esa.int/web/gaia/dpac/consortium}). Funding for the DPAC
has been provided by national institutions, in particular the institutions
participating in the {\it Gaia} Multilateral Agreement.

%%%%%%%%%%%%%%%%%%%%%%%%%%%%%%%%%%%%%%%%%%%%%%%%%%
\section*{Data Availability}

All the code and resources for this paper are available at \url{https://github.com/henrysky/astroNN\_stars\_foundation}, including the code that defines the model, to train the model, and to generate the figures. This also includes the trained model, which can be used to test the model further. The underlying data are all publicly available and all of the software used in this work is open source.



\bibliographystyle{mnras}
\bibliography{mnras_template} % if your bibtex file is called example.bib

\begin{thebibliography}{}
\makeatletter
\relax
\def\mn@urlcharsother{\let\do\@makeother \do\$\do\&\do\#\do\^\do\_\do\%\do\~}
\def\mn@doi{\begingroup\mn@urlcharsother \@ifnextchar [ {\mn@doi@} {\mn@doi@[]}}
\def\mn@doi@[#1]#2{\def\@tempa{#1}\ifx\@tempa\@empty \href {http://dx.doi.org/#2} {doi:#2}\else \href {http://dx.doi.org/#2} {#1}\fi \endgroup}
\def\mn@eprint#1#2{\mn@eprint@#1:#2::\@nil}
\def\mn@eprint@arXiv#1{\href {http://arxiv.org/abs/#1} {{\tt arXiv:#1}}}
\def\mn@eprint@dblp#1{\href {http://dblp.uni-trier.de/rec/bibtex/#1.xml} {dblp:#1}}
\def\mn@eprint@#1:#2:#3:#4\@nil{\def\@tempa {#1}\def\@tempb {#2}\def\@tempc {#3}\ifx \@tempc \@empty \let \@tempc \@tempb \let \@tempb \@tempa \fi \ifx \@tempb \@empty \def\@tempb {arXiv}\fi \@ifundefined {mn@eprint@\@tempb}{\@tempb:\@tempc}{\expandafter \expandafter \csname mn@eprint@\@tempb\endcsname \expandafter{\@tempc}}}

\bibitem[\protect\citeauthoryear{{Abdurro'uf} et~al.}{{Abdurro'uf} et~al.}{2022}]{2022ApJS..259...35A}
{Abdurro'uf} et~al., 2022, \mn@doi [\apjs] {10.3847/1538-4365/ac4414}, \href {https://ui.adsabs.harvard.edu/abs/2022ApJS..259...35A} {259, 35}

\bibitem[\protect\citeauthoryear{{Allam} \& {McEwen}}{{Allam} \& {McEwen}}{2021}]{2021arXiv210506178A}
{Allam} Tarek J.,  {McEwen} J.~D.,  2021, \mn@doi [arXiv e-prints] {10.48550/arXiv.2105.06178}, \href {https://ui.adsabs.harvard.edu/abs/2021arXiv210506178A} {p. arXiv:2105.06178}

\bibitem[\protect\citeauthoryear{{Anderson}, {Hogg}, {Leistedt}, {Price-Whelan}  \& {Bovy}}{{Anderson} et~al.}{2018}]{2018AJ....156..145A}
{Anderson} L.,  {Hogg} D.~W.,  {Leistedt} B.,  {Price-Whelan} A.~M.,   {Bovy} J.,  2018, \mn@doi [\aj] {10.3847/1538-3881/aad7bf}, \href {https://ui.adsabs.harvard.edu/abs/2018AJ....156..145A} {156, 145}

\bibitem[\protect\citeauthoryear{{Andrae}, {Rix}  \& {Chandra}}{{Andrae} et~al.}{2023}]{2023ApJS..267....8A}
{Andrae} R.,  {Rix} H.-W.,   {Chandra} V.,  2023, \mn@doi [\apjs] {10.3847/1538-4365/acd53e}, \href {https://ui.adsabs.harvard.edu/abs/2023ApJS..267....8A} {267, 8}

\bibitem[\protect\citeauthoryear{{Bahdanau}, {Cho}  \& {Bengio}}{{Bahdanau} et~al.}{2014}]{2014arXiv1409.0473B}
{Bahdanau} D.,  {Cho} K.,   {Bengio} Y.,  2014, \mn@doi [arXiv e-prints] {10.48550/arXiv.1409.0473}, \href {https://ui.adsabs.harvard.edu/abs/2014arXiv1409.0473B} {p. arXiv:1409.0473}

\bibitem[\protect\citeauthoryear{Barbary}{Barbary}{2016}]{barbary_kyle_2016_804967}
Barbary K.,  2016, extinction v0.3.0, \mn@doi{10.5281/zenodo.804967}, \url {https://doi.org/10.5281/zenodo.804967}

\bibitem[\protect\citeauthoryear{Bengio, Ducharme, Vincent  \& Janvin}{Bengio et~al.}{2003}]{10.5555/944919.944966}
Bengio Y.,  Ducharme R.,  Vincent P.,   Janvin C.,  2003, J. Mach. Learn. Res., 3, 1137–1155

\bibitem[\protect\citeauthoryear{{Blanton} et~al.,}{{Blanton} et~al.}{2017}]{2017AJ....154...28B}
{Blanton} M.~R.,  et~al., 2017, \mn@doi [\aj] {10.3847/1538-3881/aa7567}, \href {https://ui.adsabs.harvard.edu/abs/2017AJ....154...28B} {154, 28}

\bibitem[\protect\citeauthoryear{{Bovy}, {Rix}, {Green}, {Schlafly}  \& {Finkbeiner}}{{Bovy} et~al.}{2016}]{2016ApJ...818..130B}
{Bovy} J.,  {Rix} H.-W.,  {Green} G.~M.,  {Schlafly} E.~F.,   {Finkbeiner} D.~P.,  2016, \mn@doi [\apj] {10.3847/0004-637X/818/2/130}, \href {https://ui.adsabs.harvard.edu/abs/2016ApJ...818..130B} {818, 130}

\bibitem[\protect\citeauthoryear{{Bubeck} et~al.,}{{Bubeck} et~al.}{2023}]{2023arXiv230312712B}
{Bubeck} S.,  et~al., 2023, \mn@doi [arXiv e-prints] {10.48550/arXiv.2303.12712}, \href {https://ui.adsabs.harvard.edu/abs/2023arXiv230312712B} {p. arXiv:2303.12712}

\bibitem[\protect\citeauthoryear{{Carrasco} et~al.,}{{Carrasco} et~al.}{2021}]{2021A&A...652A..86C}
{Carrasco} J.~M.,  et~al., 2021, \mn@doi [\aap] {10.1051/0004-6361/202141249}, \href {https://ui.adsabs.harvard.edu/abs/2021A&A...652A..86C} {652, A86}

\bibitem[\protect\citeauthoryear{Chase}{Chase}{2022}]{Chase_LangChain_2022}
Chase H.,  2022, {LangChain}, \url {https://github.com/hwchase17/langchain}

\bibitem[\protect\citeauthoryear{Chopra, Hadsell  \& LeCun}{Chopra et~al.}{2005}]{e6377ee676a34e8eb97cfdc53cd489ef}
Chopra S.,  Hadsell R.,   LeCun Y.,  2005, in Proceedings - 2005 IEEE Computer Society Conference on Computer Vision and Pattern Recognition, CVPR 2005. Proceedings - 2005 IEEE Computer Society Conference on Computer Vision and Pattern Recognition, CVPR 2005.
IEEE Computer Society, pp 539--546, \mn@doi{10.1109/CVPR.2005.202}

\bibitem[\protect\citeauthoryear{{Ciuc{\u{a}}} \& {Ting}}{{Ciuc{\u{a}}} \& {Ting}}{2023}]{2023RNAAS...7..193C}
{Ciuc{\u{a}}} I.,  {Ting} Y.-S.,  2023, \mn@doi [Research Notes of the American Astronomical Society] {10.3847/2515-5172/acf85f}, \href {https://ui.adsabs.harvard.edu/abs/2023RNAAS...7..193C} {7, 193}

\bibitem[\protect\citeauthoryear{{Coughlin} et~al.,}{{Coughlin} et~al.}{2023}]{2023ApJS..267...31C}
{Coughlin} M.~W.,  et~al., 2023, \mn@doi [\apjs] {10.3847/1538-4365/acdee1}, \href {https://ui.adsabs.harvard.edu/abs/2023ApJS..267...31C} {267, 31}

\bibitem[\protect\citeauthoryear{{Dagli}}{{Dagli}}{2023}]{2023arXiv230405350D}
{Dagli} R.,  2023, \mn@doi [arXiv e-prints] {10.48550/arXiv.2304.05350}, \href {https://ui.adsabs.harvard.edu/abs/2023arXiv230405350D} {p. arXiv:2304.05350}

\bibitem[\protect\citeauthoryear{{De Angeli} et~al.,}{{De Angeli} et~al.}{2023}]{2023A&A...674A...2D}
{De Angeli} F.,  et~al., 2023, \mn@doi [\aap] {10.1051/0004-6361/202243680}, \href {https://ui.adsabs.harvard.edu/abs/2023A&A...674A...2D} {674, A2}

\bibitem[\protect\citeauthoryear{{Devlin}, {Chang}, {Lee}  \& {Toutanova}}{{Devlin} et~al.}{2018}]{2018arXiv181004805D}
{Devlin} J.,  {Chang} M.-W.,  {Lee} K.,   {Toutanova} K.,  2018, \mn@doi [arXiv e-prints] {10.48550/arXiv.1810.04805}, \href {https://ui.adsabs.harvard.edu/abs/2018arXiv181004805D} {p. arXiv:1810.04805}

\bibitem[\protect\citeauthoryear{{Donoso-Oliva}, {Becker}, {Protopapas}, {Cabrera-Vives}, {Vishnu}  \& {Vardhan}}{{Donoso-Oliva} et~al.}{2023}]{2023A&A...670A..54D}
{Donoso-Oliva} C.,  {Becker} I.,  {Protopapas} P.,  {Cabrera-Vives} G.,  {Vishnu} M.,   {Vardhan} H.,  2023, \mn@doi [\aap] {10.1051/0004-6361/202243928}, \href {https://ui.adsabs.harvard.edu/abs/2023A&A...670A..54D} {670, A54}

\bibitem[\protect\citeauthoryear{{Dosovitskiy} et~al.,}{{Dosovitskiy} et~al.}{2020}]{2020arXiv201011929D}
{Dosovitskiy} A.,  et~al., 2020, \mn@doi [arXiv e-prints] {10.48550/arXiv.2010.11929}, \href {https://ui.adsabs.harvard.edu/abs/2020arXiv201011929D} {p. arXiv:2010.11929}

\bibitem[\protect\citeauthoryear{{Drimmel}, {Cabrera-Lavers}  \& {L{\'o}pez-Corredoira}}{{Drimmel} et~al.}{2003}]{2003A&A...409..205D}
{Drimmel} R.,  {Cabrera-Lavers} A.,   {L{\'o}pez-Corredoira} M.,  2003, \mn@doi [\aap] {10.1051/0004-6361:20031070}, \href {https://ui.adsabs.harvard.edu/abs/2003A&A...409..205D} {409, 205}

\bibitem[\protect\citeauthoryear{{Fitzpatrick}}{{Fitzpatrick}}{1999}]{1999PASP..111...63F}
{Fitzpatrick} E.~L.,  1999, \mn@doi [\pasp] {10.1086/316293}, \href {https://ui.adsabs.harvard.edu/abs/1999PASP..111...63F} {111, 63}

\bibitem[\protect\citeauthoryear{{Gaia Collaboration} et~al.}{{Gaia Collaboration} et~al.}{2016}]{2016A&A...595A...1G}
{Gaia Collaboration} et~al., 2016, \mn@doi [\aap] {10.1051/0004-6361/201629272}, \href {https://ui.adsabs.harvard.edu/abs/2016A&A...595A...1G} {595, A1}

\bibitem[\protect\citeauthoryear{{Gaia Collaboration} et~al.}{{Gaia Collaboration} et~al.}{2023}]{2023A&A...674A...1G}
{Gaia Collaboration} et~al., 2023, \mn@doi [\aap] {10.1051/0004-6361/202243940}, \href {https://ui.adsabs.harvard.edu/abs/2023A&A...674A...1G} {674, A1}

\bibitem[\protect\citeauthoryear{{Garc{\'\i}a P{\'e}rez} et~al.,}{{Garc{\'\i}a P{\'e}rez} et~al.}{2016}]{2016AJ....151..144G}
{Garc{\'\i}a P{\'e}rez} A.~E.,  et~al., 2016, \mn@doi [\aj] {10.3847/0004-6256/151/6/144}, \href {https://ui.adsabs.harvard.edu/abs/2016AJ....151..144G} {151, 144}

\bibitem[\protect\citeauthoryear{{Green}, {Schlafly}, {Zucker}, {Speagle}  \& {Finkbeiner}}{{Green} et~al.}{2019}]{2019ApJ...887...93G}
{Green} G.~M.,  {Schlafly} E.,  {Zucker} C.,  {Speagle} J.~S.,   {Finkbeiner} D.,  2019, \mn@doi [\apj] {10.3847/1538-4357/ab5362}, \href {https://ui.adsabs.harvard.edu/abs/2019ApJ...887...93G} {887, 93}

\bibitem[\protect\citeauthoryear{Hadsell, Chopra  \& LeCun}{Hadsell et~al.}{2006}]{c441a21f627f4b0a896fa62cb143176f}
Hadsell R.,  Chopra S.,   LeCun Y.,  2006, in Proceedings - 2006 IEEE Computer Society Conference on Computer Vision and Pattern Recognition, CVPR 2006. Proceedings of the IEEE Computer Society Conference on Computer Vision and Pattern Recognition.
pp 1735--1742, \mn@doi{10.1109/CVPR.2006.100}

\bibitem[\protect\citeauthoryear{{Hauser} et~al.,}{{Hauser} et~al.}{1998}]{1998ApJ...508...25H}
{Hauser} M.~G.,  et~al., 1998, \mn@doi [\apj] {10.1086/306379}, \href {https://ui.adsabs.harvard.edu/abs/1998ApJ...508...25H} {508, 25}

\bibitem[\protect\citeauthoryear{{Hayat}, {Stein}, {Harrington}, {Luki{\'c}}  \& {Mustafa}}{{Hayat} et~al.}{2021}]{2021ApJ...911L..33H}
{Hayat} M.~A.,  {Stein} G.,  {Harrington} P.,  {Luki{\'c}} Z.,   {Mustafa} M.,  2021, \mn@doi [\apjl] {10.3847/2041-8213/abf2c7}, \href {https://ui.adsabs.harvard.edu/abs/2021ApJ...911L..33H} {911, L33}

\bibitem[\protect\citeauthoryear{{He}, {Zhang}, {Ren}  \& {Sun}}{{He} et~al.}{2015}]{2015arXiv151203385H}
{He} K.,  {Zhang} X.,  {Ren} S.,   {Sun} J.,  2015, \mn@doi [arXiv e-prints] {10.48550/arXiv.1512.03385}, \href {https://ui.adsabs.harvard.edu/abs/2015arXiv151203385H} {p. arXiv:1512.03385}

\bibitem[\protect\citeauthoryear{{He}, {Fan}, {Wu}, {Xie}  \& {Girshick}}{{He} et~al.}{2019}]{2019arXiv191105722H}
{He} K.,  {Fan} H.,  {Wu} Y.,  {Xie} S.,   {Girshick} R.,  2019, \mn@doi [arXiv e-prints] {10.48550/arXiv.1911.05722}, \href {https://ui.adsabs.harvard.edu/abs/2019arXiv191105722H} {p. arXiv:1911.05722}

\bibitem[\protect\citeauthoryear{{Hendrycks} \& {Gimpel}}{{Hendrycks} \& {Gimpel}}{2016}]{2016arXiv160608415H}
{Hendrycks} D.,  {Gimpel} K.,  2016, \mn@doi [arXiv e-prints] {10.48550/arXiv.1606.08415}, \href {https://ui.adsabs.harvard.edu/abs/2016arXiv160608415H} {p. arXiv:1606.08415}

\bibitem[\protect\citeauthoryear{{Hu}, {Shen}, {Wallis}, {Allen-Zhu}, {Li}, {Wang}, {Wang}  \& {Chen}}{{Hu} et~al.}{2021}]{2021arXiv210609685H}
{Hu} E.~J.,  {Shen} Y.,  {Wallis} P.,  {Allen-Zhu} Z.,  {Li} Y.,  {Wang} S.,  {Wang} L.,   {Chen} W.,  2021, \mn@doi [arXiv e-prints] {10.48550/arXiv.2106.09685}, \href {https://ui.adsabs.harvard.edu/abs/2021arXiv210609685H} {p. arXiv:2106.09685}

\bibitem[\protect\citeauthoryear{{Ivezi{\'c}} et~al.}{{Ivezi{\'c}} et~al.}{2019}]{2019ApJ...873..111I}
{Ivezi{\'c}} {\v{Z}}.,  et~al., 2019, \mn@doi [\apj] {10.3847/1538-4357/ab042c}, \href {https://ui.adsabs.harvard.edu/abs/2019ApJ...873..111I} {873, 111}

\bibitem[\protect\citeauthoryear{{Khosla} et~al.,}{{Khosla} et~al.}{2020}]{2020arXiv200411362K}
{Khosla} P.,  et~al., 2020, \mn@doi [arXiv e-prints] {10.48550/arXiv.2004.11362}, \href {https://ui.adsabs.harvard.edu/abs/2020arXiv200411362K} {p. arXiv:2004.11362}

\bibitem[\protect\citeauthoryear{{Kingma} \& {Ba}}{{Kingma} \& {Ba}}{2014}]{2014arXiv1412.6980K}
{Kingma} D.~P.,  {Ba} J.,  2014, \mn@doi [arXiv e-prints] {10.48550/arXiv.1412.6980}, \href {https://ui.adsabs.harvard.edu/abs/2014arXiv1412.6980K} {p. arXiv:1412.6980}

\bibitem[\protect\citeauthoryear{{Kollmeier} et~al.,}{{Kollmeier} et~al.}{2017}]{2017arXiv171103234K}
{Kollmeier} J.~A.,  et~al., 2017, \mn@doi [arXiv e-prints] {10.48550/arXiv.1711.03234}, \href {https://ui.adsabs.harvard.edu/abs/2017arXiv171103234K} {p. arXiv:1711.03234}

\bibitem[\protect\citeauthoryear{{Laroche} \& {Speagle}}{{Laroche} \& {Speagle}}{2023}]{2023arXiv230706378L}
{Laroche} A.,  {Speagle} J.~S.,  2023, \mn@doi [arXiv e-prints] {10.48550/arXiv.2307.06378}, \href {https://ui.adsabs.harvard.edu/abs/2023arXiv230706378L} {p. arXiv:2307.06378}

\bibitem[\protect\citeauthoryear{{Laureijs} et~al.}{{Laureijs} et~al.}{2011}]{2011arXiv1110.3193L}
{Laureijs} R.,  et~al., 2011, \mn@doi [arXiv e-prints] {10.48550/arXiv.1110.3193}, \href {https://ui.adsabs.harvard.edu/abs/2011arXiv1110.3193L} {p. arXiv:1110.3193}

\bibitem[\protect\citeauthoryear{{Lei Ba}, {Kiros}  \& {Hinton}}{{Lei Ba} et~al.}{2016}]{2016arXiv160706450L}
{Lei Ba} J.,  {Kiros} J.~R.,   {Hinton} G.~E.,  2016, \mn@doi [arXiv e-prints] {10.48550/arXiv.1607.06450}, \href {https://ui.adsabs.harvard.edu/abs/2016arXiv160706450L} {p. arXiv:1607.06450}

\bibitem[\protect\citeauthoryear{{Leung} \& {Bovy}}{{Leung} \& {Bovy}}{2019a}]{2019MNRAS.483.3255L}
{Leung} H.~W.,  {Bovy} J.,  2019a, \mn@doi [\mnras] {10.1093/mnras/sty3217}, \href {https://ui.adsabs.harvard.edu/abs/2019MNRAS.483.3255L} {483, 3255}

\bibitem[\protect\citeauthoryear{{Leung} \& {Bovy}}{{Leung} \& {Bovy}}{2019b}]{2019MNRAS.489.2079L}
{Leung} H.~W.,  {Bovy} J.,  2019b, \mn@doi [\mnras] {10.1093/mnras/stz2245}, \href {https://ui.adsabs.harvard.edu/abs/2019MNRAS.489.2079L} {489, 2079}

\bibitem[\protect\citeauthoryear{{Leung}, {Bovy}, {Mackereth}, {Hunt}, {Lane}  \& {Wilson}}{{Leung} et~al.}{2023a}]{2023MNRAS.519..948L}
{Leung} H.~W.,  {Bovy} J.,  {Mackereth} J.~T.,  {Hunt} J. A.~S.,  {Lane} R.~R.,   {Wilson} J.~C.,  2023a, \mn@doi [\mnras] {10.1093/mnras/stac3529}, \href {https://ui.adsabs.harvard.edu/abs/2023MNRAS.519..948L} {519, 948}

\bibitem[\protect\citeauthoryear{{Leung}, {Bovy}, {Mackereth}  \& {Miglio}}{{Leung} et~al.}{2023b}]{2023MNRAS.522.4577L}
{Leung} H.~W.,  {Bovy} J.,  {Mackereth} J.~T.,   {Miglio} A.,  2023b, \mn@doi [\mnras] {10.1093/mnras/stad1272}, \href {https://ui.adsabs.harvard.edu/abs/2023MNRAS.522.4577L} {522, 4577}

\bibitem[\protect\citeauthoryear{{Lindegren} et~al.,}{{Lindegren} et~al.}{2021}]{2021A&A...649A...4L}
{Lindegren} L.,  et~al., 2021, \mn@doi [\aap] {10.1051/0004-6361/202039653}, \href {https://ui.adsabs.harvard.edu/abs/2021A&A...649A...4L} {649, A4}

\bibitem[\protect\citeauthoryear{{Loshchilov} \& {Hutter}}{{Loshchilov} \& {Hutter}}{2016}]{2016arXiv160803983L}
{Loshchilov} I.,  {Hutter} F.,  2016, \mn@doi [arXiv e-prints] {10.48550/arXiv.1608.03983}, \href {https://ui.adsabs.harvard.edu/abs/2016arXiv160803983L} {p. arXiv:1608.03983}

\bibitem[\protect\citeauthoryear{{Loshchilov} \& {Hutter}}{{Loshchilov} \& {Hutter}}{2017}]{2017arXiv171105101L}
{Loshchilov} I.,  {Hutter} F.,  2017, \mn@doi [arXiv e-prints] {10.48550/arXiv.1711.05101}, \href {https://ui.adsabs.harvard.edu/abs/2017arXiv171105101L} {p. arXiv:1711.05101}

\bibitem[\protect\citeauthoryear{{Majewski} et~al.,}{{Majewski} et~al.}{2017}]{2017AJ....154...94M}
{Majewski} S.~R.,  et~al., 2017, \mn@doi [\aj] {10.3847/1538-3881/aa784d}, \href {https://ui.adsabs.harvard.edu/abs/2017AJ....154...94M} {154, 94}

\bibitem[\protect\citeauthoryear{{Marshall}, {Robin}, {Reyl{\'e}}, {Schultheis}  \& {Picaud}}{{Marshall} et~al.}{2006}]{2006A&A...453..635M}
{Marshall} D.~J.,  {Robin} A.~C.,  {Reyl{\'e}} C.,  {Schultheis} M.,   {Picaud} S.,  2006, \mn@doi [\aap] {10.1051/0004-6361:20053842}, \href {https://ui.adsabs.harvard.edu/abs/2006A&A...453..635M} {453, 635}

\bibitem[\protect\citeauthoryear{{Maynez}, {Narayan}, {Bohnet}  \& {McDonald}}{{Maynez} et~al.}{2020}]{2020arXiv200500661M}
{Maynez} J.,  {Narayan} S.,  {Bohnet} B.,   {McDonald} R.,  2020, \mn@doi [arXiv e-prints] {10.48550/arXiv.2005.00661}, \href {https://ui.adsabs.harvard.edu/abs/2020arXiv200500661M} {p. arXiv:2005.00661}

\bibitem[\protect\citeauthoryear{{Micikevicius} et~al.,}{{Micikevicius} et~al.}{2017}]{2017arXiv171003740M}
{Micikevicius} P.,  et~al., 2017, \mn@doi [arXiv e-prints] {10.48550/arXiv.1710.03740}, \href {https://ui.adsabs.harvard.edu/abs/2017arXiv171003740M} {p. arXiv:1710.03740}

\bibitem[\protect\citeauthoryear{{Mikolov}, {Chen}, {Corrado}  \& {Dean}}{{Mikolov} et~al.}{2013}]{2013arXiv1301.3781M}
{Mikolov} T.,  {Chen} K.,  {Corrado} G.,   {Dean} J.,  2013, \mn@doi [arXiv e-prints] {10.48550/arXiv.1301.3781}, \href {https://ui.adsabs.harvard.edu/abs/2013arXiv1301.3781M} {p. arXiv:1301.3781}

\bibitem[\protect\citeauthoryear{{Montegriffo} et~al.,}{{Montegriffo} et~al.}{2023}]{2023A&A...674A...3M}
{Montegriffo} P.,  et~al., 2023, \mn@doi [\aap] {10.1051/0004-6361/202243880}, \href {https://ui.adsabs.harvard.edu/abs/2023A&A...674A...3M} {674, A3}

\bibitem[\protect\citeauthoryear{{Moreno-Cartagena}, {Cabrera-Vives}, {Protopapas}, {Donoso-Oliva}, {P{\'e}rez-Carrasco}  \& {C{\'a}diz-Leyton}}{{Moreno-Cartagena} et~al.}{2023}]{2023arXiv230806404M}
{Moreno-Cartagena} D.,  {Cabrera-Vives} G.,  {Protopapas} P.,  {Donoso-Oliva} C.,  {P{\'e}rez-Carrasco} M.,   {C{\'a}diz-Leyton} M.,  2023, arXiv e-prints, \href {https://ui.adsabs.harvard.edu/abs/2023arXiv230806404M} {p. arXiv:2308.06404}

\bibitem[\protect\citeauthoryear{{OpenAI}}{{OpenAI}}{2023}]{2023arXiv230308774O}
{OpenAI} 2023, \mn@doi [arXiv e-prints] {10.48550/arXiv.2303.08774}, \href {https://ui.adsabs.harvard.edu/abs/2023arXiv230308774O} {p. arXiv:2303.08774}

\bibitem[\protect\citeauthoryear{{Paszke} et~al.,}{{Paszke} et~al.}{2019}]{2019arXiv191201703P}
{Paszke} A.,  et~al., 2019, \mn@doi [arXiv e-prints] {10.48550/arXiv.1912.01703}, \href {https://ui.adsabs.harvard.edu/abs/2019arXiv191201703P} {p. arXiv:1912.01703}

\bibitem[\protect\citeauthoryear{{Peters}, {Neumann}, {Iyyer}, {Gardner}, {Clark}, {Lee}  \& {Zettlemoyer}}{{Peters} et~al.}{2018}]{2018arXiv180205365P}
{Peters} M.~E.,  {Neumann} M.,  {Iyyer} M.,  {Gardner} M.,  {Clark} C.,  {Lee} K.,   {Zettlemoyer} L.,  2018, \mn@doi [arXiv e-prints] {10.48550/arXiv.1802.05365}, \href {https://ui.adsabs.harvard.edu/abs/2018arXiv180205365P} {p. arXiv:1802.05365}

\bibitem[\protect\citeauthoryear{{Pimentel}, {Est{\'e}vez}  \& {F{\"o}rster}}{{Pimentel} et~al.}{2023}]{2023AJ....165...18P}
{Pimentel} {\'O}.,  {Est{\'e}vez} P.~A.,   {F{\"o}rster} F.,  2023, \mn@doi [\aj] {10.3847/1538-3881/ac9ab4}, \href {https://ui.adsabs.harvard.edu/abs/2023AJ....165...18P} {165, 18}

\bibitem[\protect\citeauthoryear{{Radford} et~al.,}{{Radford} et~al.}{2021}]{2021arXiv210300020R}
{Radford} A.,  et~al., 2021, \mn@doi [arXiv e-prints] {10.48550/arXiv.2103.00020}, \href {https://ui.adsabs.harvard.edu/abs/2021arXiv210300020R} {p. arXiv:2103.00020}

\bibitem[\protect\citeauthoryear{{Riello} et~al.,}{{Riello} et~al.}{2021}]{2021A&A...649A...3R}
{Riello} M.,  et~al., 2021, \mn@doi [\aap] {10.1051/0004-6361/202039587}, \href {https://ui.adsabs.harvard.edu/abs/2021A&A...649A...3R} {649, A3}

\bibitem[\protect\citeauthoryear{{Rix} et~al.,}{{Rix} et~al.}{2022}]{2022ApJ...941...45R}
{Rix} H.-W.,  et~al., 2022, \mn@doi [\apj] {10.3847/1538-4357/ac9e01}, \href {https://ui.adsabs.harvard.edu/abs/2022ApJ...941...45R} {941, 45}

\bibitem[\protect\citeauthoryear{{R{\'o}{\.z}a{\'n}ski}, {Ting}  \& {Jab{\l}o{\'n}ska}}{{R{\'o}{\.z}a{\'n}ski} et~al.}{2023}]{2023arXiv230615703R}
{R{\'o}{\.z}a{\'n}ski} T.,  {Ting} Y.-S.,   {Jab{\l}o{\'n}ska} M.,  2023, \mn@doi [arXiv e-prints] {10.48550/arXiv.2306.15703}, \href {https://ui.adsabs.harvard.edu/abs/2023arXiv230615703R} {p. arXiv:2306.15703}

\bibitem[\protect\citeauthoryear{Salton}{Salton}{1962}]{10.1145/1461518.1461544}
Salton G.,  1962, in Proceedings of the December 4-6, 1962, Fall Joint Computer Conference. AFIPS '62 (Fall).
Association for Computing Machinery, New York, NY, USA, p. 234–250, \mn@doi{10.1145/1461518.1461544}, \url {https://doi.org/10.1145/1461518.1461544}

\bibitem[\protect\citeauthoryear{{Sanders} \& {Matsunaga}}{{Sanders} \& {Matsunaga}}{2023}]{2023MNRAS.521.2745S}
{Sanders} J.~L.,  {Matsunaga} N.,  2023, \mn@doi [\mnras] {10.1093/mnras/stad574}, \href {https://ui.adsabs.harvard.edu/abs/2023MNRAS.521.2745S} {521, 2745}

\bibitem[\protect\citeauthoryear{{Schlafly} \& {Finkbeiner}}{{Schlafly} \& {Finkbeiner}}{2011}]{2011ApJ...737..103S}
{Schlafly} E.~F.,  {Finkbeiner} D.~P.,  2011, \mn@doi [\apj] {10.1088/0004-637X/737/2/103}, \href {https://ui.adsabs.harvard.edu/abs/2011ApJ...737..103S} {737, 103}

\bibitem[\protect\citeauthoryear{{Schlegel}, {Finkbeiner}  \& {Davis}}{{Schlegel} et~al.}{1998}]{1998ApJ...500..525S}
{Schlegel} D.~J.,  {Finkbeiner} D.~P.,   {Davis} M.,  1998, \mn@doi [\apj] {10.1086/305772}, \href {https://ui.adsabs.harvard.edu/abs/1998ApJ...500..525S} {500, 525}

\bibitem[\protect\citeauthoryear{{Skrutskie} et~al.,}{{Skrutskie} et~al.}{2006}]{2006AJ....131.1163S}
{Skrutskie} M.~F.,  et~al., 2006, \mn@doi [\aj] {10.1086/498708}, \href {https://ui.adsabs.harvard.edu/abs/2006AJ....131.1163S} {131, 1163}

\bibitem[\protect\citeauthoryear{{Slijepcevic}, {Scaife}, {Walmsley}, {Bowles}, {Wong}, {Shabala}  \& {White}}{{Slijepcevic} et~al.}{2023}]{2023arXiv230516127S}
{Slijepcevic} I.~V.,  {Scaife} A. M.~M.,  {Walmsley} M.,  {Bowles} M.,  {Wong} O.~I.,  {Shabala} S.~S.,   {White} S.~V.,  2023, \mn@doi [arXiv e-prints] {10.48550/arXiv.2305.16127}, \href {https://ui.adsabs.harvard.edu/abs/2023arXiv230516127S} {p. arXiv:2305.16127}

\bibitem[\protect\citeauthoryear{{Smith} \& {Geach}}{{Smith} \& {Geach}}{2023}]{2023RSOS...1021454S}
{Smith} M.~J.,  {Geach} J.~E.,  2023, \mn@doi [Royal Society Open Science] {10.1098/rsos.221454}, \href {https://ui.adsabs.harvard.edu/abs/2023RSOS...1021454S} {10, 221454}

\bibitem[\protect\citeauthoryear{Sohn}{Sohn}{2016}]{10.5555/3157096.3157304}
Sohn K.,  2016, in Proceedings of the 30th International Conference on Neural Information Processing Systems. NIPS'16.
Curran Associates Inc., Red Hook, NY, USA, p. 1857–1865

\bibitem[\protect\citeauthoryear{Srivastava, Hinton, Krizhevsky, Sutskever  \& Salakhutdinov}{Srivastava et~al.}{2014}]{10.5555/2627435.2670313}
Srivastava N.,  Hinton G.,  Krizhevsky A.,  Sutskever I.,   Salakhutdinov R.,  2014, J. Mach. Learn. Res., 15, 1929–1958

\bibitem[\protect\citeauthoryear{{Stein}, {Blaum}, {Harrington}, {Medan}  \& {Luki{\'c}}}{{Stein} et~al.}{2022}]{2022ApJ...932..107S}
{Stein} G.,  {Blaum} J.,  {Harrington} P.,  {Medan} T.,   {Luki{\'c}} Z.,  2022, \mn@doi [\apj] {10.3847/1538-4357/ac6d63}, \href {https://ui.adsabs.harvard.edu/abs/2022ApJ...932..107S} {932, 107}

\bibitem[\protect\citeauthoryear{{Vaswani}, {Shazeer}, {Parmar}, {Uszkoreit}, {Jones}, {Gomez}, {Kaiser}  \& {Polosukhin}}{{Vaswani} et~al.}{2017}]{2017arXiv170603762V}
{Vaswani} A.,  {Shazeer} N.,  {Parmar} N.,  {Uszkoreit} J.,  {Jones} L.,  {Gomez} A.~N.,  {Kaiser} L.,   {Polosukhin} I.,  2017, \mn@doi [arXiv e-prints] {10.48550/arXiv.1706.03762}, \href {https://ui.adsabs.harvard.edu/abs/2017arXiv170603762V} {p. arXiv:1706.03762}

\bibitem[\protect\citeauthoryear{{Walmsley}, {Slijepcevic}, {Bowles}  \& {Scaife}}{{Walmsley} et~al.}{2022}]{2022mla..confE..29W}
{Walmsley} M.,  {Slijepcevic} I.,  {Bowles} M.~R.,   {Scaife} A.,  2022, in Machine Learning for Astrophysics. p.~29 (\mn@eprint {arXiv} {2206.11927}), \mn@doi{10.48550/arXiv.2206.11927}

\bibitem[\protect\citeauthoryear{{Wei} et~al.,}{{Wei} et~al.}{2022}]{2022arXiv220607682W}
{Wei} J.,  et~al., 2022, \mn@doi [arXiv e-prints] {10.48550/arXiv.2206.07682}, \href {https://ui.adsabs.harvard.edu/abs/2022arXiv220607682W} {p. arXiv:2206.07682}

\bibitem[\protect\citeauthoryear{{Wilson} et~al.,}{{Wilson} et~al.}{2019}]{2019PASP..131e5001W}
{Wilson} J.~C.,  et~al., 2019, \mn@doi [\pasp] {10.1088/1538-3873/ab0075}, \href {https://ui.adsabs.harvard.edu/abs/2019PASP..131e5001W} {131, 055001}

\bibitem[\protect\citeauthoryear{{Zhang}, {Lipton}, {Li}  \& {Smola}}{{Zhang} et~al.}{2021}]{2021arXiv210611342Z}
{Zhang} A.,  {Lipton} Z.~C.,  {Li} M.,   {Smola} A.~J.,  2021, \mn@doi [arXiv e-prints] {10.48550/arXiv.2106.11342}, \href {https://ui.adsabs.harvard.edu/abs/2021arXiv210611342Z} {p. arXiv:2106.11342}

\bibitem[\protect\citeauthoryear{{Zhang}, {Green}  \& {Rix}}{{Zhang} et~al.}{2023}]{2023MNRAS.524.1855Z}
{Zhang} X.,  {Green} G.~M.,   {Rix} H.-W.,  2023, \mn@doi [\mnras] {10.1093/mnras/stad1941}, \href {https://ui.adsabs.harvard.edu/abs/2023MNRAS.524.1855Z} {524, 1855}

\bibitem[\protect\citeauthoryear{{Zou}, {Zhu}, {Xu}  \& {Luo}}{{Zou} et~al.}{2020}]{2020PASP..132d4503Z}
{Zou} Z.,  {Zhu} T.,  {Xu} L.,   {Luo} A.~L.,  2020, \mn@doi [\pasp] {10.1088/1538-3873/ab7548}, \href {https://ui.adsabs.harvard.edu/abs/2020PASP..132d4503Z} {132, 044503}

\makeatother
\end{thebibliography}

%\begin{thebibliography}{99}
%\end{thebibliography}

%%%%%%%%%%%%%%%%%%%%%%%%%%%%%%%%%%%%%%%%%%%%%%%%%%

%%%%%%%%%%%%%%%%% APPENDICES %%%%%%%%%%%%%%%%%%%%%

\appendix

\section{Technical Details of the Training Process}\label{sec:technical_dl}

\begin{figure}
\centering
\includegraphics[width=0.475 \textwidth]{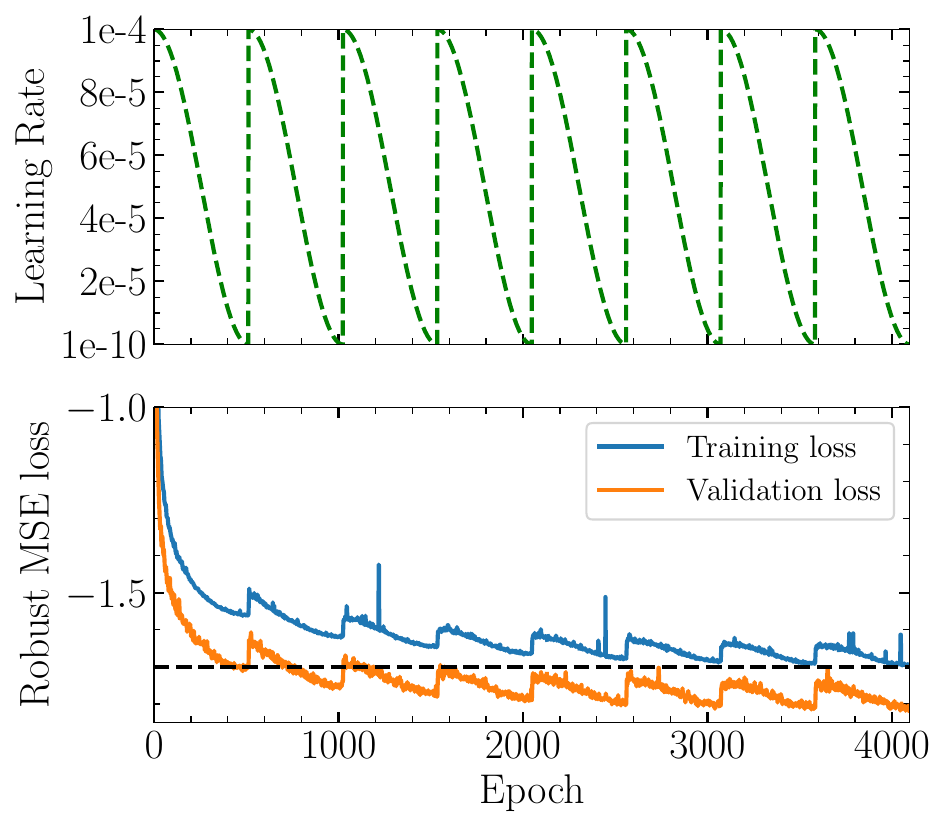}
\caption{Learning rate schedule and robust mean squared error (MSE) loss (see \eqnname\ \ref{eq:robust_mse}) for both the training and validation set during the training process. The black dashed line in the bottom panel shows the approximate loss level for the validation set at the end of the first cosine learning rate cycle as a reference line.}
\label{fig:loss}
\end{figure}

% Make this figure have label B1
\renewcommand{\thefigure}{B\arabic{figure}}
\setcounter{figure}{0}
\begin{figure*}
\centering
\begin{tikzpicture}
\node (n1) [draw=spec2paramcolor, line width=\figboxlw,rounded corners=\figboxcorner, inner xsep=\figboxsep, inner ysep=\figboxsep]
{\includegraphics[width=\textwidth]{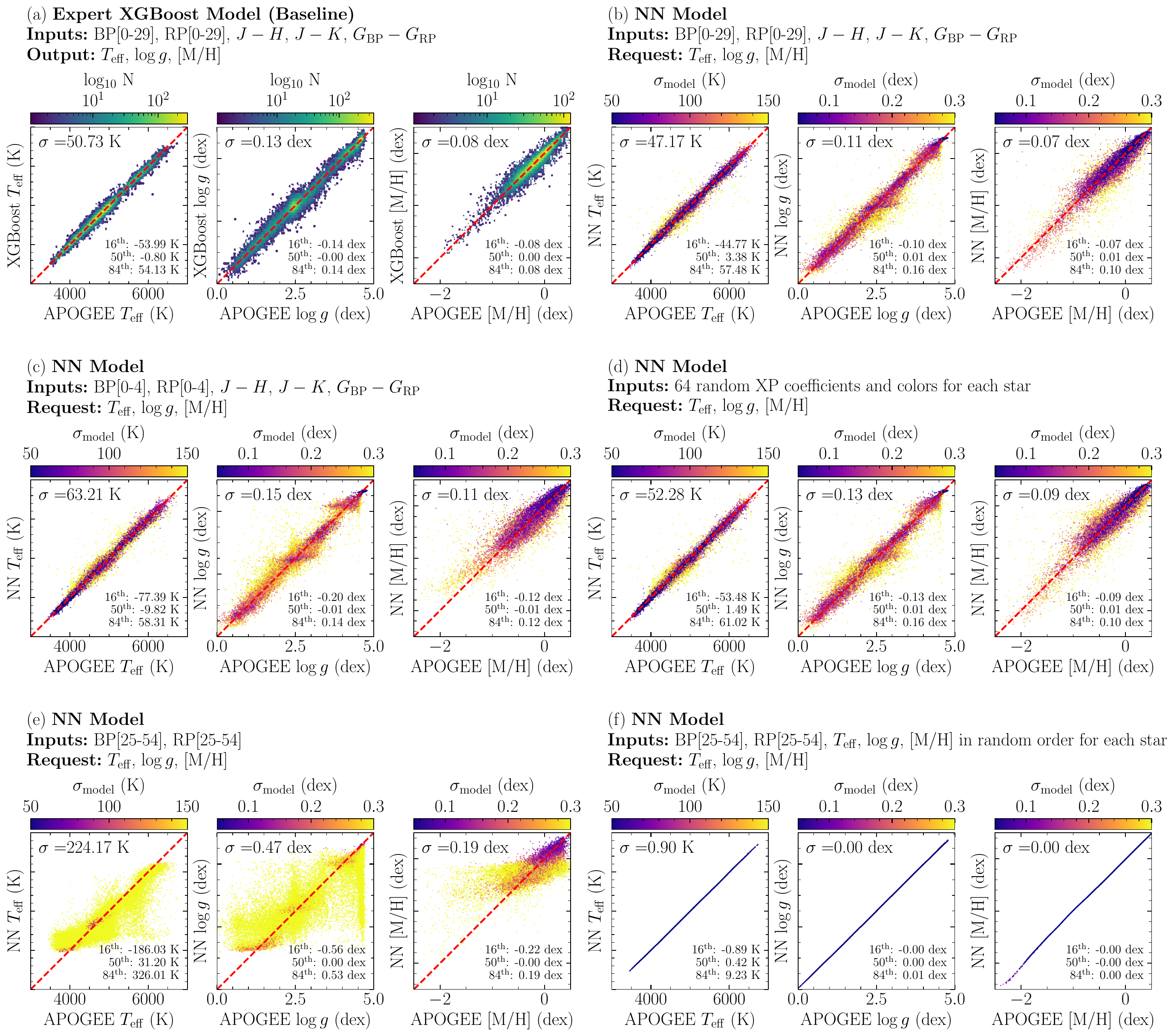}};
\node[font=\fontsize{\figheadfontsize}{0}\selectfont, text=spec2paramcolor] at (-4.5, 8.3) {Task: Stellar Spectra to Stellar Parameters};
\draw[spec2paramcolor, line width=\figboxlw] (-0.04, 8.0)  -- (-0.04, -8.0);
\draw[spec2paramcolor, line width=\figboxlw] (-9.05, 2.7)  -- (9.05, 2.7);
\draw[spec2paramcolor, line width=\figboxlw] (-9.05, -2.7)  -- (9.05, -2.7);
\end{tikzpicture}
\caption{More results for predicting stellar parameters from different combinations of XP spectra and photometry. This figure is an extended version of \figurename\ \ref{fig:teff}, now showing the result of \teff, \logg\ and\xh{M} for all of the cases considered in \figurename\ \ref{fig:teff}. The $16^\mathrm{th}$, $50^\mathrm{th}$, $84^\mathrm{th}$ percentile of the difference between the model prediction and the ground truth is indicated in the bottom-right corner of each panel.}
\label{fig:teff_logg_mh}
\end{figure*}

\setcounter{figure}{1}
\begin{figure*}
\centering
\begin{tikzpicture}
\node (n1) [draw=spec2paramcolor, line width=\figboxlw,rounded corners=\figboxcorner, inner xsep=\figboxsep, inner ysep=\figboxsep]
{\includegraphics[width=\textwidth]{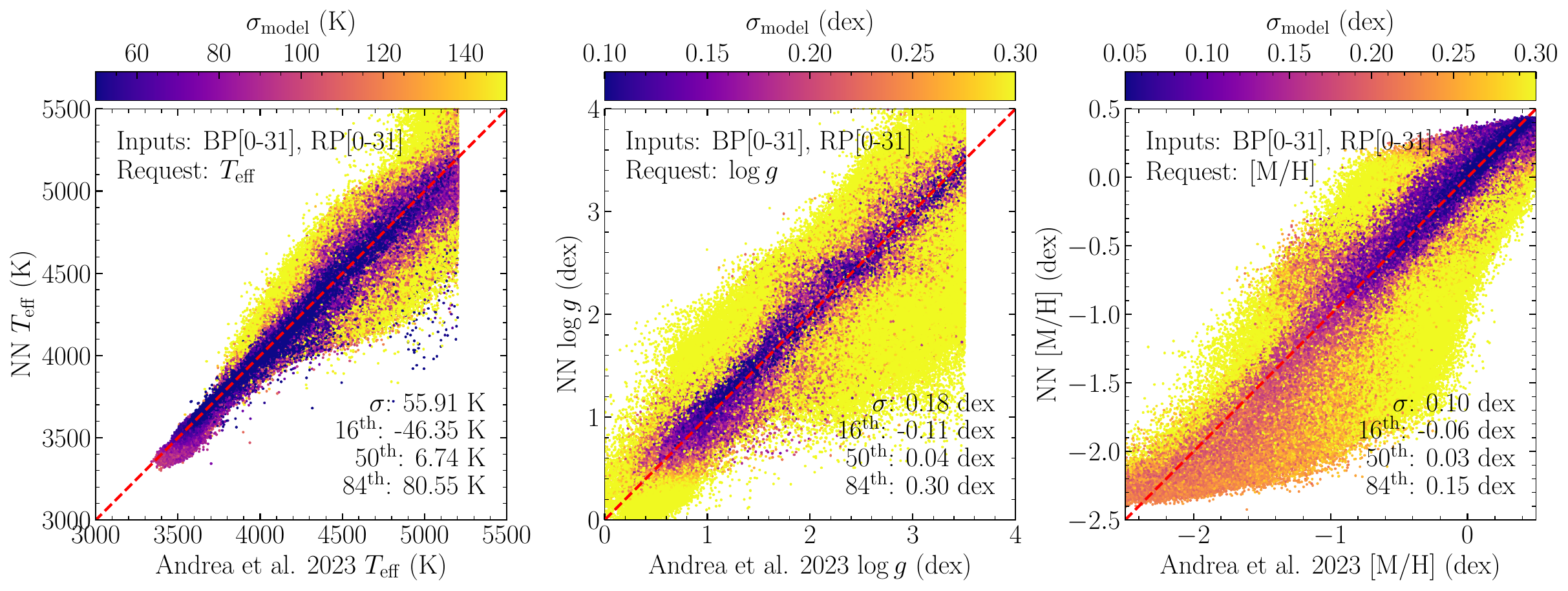}};
\node[font=\fontsize{\figheadfontsize}{0}\selectfont, text=spec2paramcolor] at (-4.5, 3.8) {Task: Stellar Spectra to Stellar Parameters};
\end{tikzpicture}
\caption{Comparison between our model's prediction of \teff, \logg\ and \xh{M} based on the first 32 BP/RP coefficients and the results from \citet{2023ApJS..267....8A} for the ``vetted'' giants sample identified by \citet{2023ApJS..267....8A} of $\approx 15.6$ millions stars. The color-coding is our model's predictive uncertainty. Our results compare well to those from \citet{2023ApJS..267....8A} on this sample that is largely outside of the APOGEE survey footprint and our model returns realistic uncertainties where stars with largest disagreement also have large uncertainties.}
\label{fig:andrae}
\end{figure*}

\renewcommand{\thefigure}{\Alph{section}\arabic{figure}} % Resetting the figure counter to normal

The training of our model requires specific software and hardware configurations to train the model reasonably quickly on a modern home desktop computer with a consumer-grade graphics processing unit (GPU) without the need for a data-center-grade GPU or cloud GPU computing. We will discuss the technical details of the training process we have employed to train our model on a $\approx \$2500$ CAD home desktop computer. 

The model is trained on a Windows 11 desktop computer with an Intel $12^\mathrm{th}$ generation i7 CPU and a NVIDIA RTX4070 Ti GPU, where the GPU theoretical \texttt{float16} and \texttt{float32} performance are both 40.1 TFLOPS with \texttt{Python} 3.11, \texttt{PyTorch} 2.1, and \texttt{CUDA} runtime 12.2. Training took $\approx 22.5$ hours (compared to an estimated $\approx 1,700$ hours of training time on the CPU on the same computer) requiring $\lesssim 12$ GB video memory. That is, a GPU with at least 12 GB video memory solely dedicated to the training process is sufficient if you follow the exact same model configuration and \texttt{PyTorch} hardware setting as this paper without needing to swap to physical memory, which would severely increase the training time. Mixed precision (mixing \texttt{float32} and \texttt{float16}; \citealt{2017arXiv171003740M}) and enabling rounding the traditional \texttt{float32} to the \texttt{TensorFloat32} float format with a 10-bit mantissa (the same as for a \texttt{float16}, thus, sharing the same precision), but using an 8-bit exponent (same as a \texttt{float32}, thus sharing the same numeric range) for matrix multiplication is needed to reduce the video memory requirement and to reduce the training time. Operation-level numerical stability, where instability can be caused by using the lower precision \texttt{float16}, is handled automatically by \texttt{PyTorch}, which determines the best float format to use for each operation. An example is the \texttt{log} operation, which uses \texttt{float32} precision to prevent numerical instability, while simple \texttt{linear} operations use \texttt{float16} precision without user intervention. Automatic gradient scaling is applied during training to ensure the stability of gradients when using mixed precision. The learning rate schedule as well as the training and validation loss during the training process can be seen in \figurename\ \ref{fig:loss}.

The use of mixed precision and of \texttt{TensorFloat32} requires specialized floating-point accelerators (specifically the NVIDIA $3^\mathrm{rd}$ generation \texttt{TensorCores} or later) in hardware, meaning that training the model requires a GPU with NVIDIA Ampere architecture, which was released in 2020, or later architectures (Ada Lovelace architecture and Hopper architecture by the time of this paper). Older GPUs with enough video memory or CPU could be used to train our model with longer training time, but should yield similar results. For inference, the model requires less than 512 MB video memory without the use of mixed precision on the GPU or inference can be performed on any reasonably-modern CPU.

\section{Detailed performance on discriminative and generative tasks}\label{sec:performance}

\setcounter{figure}{2}
\begin{figure*}
\centering
\begin{tikzpicture}
\node (n1) [draw=param2speccolor, line width=\figboxlw,rounded corners=\figboxcorner, inner xsep=\figboxsep, inner ysep=\figboxsep]
{\includegraphics[width=\textwidth]{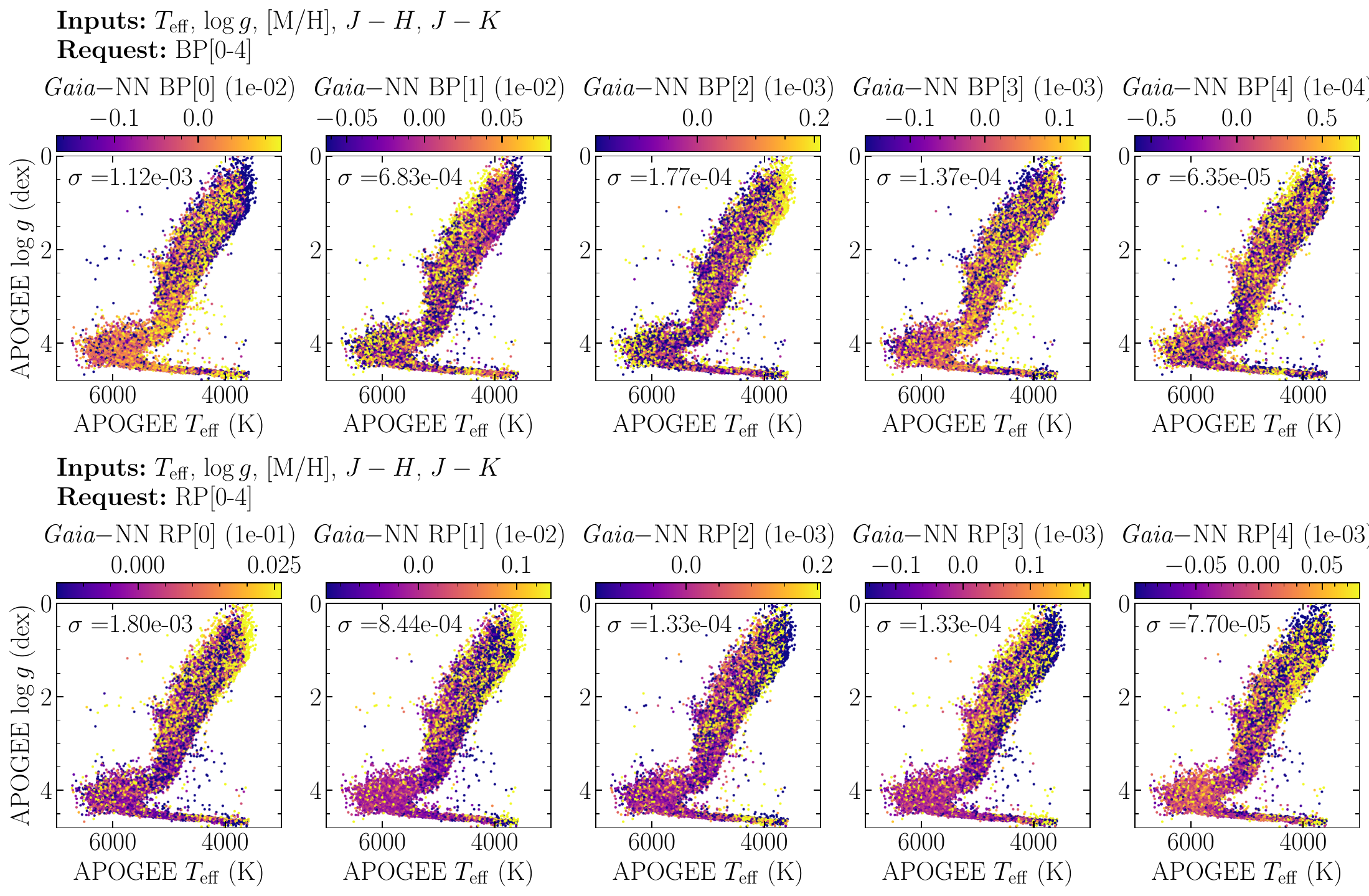}};
\node[font=\fontsize{\figheadfontsize}{0}\selectfont, text=param2speccolor] at (-4.5, 6.15) {Task: Stellar Parameters to Stellar Spectra};
\draw[param2speccolor, line width=\figboxlw] (-9.05, 0.)  -- (9.05, 0.);
\end{tikzpicture}
\caption{Predicting the first five BP (top row) and RP (bottom row) coefficients from \teff, \xh{M}, \logg, \jh, and \jk. We show the difference between the model's prediction and the ground-truth as the color-coding of the stars shown in the \teff-\logg\ diagram. The colorbar is set to the $16^\mathrm{th}$ and $84^\mathrm{th}$ percentile of the difference. We see that our model is generally able to predict the BP and RP spectra accurately and precisely across all type of stars covered by APOGEE.}
\label{fig:xp_coeffs}
\end{figure*}

To allow a better assessment of the performance of our model discussed in \secname\ \ref{sec:result}, we show additional performance tests in this appendix for the discriminative task of predicting \teff, \logg\ and \xh{M} from \gaia\ XP spectra in \figurename\ \ref{fig:teff_logg_mh} and for the generative task of reconstructing \gaia\ XP spectra from stellar parameters in \figurename\ \ref{fig:xp_coeffs}.

\figurename\ \ref{fig:teff_logg_mh} is an extended version of \figurename\ \ref{fig:teff}, whose \teff\ results it repeats for completeness in addition to showing the same precision and accuracy tests for \logg\ and \xh{M}. The performance of our model for these two additional parameters follows the general trend of the \teff\ results in \figurename\ \ref{fig:teff}, that is, our model's prediction of \teff, \logg, and \xh{M} all slightly outperform an expert \texttt{XGBoost} model trained for the same settings (i.e., based on the same inputs). \figurename\ \ref{fig:andrae} provides an additional check of the model performance by comparing to \citet{2023ApJS..267....8A} results for their ``vetted'' giants sample of $\approx 15.6$ millions stars largely outside of the APOGEE survey's footprint that we restricted ourselves to in both our training and test sets. Our model's prediction of \teff, \logg, and \xh{M} on the stars in \figurename\ \ref{fig:andrae} performs slightly worse compared to our model's performance on the test set, but here we only provide the model with XP coefficients without any colors (this is to avoid having to perform cross-matching to find the corresponding \tmass\ colors) while \citet{2023ApJS..267....8A} predictions comes from XP coefficients, synthesized colors, and WISE photometry with \texttt{XGBoost}. Even though the two predictions disagree with each other for many stars, our model assigns a large uncertainty for most of these, indicating that our model is very uncertain in those predictions.

\figurename\ \ref{fig:xp_coeffs} provides a more direct assessment than that of \figurename\ \ref{fig:spec_recon_teff_mh} of our model's performance on the generative task of predicting the \gaia\ XP spectra from stellar parameters, because our model directly outputs only the XP coefficients, not the XP spectra. \figurename\ \ref{fig:xp_coeffs} thus displays the model's prediction for the first few BP and RP coefficients for stars in the test set and these coefficients represent the large- and medium-scale features in the XP spectra. We see that the model's predictions are accurate and precise.

\section{Demonstration of our model as a Foundation Model}\label{sec:foundationdemo}

\begin{figure*}
\centering
\includegraphics[width= \textwidth]{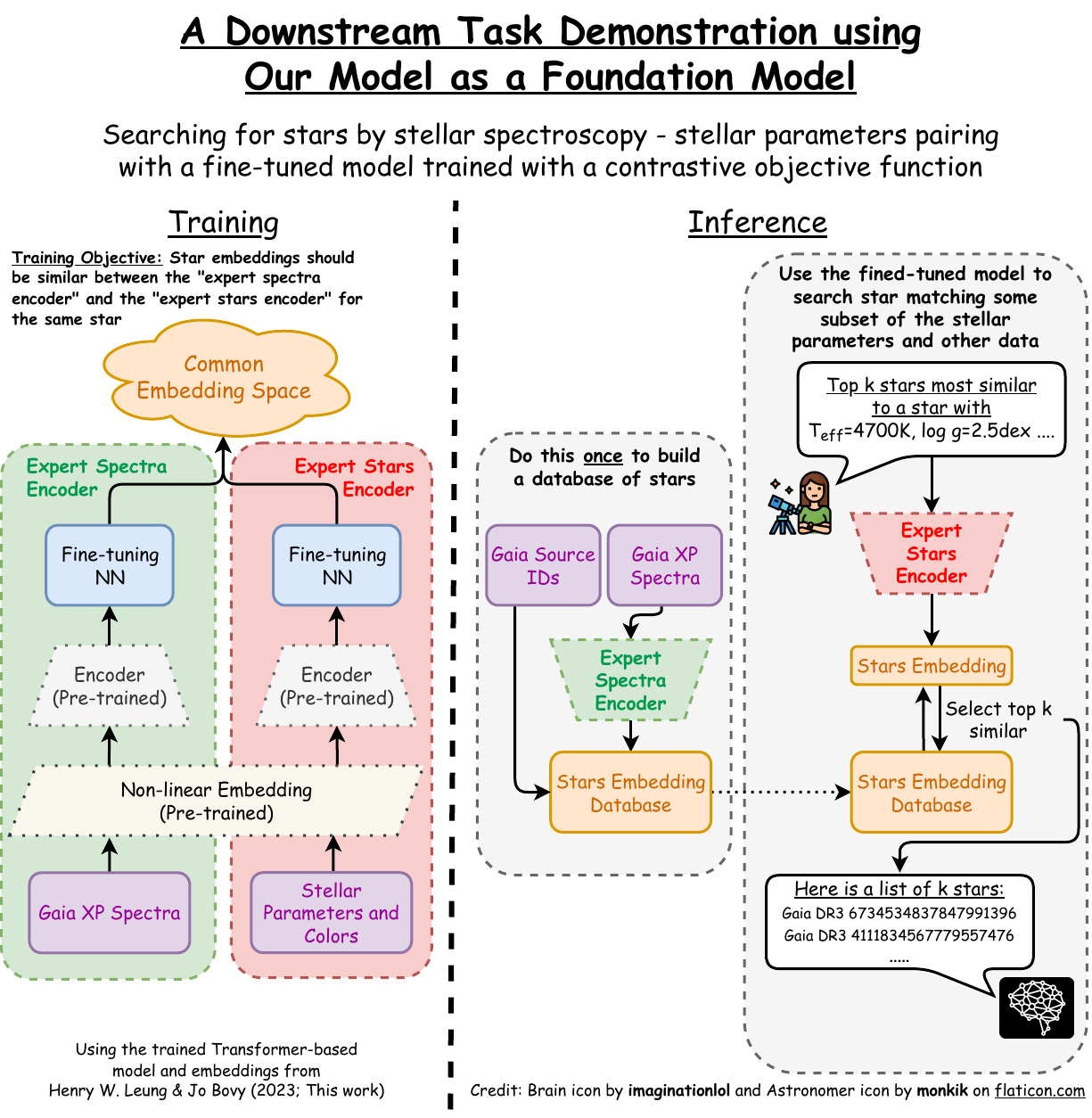}
\caption{Example of using our model as a foundation model on a downstream task. This diagram gives a high-level overview of how a downstream task can use our pre-trained Transformer-based model by showcasing an application of searching for stars in a database with stellar parameters close to a chosen set of values. This is achieved by fine-tuning our model using a contrastive objective (left part) to produce a common embedding space that can be used to build a database of embeddings (middle part) and then searching this database for stars with embeddings similar to those of the chosen stellar parameters (right part). In more detail, the left part of the figure shows the training process which uses two encoders, one an expert stellar-spectra encoder and the other one an expert stellar-parameter/other-observations encoder, which in this case are both the same and given by our trained model's encoder and non-linear embedding, but they could be different. The training objective is to generate a common embedding space for the two expert encoders by using a contrastive loss that encourages the model to generate similar embeddings from the two encoders for the same stars. The right part of the figure shows the inference process where stellar spectra are processed by the fine-tuned expert encoder to generate a database of embeddings, which can be queried by top-$k$ similarity for the embedding corresponding to the chosen set of stellar parameters (and possibly other observations, such as the \tmass\ colors).}
\label{fig:foundationdemo}\label{lastpage} %Put the lastpage label here, because otherwise the page with the figure is missed
\end{figure*}

To demonstrate how one can use our trained Transformer-based model as a foundation model to perform downstream tasks, we implement an example of a downstream task in this appendix. An overview of the tasks and our approach to it is given in \figurename\ \ref{fig:foundationdemo}. The task we pick does not have a strong scientific motivation currently, we mainly choose it to show off the versatility of our model. All code for this example is available at the same GitHub repository for the rest of this paper.

The downstream task that we choose as an example is the problem of searching for stars in a stellar database based on an approximate match to chosen stellar parameters and/or colors. We approach this problem using a contrastive objective function while re-using and fine-tuning trained components of our main model. A contrastive objective function \citep{e6377ee676a34e8eb97cfdc53cd489ef, c441a21f627f4b0a896fa62cb143176f} generally refers to an objective function that is designed to determine whether two samples are similar (e.g., from the same class) or not without involving reconstructing the data as is done in an auto-encoder. Once we train new encoders of stellar spectra and stellar parameters/colors by fine-tuning our main model, we can build a database of embeddings of the stellar spectra of stars which one can then query to determine which stars in the database are most similar in terms of the embedding to a given set of stellar parameters and colors. The new model that we train to generate the new embeddings took only $\approx 10$ minutes to train for a total of 32 epochs. The training procedure that we use is similar to that implemented in OpenAI's Contrastive Language-Image Pre-Training (CLIP; \citealt{2021arXiv210300020R}) using a contrastive objective function within a mini-batch as first proposed by \citet{10.5555/3157096.3157304}.

We use a constrastive objective function in order to encourage the embeddings produced by two expert encoders---one encoding the spectra and the other encoding the stellar parameters and colors---to be similar for stars that are similar as considered by both expert encoders without the need for ground-truth matches or forcing only the same exact stars to be considered to be similar. That is, stars can have similar embeddings without being the same stars physically as long as their properties are similar (this is different from the usual constrastive objective function such as the one implemented in \citet{2021arXiv210300020R} and will return to this point later).

We denote the embeddings generated from the expert spectra encoder as $\hat{y}_j$, which has dimension $m \times d$, and that from the expert stellar parameters/colors encoder as $\hat{y}_k$, which also has dimension  $m \times d$, where $m$ represents the mini-batch size (equal to the number of stars in this mini-batch) and $d$ represents that embedding dimension for the common embedding space (not to be confused with the $d_K$ that we introduced in \secname\ \ref{subsec:attention}). The logits $z$---logits refer to the unnormalized logarithmic probabilities which usually have dimensions of $m \times c$ where $c$ is the number of classes before the application of \texttt{Softmax}---in this application are defined as $z_{jk}$
\begin{equation}\label{eq:contrastivelogits}
  z_{jk} = \hat{y}_j {\hat{y}_k^T}\,,
\end{equation}
which have dimension $m \times m$ representing the dot-product similarity between the embeddings from the two expert encoders (so number of classes $c = m$, because each star is its own class in a mini-batch). These are normalized by applying the \texttt{Softmax} function that takes in logits $z$ and normalizes them such that they sum to one along the dimension representing classes (which is the second dimension if the logits dimension is $m \times c$ as we have discussed). Rather than using the usual \texttt{Softmax} function, we employ a more general \texttt{Softmax} function that has an additional hyper-parameter $\tau$ that is called the temperature, because it controls the entropy of the output probability distribution. The generalized \texttt{Softmax} function for each $z_i$ in $z$ is given by
\begin{equation}\label{eq:softmax}
  \text{Softmax}(z_i, \tau) = \frac{\exp{\big(\frac{z_i}{\tau}}\big)}{\sum_{h=1}^c\exp{\big(\frac{z_h}{\tau}\big)}}\,,
\end{equation}
where the conventional \texttt{Softmax} function corresponds to $\tau=1$. For high temperature (which has high entropy), this generalized \texttt{Softmax} output probability distribution is more uniform and vice versa for lower temperature. In the present application, we use a cold temperature $\tau=0.05$.  Thus, we effectively require stars to be very similar for them to be considered similar by our objective function, because the output probability distribution has low entropy. To create a contrastive objective function from the \texttt{Softmax} outputs, we use the \texttt{CrossEntropy} function between two probability distributions over a mini-batch, defined as
\begin{equation}\label{eq:crossentropy}
  \text{CrossEntropy}(P, Q) = - \frac{1}{m} \sum_{i=1}^m \sum_{h=1}^c p_{i, h}\ln {q_{i,h}}\,,
\end{equation}
where $P$ will be the logits and $Q$ the ground truth probabilities both with dimension of $m \times c$.

While the procedure that we use is overall  similar to that in \citet{2021arXiv210300020R},  we do not force only the exact same stars to be considered similar (i.e., setting the ground-truth matrix to be a diagonal matrix). Instead, we define the ``ground truth'' as the average of the dot product similarity from the two encoders, which are $\hat{y}_j$ to itself (i.e., what is the similarity of the embedding $\hat{y}_j$ of the spectra for each star with the embedding of the spectra of every other star) and $\hat{y}_k$ to itself (i.e. ,what is the similarity of the embedding $\hat{y}_k$ of the stellar parameters/colors for each star with the same embedding for every other star), respectively. Hence the ``ground truth'' is, 
\begin{equation}\label{eq:contrastivelossGT}
  y = \text{Softmax} \bigg(\frac{y_j {y_j}^T +  y_k {y_k}^T}{2}, \tau\bigg)\,.
\end{equation}

where $y$ has dimension of $m \times m$.

Now that we have defined the ``ground truth'' and the logits, we can calculate cross-entropy loss for the two encoders,
\begin{equation}\label{eq:contrastivelossCE}
\begin{split}
  J_j &= \text{CrossEntropy} \big(z_{jk}, y\big)\,, \\
  J_k &= \text{CrossEntropy} \big({z_{jk}}^T, {y}^T\big)\,,
\end{split}
\end{equation}
where $J_j$ refers to the constrastive loss from stellar spectra to stellar parameters and $J_k$ refers to the constrastive loss from stellar parameters to stellar spectra. The final loss $J$ is defined as the average of two losses, that is
\begin{equation}\label{eq:contrastiveloss}
J = \frac{(J_j + J_k)}{2}\,.
\end{equation}
We use this objective function to train the ``Fine-tuning NN'' parts of the encoders shown in the left part of \figurename\ \ref{fig:foundationdemo}.

After training the embeddings, we can evaluate the performance of the embeddings using a metric called the top-$N$ recovery rate. The top-$N$ recovery rate refers to the percentage of times that a given star is within the top-$N$ most similar matches when comparing to the entire database using the dot-product similarity of its embedding. The top-100 recovery rate using stellar parameters and colors (excluding \bprp\ to prevent information leakage) among a test set of embeddings of \gaia\ XP spectra of $\approx 25,000$ stars is $\approx12\%$ (compared to the completely random baseline of $0.4\%$). Thus, our fine-tuned model is able to match up stellar parameters/colors and stellar spectra.

Furthermore, we do not have to give a certain set of inputs like a traditional neural network model, because the downstream task can benefit from the flexibility of the pre-trained Transformer-based model and we do not need to use the stellar parameters for real stars. For example, we can simply give $\teff=4,700\,\mathrm{K}$ and $\logg=2.5\,\mathrm{dex}$ and request the spectra for the top-100 stars in the database with embeddings most similar to those of the given stellar parameters. For these stars, we can check against the APOGEE ground truth and we find that the list of stars have observed \teff\ and \logg\ around 4700 K and 2.5 dex. Or one can simply input $\jh=0.6$, for which the model effectively needs to internally assume parameters such as \teff\ and the reddening, and similarly obtain the top-100 spectra with most similar embeddings and we get stars with \jh\ around 0.6 mag. Thus, this fine-tuned model allows one to search for stars with approximate matches to stellar parameters and colors in a database of spectra.

By showing this example of how to use our pre-trained Transformation-based model to perform a downstream task, we hope that the reader has a better idea of what the existence of a foundation model can mean in practice for downstream users.

%%%%%%%%%%%%%%%%%%%%%%%%%%%%%%%%%%%%%%%%%%%%%%%%%%

% Don't change these lines
\bsp	% typesetting comment
\end{document}